# TITLE: MAPPING R&D SUPPORT INFRASTRUCTURES: A SCIENTOMETRIC AND WEBOMETRIC STUDY OF UK SCIENCE PARKS

David Minguillo, M.Phil.

A thesis submitted in partial fulfilment of the
requirements of the University of Wolverhampton
for the degree of Doctor of Philosophy

October 2013

This work or any part thereof has not previously been presented in any form to the University or to any other body whether for the purposes of assessment, publication or for any other purpose (unless otherwise indicated). Save for any express acknowledgments, references and/or bibliographies cited in the work, I confirm that the intellectual content of the work is the result of my own efforts and of no other person.

The right of David Minguillo to be identified as author of this work is asserted in accordance with ss.77 and 78 of the Copyright, Designs and Patents Act 1988. At this date copyright is owned by the author.

Signature …………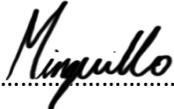……………………..

Date ……………17/10/2013……………....

# Abstract

Academic research is increasingly considered to be a fundamental socio-economic asset. This is turning academia into the central actor of a complex network in which interactions with industry and government are fundamental to exploit research in order to create an organizational system of innovation. To facilitate this difficult task, innovation support infrastructures known as science parks (SPs) provide micro-environments containing diverse organisations fostering collaborative networks and boosting innovation. However, there is no evidence about SPs' roles in supporting the commercialization of academic knowledge, their effects on the entrepreneurial identity of Higher Education Institutions (HEIs), or their intermediary roles in linking together diverse sets of organisations. In response, this thesis analyses UK SPs with a scientometric and webometric approach to study (1) the role of public science and HEIs in research and development (R&D) networks associated with SPs, and (2) the web-based patterns that reflect the configuration of R&D support infrastructures associated with SPs.

Bibliographic data from Scopus and other secondary data sources were used to investigate: (a) the degree of R&D within SPs, (b) the role played by HEIs in these R&D collaboration networks, and (c) SPs' ability to link academia and on-park firms. R&D production was found to be highly concentrated in terms of infrastructures and regions. HEIs were the main external sources of knowledge, higher quality HEIs interacted more with on-park firms, and HEIs' third stream activities resulting from a strong research base strongly associated with their R&D capacity. Nevertheless, HEIs' formal associations with SPs did not strengthen their R&D capacity or U-I synergy so SPs might not be the right policy tools to create U-I synergy across the country.

Hyperlinks were used to explore: (a) a more reliable method to identify R&D networks, and (b) if links can reveal potential offline behaviours of the different actors involved around SPs. Link analysis was not found to be a reliable method although the new webometric method partially reduced the bias inherent in webometric studies. The interlinking networks created were able to reflect offline SP structures to some extent and the presence and potential behaviours of institutional agents involved in SPs could also be identified.

In conclusion, informetric approaches can uncover otherwise hidden dimensions of SPs, providing greater insights into the multi-level and complex interactions associated with them. They can identify interactions between academia and on-park industry, the different types of actors involved, and the networks established by the broad ties between the diverse organisations that operate in innovation systems.

# Table of contents

## List of tables

## List of figures

## Abbreviations

| | |
|---|---|
| AMP | Advanced Manufacturing Park |
| AMRC | Advanced Manufacturing Research Centre |
| BPs | Business Parks |
| ERDF | European Regional Development Fund |
| HEFCE | Higher Education Founding Council for England |
| HEIs | Higher Education Institutions |
| LIC | Leeds Innovation Centre |
| NTFBs | New technology-based businesses |
| PROs / RIs | (Public) Research institutes |
| R&T | Research and technology |
| RDA | Regional development agency |
| SPs | Science parks |
| SMEs | Small and medium size enterprises |
| SNA | Social Network Analysis |
| TH | Triple Helix |
| UKSPA | United Kingdom Science Park Association |
| U-I-G | University-Industry-Government |
| YSP | York Science Park |
| WAINNOVA | World Alliance for Innovation |

# Publications from this Thesis

**Journal papers:**

- Minguillo, D. & Thelwall, M. (in preparation). Mapping the university-industry ties in the UK: role of public science for science parks.
- Minguillo, D. & Thelwall, M. (under review). Are science parks encouraging a more entrepreneurial University in the UK?.
- Minguillo, D.; Tijssen, R.; Thelwall, M. (under review). Are Science Parks involved in research and technology production?: Overview of scientific output associated with UK science parks.
- Minguillo, D. & Thelwall, M. (2012). Mapping the network structure of science parks: An exploratory study of cross-sectoral interactions reflected on the web. *Aslib Proceedings, Vol. 64 Iss: 4,* 332 – 357.

**Conference proceedings:**

- Minguillo, D. & Thelwall, M. (2013). Industry research production and linkages with academia: Evidence from UK science parks. *ISSI 2013 Conference - 14th International Conference of the International Society for Scientometrics & Informetrics.* (Vienna, Austria, July 15-18).
- Minguillo, D. & Thelwall, M. (2011). "The entrepreneurial role of the University: a link analysis of York Science Park", in: Noyons, E., Ngulube, P., Leta, J. (Eds.), *Proceedings of the ISSI 2011 Conference - 13th International Conference of the International Society for Scientometrics & Informetrics*, Durban, South Africa, July 4-8. South Africa, pp. 570-583.
- Minguillo, D. & Thelwall, M. (2011). Mapping the network structure of science parks: An exploratory study of S&T and the cross-sectoral interactions of dynamic regions on the web. *6th International Conference on Webometrics Informetrics and Scientometrics & 11th COLLNET Meeting* (Mysore, India, October 19-22).

# Acknowledgments

I would like to thank Prof. Mike Thelwall for being an extraordinary supervisor and unique person. All his valuable experience, guidance, advice and encouragement have made this long process much easier for me. I feel very fortunate to have had the opportunity to work with such a brilliant and humble researcher.

I would also like to thank all my colleagues and members of the Statistical Cybermetrics Research Group for their very helpful, inspirational and constructive feedback at the different stages of my project.

My deepest gratitude also goes to my examiners Prof. Ismael Rafols and Kim Holmberg for their insightful comments.

# Chapter 1: Introduction

## 1.1 Background

The production of scientific knowledge and its technological application has become a critical source of power and productivity for the modern networked society (Willetts, 2013). New models of production and economic development based on knowledge-intensive high technology industries (Castells, 2000; Etzkowitz & Leydesdorff, 2000; Nowotny, Scott, & Gibbons, 2001) have emerged to tackle this new environment which is shaped by the complexity and uncertainty of a global and highly competitive economy. This need to translate academic knowledge into technological applications as a source of social and economic development has transformed the traditional teaching and research role of the university. This new social mission is turning academia into the central actor of a complex network in which interactions with industrial and governmental spheres are fundamental to commercialize research and create an organizational system of innovation (Etzkowitz, 2006; Godin & Gingras, 2000). In this context, government support can be seen in the adoption of programs and policies which promote and strengthen mutual dependence between research institutions and enterprises, leading to the decline of institutional boundaries and the emergence of a triadic sub-network of actors which work together for a common benefit (Etzkowitz, 2008). The growth of cross-sectoral interactions is promoted by science, technology and innovation policies which create organizational mechanisms for innovation.

The UK, like most developed countries, has increasingly adopted research and development (R&D) programs to promote and strengthen interactions between research institutions and companies in order to exploit the excellent research and technology base of the country (Lambert, 2003; Sainsbury, 1999; Wilson, 2012). Some of these initiatives have created physical spaces, such as science parks, research parks, technology parks, incubators or similar support infrastructures, which provide a micro-environment to agglomerate and interconnect organisations belonging to the public, private and academic sectors (Gordon & McCann, 2000). These artificial environments bridge together a wide range of interdependent actors, such as academic and research institutions, hospitals, research groups, spin-offs, investors, public and non-profit organizations, and also offer the competence and resources to facilitate the transfer and application of knowledge, which promotes the development and commercialisation of innovative services and products. All this should eventually lead to more sustainable socio-economic development at regional and national levels (Breschi & Catalini, 2010; Etzkowitz, 2008; Phan, 2005).

Science parks (SPs) started to appear in most developed countries in the 1980s and have since been rapidly adopted by emerging economies. Policy makers increasingly consider SPs to be a key policy tool for raising the level of technological sophistication and reactivating local economies, thus moving towards a knowledge intensive economy (Cumbers & Mackinnon, 2004; Porter, 2000). As a source of research-intensive firms that attract foreign investments, SPs are expected to nurture the transition to a knowledge intensive economy, catalyse regional economic growth and social well-being, and increase the technological sophistication of local industries. Consequently, the networks formed by SPs, in vibrant regions, tend to be embedded in larger infrastructures, called clusters, which specialize in a high-technology area (Saublens, 2007). Industrial clusters are larger and more complex networks that concentrate the resources and agents necessary to gain international competitiveness and visibility (Cooke & Leydesdorff, 2006a; Porter, 1998).

In the UK, the SP building boom started in the 1980s with investment in the construction of the parks, but with a parallel reduction in research funding, undermining the strengths that SPs were designed to exploit. This was a policy initiative from the government to promote an industrial resurgence and create job opportunities to overcome the severe recession and unemployment of 79-81 (Quintas, et al., 1992). The transformation of polytechnics into universities occurred at the same time and with similar goals (DS Siegel, Westhead, & Wright, 2003). Today, SPs play a central role in the innovation ecosystem and national science policy and they are expected to foster the creation of local and regional high technology clusters that can help to rebalance the economy and narrow the economic gap between the UK North and South.

## 1.2 Research problem

Despite significant academic and public expenditure on SPs, their economic value is still unproven (Hassink & Hu, 2012; Löfsten & Lindelöf, 2002; Quintas et al., 1992; Soetanto & Jack, 2011). Most literature assessing SPs focuses on a single dimension of the phenomenon, such as the performance of companies in SPs, knowledge transfer, or an abstract system-level analysis. There is thus a growing need to investigate SPs' roles in supporting the commercialization of academic knowledge and technology, their effect on the entrepreneurial mission of Higher Education Institutions (HEIs), and the promotion of social wellbeing (A. N. Link & Scott, 2003). Another important dimension that has been neglected is the intermediary role of SPs in linking together the heterogeneous set of organisations necessary to foster open innovation (Bøllingtoft & Ulhøi, 2005).

The lack of an assessment framework that takes into account both the impact of SPs on HEIs' entrepreneurial missions and the relational and multi-dimensional nature of SPs makes it more difficult to understand and assess their capacity to strengthen the cross-sectoral connections that are necessary to generate highly innovative environments. In order to fill this gap and gain a fuller understanding of these support infrastructures, this thesis introduces a scientometric and webometric approach to explore: the degree of research activity within SPs and the interactions fostered between on-park firms and HEIs; and the web-based links between the different organisations associated with SPs, and their relationship with off-line interactions.

Scientometric techniques, including quantitative indicators of research performance from the analysis of scholarly communication in form of publications and patents (Moed, 2005), can provide evidence about the research intensity of SPs and U-SP interactions. This makes it possible to add quantitative evidence to the mostly qualitative evidence that can be found about the knowledge transfer associated with SPs. Webometrics, on the other hand, is a new research area whose main focus is on analyzing and understanding patterns of scientific and social interactions and behaviours on the Web (Thelwall, 2004a). It can help to gain a first overview of SPs' ability to foster cross-sectoral and cross-disciplinary interactions, which are likely aid the survival and success of on-park companies. Although webometric techniques are still experimental and provide evidence that lacks sufficient reliability for the decision making process, this approach can shed light on complex networks that are difficult to track and analyse through traditional scientometric indicators.

## 1.3 Aims and objectives

The main aim of this thesis is to assess the use of webometric and scientometric methods to analyse UK SPs and similar research-intensive environments. This aim is guided by the following key research questions: (1) what insights can scientometric methods give into the role of public science and HEIs in R&D networks associated with SPs?; (2) can web-based patterns reflect the configuration of R&D support infrastructures associated with science parks?.

The main objectives are as follows:

1. ***To use scientific publications to map R&D fostered by SPs across the UK.***
   The analysis of research publications produced by on-park organisations makes it possible to identify: (1) what types of innovation infrastructures are established across the country and which physical and organisational infrastructures are the most research-intensive; (2) where these infrastructures are established and which type of organisations engage in these environments; (3) how on-site organisations collaborate, whether collaborations occur between on-park firms and knowledge producers, and how these linkages extend across the country and beyond; (4) which research areas attract most on-park research and the contribution of the geographic regions and U-I collaboration across different areas; (5) whether the research production associated with SPs has higher quality and impact than the average for research across the different areas. (Dataset & Methods: Section 3.3.2.; Results & Discussion: Section 4.2)

2. ***To examine the structural organisation and role played by HEIs in R&D collaboration networks at different levels.***
   This will help to identify (1) the main structural properties of the inter-institutional collaboration network of on-park organisations; (2) the main structural properties and role of HEIs in the inter-institutional collaboration network of on-park and off-park organisations; (3) the main structural properties and role of HEIs, as the main external actors, in the firms-HEIs-RIs inter-institutional collaboration network.

3. ***To determine whether HEIs' research performances are related to the strength of U-I collaborations.***
   This will help to determine whether there is an association between HEI prestige and research excellence and stronger links with on-park firms. (Dataset: Section 3.3.2.; Methods: Section 3.3.5. & 5.2; Results & Discussion Section: 5.3-4)

4. ***To determine whether UK SPs are able to bridge between academia and on-park firms.***
   This will help to uncover (1) how long after SPs are established they start to promote research production and U-I cross-fertilization; (2) the factors which help R&D production and U-I collaboration within SPs; (3) whether HEIs with a formal association with SPs have a greater capacity to produce R&D and collaborate with firms, and whether other factors have a greater influence on U-I research interaction. (Dataset: Section 3.3.2-3.; Methods: Section 3.3.4; Results & Discussion Section: 6.2-3)

5. ***To develop a methodology for collecting and analysing web-based interlinking structures to identify R&D networks that are supported and tied together by SPs.***
   This will help to determine to what extent hyperlinks can detect the mutual and diversified interactions that exist within organizational innovation structures, and how these heterogeneous organisations are interconnected. (Section 3.3.6-8. & 7.3.1-2)

6. ***To compare the key features of the web-based structures with off-line features which are widely applied to evaluate SPs and the R&D conditions regionally in the UK.***
   This will help to determine the validity and viability of the webometric approach to represent and analyse the mutual interactions among the heterogeneous organisations involved in R&D networks. (Dataset: Section 3.3.6.; Methods: Section 3.3.7-8.; Results & Discussion Section: 7.3-4)

7. ***To design the SP actor framework.***

    This new framework, based on the Triple Helix model, lists the key actors that should be involved in the SPs and identifies their missions, functions, and potential interactions.

8. ***To compare this framework with the network created by the hyperlinks among the actors within SPs.***

    This will help to determine if links can reveal potential offline behaviours and whether the identified patterns are in line with the results obtained with traditional indicators. The latter makes sense because hyperlinks can reflect social interactions and represent an underlying social structure (Reid, 2003). (Dataset: Section 3.3.6.; Methods: Section 3.3.7-.8. & 8.2; Results & Discussion: Section 8.4)

The evidence obtained from this doctoral project represents the first step toward extending the analytical framework used to evaluate SPs, and will provide evidence that can reduce the risks involved in the development of policies oriented to develop an extensive network of SPs in the UK. Furthermore, this may help to broaden the indicators used to assess the entrepreneurial and social functions of universities (M. T. Larsen, 2010) and hence, fill the gap in traditional indicators to evaluate the inherent multidimensional and synergetic nature of triple helix links (Etzkowitz, 2008).

## 1.4 Contribution to knowledge

The main novelty and contribution of this thesis is the application and assessment of an informetric (scientometric and webometric) approach to the analysis of innovation support infrastructures. At the same time this project seeks to extend the application of webometric and bibliometric indicators beyond the Library and Information research community.

## 1.5 Thesis structure

This thesis is divided into four main parts:

- General introduction and literature review (Chapter 1 and 2)
- Global research questions and methodology (Chapter 3)
- Empirical chapters (Chapters 4, 5, 6, 7, 8)
- Main conclusions (Chapter 9)

### 1.5.1. General introduction

The first part of the thesis is the introduction and literature review. The introduction (chapter 1) presents the background, research problem, and the different objectives established to reach the main aims of this project. The literature review (chapter 2) briefly discusses the research that is related to the thesis and the empirical evidence that leads to the research design and empirical studies. The literature review is framed within the importance of public science as a source of economic competence, the university third mission, university and industry collaboration, innovation support infrastructures (SPs), SPs' role as facilitators of university and industry interaction and R&D transfer, and SPs' roles as connectors of heterogeneous actors involved in innovation systems. Finally the webometric literature dedicated to the study of R&D networks and link analysis is also discussed.

### 1.5.2. Global research questions and methodology

Chapter 3 is divided into two main sections. The first section lists the rationale behind the different research questions that arise from the overall aim and objectives and that shape the investigations of this thesis. The second section describes the main methods adopted in the different empirical studies to best answer the different research questions. It also describes the population of study, how the structured and unstructured data was collected, processed and cleaned, the secondary data sources used, the statistical and other techniques applied, and the method developed to create a more robust interlink network.

### 1.5.3. Empirical chapters

The empirical chapters form the principal part of this thesis. They provide the results, discussion and conclusions from four published and three under review studies that have been undertaken to meet the different objectives of this project.

Chapter 4 is derived from two publications (Minguillo & Thelwall, 2013; Minguillo, Tijssen, & Thelwall, under review) that employ structured data to map the capacity of the UK SP movement to encourage and generate R&D and then, start examining whether publications could be used to monitor the R&D ability fostered by support infrastructures.

Chapter 5 is derived from one publication (Minguillo & Thelwall, under review-a) that is based on collaborative links to represent the relational structure and extent SPs rely on public science to get involved in R&D activities at three different levels (on-park organisations; SP and off-park organisations; firms-HEIs-RIs). It also investigates whether HEI research quality strengthens collaborative linkages with on-park tenants.

Chapter 6 is derived from one publication (Minguillo & Thelwall, under review-b) that focuses on determining the time required by SPs to promote research activities and the factors that may influence it, and whether HEIs formal association with SPs benefit HEIs' R&D and knowledge transfer activities.

Chapter 7 is derived from two publications (Minguillo & Thelwall, 2011a, 2012) that employ unstructured data to carry out an exploratory study of three science parks from Yorkshire and the Humber in the UK. Here a new methodology is developed and applied for collecting and analysing web-based interlinking structures to identify R&D networks that are supported and tied together by SPs and also determine whether this webometric approach gives plausible results.

Chapter 8 is derived from one publication (Minguillo & Thelwall, 2011b) that introduces a new framework which lists the key actors that should be involved in the SPs and identifies their missions, functions, and potential interactions. This framework is compared with the hyperlink network of organisations associated with York Science Park to determine if the web links reveal potential offline behaviours and whether the identified patterns are in line with the results obtained with traditional indicators.

### 1.5.4. Main conclusions

The last part of the thesis (chapter 9) draws conclusions about the ability of an informetric approach to provide reliable information on SPs' R&D activities and networking. Finally, the main limitations and future areas of research are also described in this section.

# Chapter 2: Literature review

## 2.1 Introduction

Regional innovation systems are formed by the interplay of two subsystems; a subsystem of knowledge generation and a subsystem of knowledge exploitation. The interaction within and outside this localised system by means of resources, personnel, and administrative authority creates collaborative associations while facilitating knowledge transfer between both subsystems. It forms network linkages among distinctive communities rooted in academia, government and industry in a given territory to build regional knowledge capabilities and systemic innovation strengths that reduce the significant risk in accessing knowledge sources and overcome intra-firm knowledge asymmetries (Cooke, 2005a). These systems are actively promoted because innovation is not a linear or uniform process but is a knowledge spiral stimulated by regional capabilities that derive from dynamic knowledge networking capabilities comprising a wide set of institutional structures, interactions (like governance, knowledge, productive) and forces (like macro-economic, political, social and ecological). This interactive process transforms a learning region (i.e. system of collaborative learning in which dynamic knowledge networking capabilities of a wide range of institutional mechanisms and social conventions play a role) into a regional system of innovation (Caraça, Lundvall, & Mendonça, 2009; Cooke, 2005b). In today's knowledge-based economy these dynamic and collaborative efforts to stimulate regional learning networks are usually realised at local and regional levels with national-level support. They are intended to respond in a more adaptable, competitive and innovative way to the fast pace of change while taking advantage of the agglomeration of specific knowledge expertise, resources, entrepreneurship to become a competitive region and a source of regional renewal and development (Cooke & Leydesdorff, 2006b; Cooke, 2005a; Porter, 2000).

The dynamics of regional innovation networks have also been conceptualised by the Triple Helix model developed by Etzkowitz and Leydesdorff (2000), who describe the prominent economic role of knowledge producing institutions and how close and systematic collaboration between actors with roots in the University, Industry, and Government (U-I-G) sectors leads them to re-invent and re-combine their missions and functions. This hybridization process among the institutional spheres promotes the linkages and cooperation necessary to encourage the commercialization of scientific research and the development of an organizational innovation structure. Similarly, from the network society perspective Castells (2009) states that the different institutional spheres are considered to be programmed networks with their own goals, values, communication standards, and interests. These networks "are open-ended and multi-edged, and their expansion or contraction depends on the compatibility or competition between the interest and values programmed into each network and the interests and values programmed into the networks they come into contact with in their expansionary movement." (2009: 19). They arise from local conditions but need to adapt themselves to integrate into a global network in order to prevent being globally excluded, being local and global at the same time. The particularities embedded in these networks hinder an active interchange and cooperation and thus each co-exist and co-evolve with weak links to each other. They do not merge, instead "they engage in strategies of partnerships and competition, practicing cooperation and competition simultaneously by forming ad hoc networks around specific projects and by changing partners depending on their interest in each context and in each moment in time" (2009: 426).

As a result of globalised competitive forces, research-intensive firms are forced to get involved with innovation processes despite the fact that they often cannot afford large-scale research infrastructures nor rely on their limited in-house research. Research-intensive firms become largely interdependent on the surrounding environment and particularly on knowledge producers, such as specialised R&D companies, Higher Education Institutions (HEIs) and RIs (Nowotny, Scott, & Gibbons, 2001; 71-2). Significant efforts from policymakers, along with HEIs and investors, to support the growth of strong high-technology industries have led to the systematic establishment of supporting infrastructures (i.e. science, research, technology parks or incubators) in the UK (Dyson, 2010; Hauser, 2010; Lambert, 2003). These infrastructures, widely known as science parks (SPs), are physical environments associated with a research institution which appear at the interface of these triadic (U-I-G) relationships as platforms of interaction to support a complex synergy between the three institutional spheres with the help of a set of intermediary organizations. SPs are supposed (1) to actively forge and strengthen collaboration between HEIs and on-park organisations to boost research and technology exploitation, and (2) to provide support and networking activities to promote a close institutional collaboration among diverse actors which operate in complex and heterogeneous R&D networks. This intermediary role filled by SPs in regional innovation systems and complex R&D networks is expected to promote the growth of cutting-edge industries and a socio-economic development at local and regional levels.

However, the few studies on the capacity of these intermediary infrastructures to strengthen the university and industry (U-I) interaction tend to focus on only one of the partners, namely industry, and the evidence found mainly suggests that SPs are not effectively fostering research transfer (Bakouros, et al., 2002; Quintas, et al., 1992; Radosevic & Myrzakhmet, 2009; Siegel, et al., 2003; Westhead & Storey, 1995). On the other hand, the lack of data available and the lack of a systematic framework to study the complexity and diversity of linkages established within and across stakeholders like industry, HEIs, government and other intermediary organisations makes it difficult to uncover the relational structure fostered by SPs (Bøllingtoft & Ulhøi, 2005, 2010; Bøllingtoft, 2012; Hansen, et al., 2010; Phan, 2005). Therefore, this thesis seeks to fill these two important gaps in the literature through the introduction of scientometric and webometric approaches in the study of SPs. First, scientometric methods are applied to analyse the knowledge created within SPs and transferred to them. More specifically, this part of the thesis aims to assess the extent to which SPs promote R&T activities and strengthen U-I linkages, as well as the role of public science and HEIs in R&D networks associated to SPs. Second, a webometric approach is applied to tackle the complexity of the networks established by SPs and uncover SPs' intermediary role by identifying the networking function of SPs and determine whether web-based networks could reflect the offline configuration of R&D support infrastructures associated with SPs.

## 2.2 Public science as a source of economic competence

While cutting-edge sciences and technologies are becoming increasingly important for regional competitiveness and economic growth, the relationships between producers and consumers of research-based knowledge and skills are becoming a central issue for the development of policies and regional innovation strategies that aim to increase the absorptive capacity and application of scientific research by fostering closer linkages between university and industry (Cooke, 2002; Etzkowitz, 2008). For instance, the EU's Competitiveness and Innovation Framework Programme (CIP) aims to foster successful engagements between both communities to enhance competitiveness and integration in the region (European Commission, 2012). In the case of the United Kingdom (UK), with its shrinking manufacturing sector and high dependence on the service sector, there is a need for upgrading and diversifying the manufacturing sector to compete with advanced and globalising economies (Porter & Ketels, 2003). UK efforts are currently focusing on exploiting its prominent position in high-value added sectors of manufacturing, such as aerospace, pharmaceuticals and biotechnology (Willetts, 2013). Another important focal point is to improve the process of translating impact and output of the UK science base into economic value (Dyson, 2010; Hauser, 2010; Lambert, 2003).

This social and institutional demand for research in the learning economy (i.e. the new economy that is characterised by a speed up in the rate of change giving a stronger importance to the capacity of people, organisations, networks and regions to use the full potential of new knowledge and technologies to develop learning processes that facilitates economic performance (Lundvall & Johnson, 1994)) has in turn led universities to actively get involved in entrepreneurial activities to capitalise on their research activities (Clarysse, et al., 2011; Larsen, 2011; Lundvall, et al., 2002), and increasingly also to operate as a bridge to new technology-based firms (George, et al., 2002). Universities have become more open to partnerships with private organisations to gain social recognition and access to external funds (PACEC, 2009). Moreover, the limited in-house knowledge of most businesses suggests that their innovative capacity will increasingly be dependent upon *open innovation* (Chesbrough, 2003; Narin & Hamilton, 1997): harnessing external knowledge and skills in conjunction with their internal R&D to commercialise new technological advances. This can occur through pathways inside or outside of a specific business and means that the boundary between a firm and its surrounding environment has become more porous and the output of the R&D activities more diversified. In this context, intermediary infrastructures established to foster dynamism and synergy among and between firms, knowledge producers and government, such as science- research- and technology parks, seem to be ideal seedbeds for innovation (Felsenstein, 1994; Squicciarini, 2009). The study of these innovation infrastructures is therefore fundamental to assess to what extent they facilitate U-I collaboration to help boost technological innovation and generate local socio-economic growth.

### 2.2.1. University Third Mission

High competition and rapid technology development have led to academic research being regarded as an attractive foundation for innovation in knowledge intensive industries. The need for specialised, localised and diversified knowledge accumulation for regional development is based on the rationale that new knowledge is the building block of regional growth in the form of a concentration of research resources on a particular topic from which technological ideas can be commercialized (Cooke, 2005; Etzkowitz, et al., 2000; Etzkowitz, 2008; Nowotny, et al., 2001). This has resulted in a growing interest from research councils and governments in fostering the establishment of close links between university and industry (U-I) to facilitate effective research and technology transfer (Wilson, 2012). This central role of HEIs, as the main source of knowledge, is particularly emphasised in the conceptualizations of knowledge-based innovation production systems like Mode 2 (Nowotny et al., 2001) and the Triple Helix model (Etzkowitz & Leydesdorff, 2000; Etzkowitz, 2008).

The new production of knowledge described as Mode 2 views science as the main driving force in the transformation of society during the past two centuries. This unidirectional interaction has not only led to a scientification of society but also a socialisation of science. Consequently, science and society have become transgressive institutional spheres, following co-evolutionary trends. Universities are no longer intimately associated with the dissemination of new knowledge and production of professional elites. Instead, universities have now embraced more 'transdisciplinary' and contextualised forms of research as a result of scientific communities becoming more diffused, the boundaries with R&D systems being more porous, and a much wider range of social and economic activities having research components. This socio-economic demand means that universities have to be flexible and accommodate themselves to new configurations of knowledge by establishing novel alliances with other research-intensive organisations. Universities must also become both open and synergetic to be able to exploit academic knowledge (Nowotny et al., 2001). Similarly, the Triple Helix (TH) model, for example, states that the university is the central actor of a complex network in which intensive interactions and overlapping roles with industrial and governmental spheres are fundamental to create an organizational system of innovation that promotes knowledge-based economic development. The trilateral relationships between the institutional spheres are reshaped and supported by intermediary organizations with a quasi-governmental, quasi-industrial and quasi-academic nature that perform hybrid tasks and functions. They are a support mechanism specialised in fostering strong and mutual knowledge flows and correcting any local conditions and elements that may hinder the innovation process (Etzkowitz & Leydesdorff, 2000; Etzkowitz, 2006, 2008). Thus, the more these institutional spheres establish interactive relations between them the greater the need for intermediary and hybrid organizations, as shown by the growing number of SPs arising around the globe.

Even though these two conceptualisations to some extent rely on a linear model of innovation and are widely accepted in the informetric research community, systematic efforts from the management and business community to understand innovation systems focus on the learning capability and the competence building of individuals, organisations and regions based on external knowledge (Chesbrough & Appleyard, 2007; Cooke, 2001a; Lundvall et al., 2002; Ikujirō Nonaka & Takeuchi, 1995; Wenger, 2000) rather than on the central role of academic knowledge in innovations systems.

### 2.2.2.    University and Industry (U-I) Collaboration

A growing scholarly interest in understanding U-I links has led to a wide application of scientometric techniques to investigate U-I relationships. As Toivanen and Ponomariov (2011: 473) states, "the primary factor behind the importance of scientific collaboration lays simply in its role as a channel of knowledge flows between scientists. Innovation and creativity are dependent on the availability of ideas which can then be recombined and developed into new knowledge, and collaboration—individual, institutional, and international—is a primary setting for harnessing and developing useful ideas". Thus, U-I interaction can be considered as an evolving trend for advancing knowledge and new technologies which emerge as result of an interdependent process of knowledge exchange and collaboration (Santoro & Chakrabarti, 2002). Science and technology publications are widely employed to study and track the joint collaboration of private and academic researchers through co-authorships, representing a form of direct interaction and knowledge transfer between both communities (Abramo, et al., 2011; Cockburn & Henderson, 1998; Liang, et al., 2011; Santoro & Chakrabarti, 2002; Tijssen, et al., 2009; Zheng et al., 2011). Similarly, the analysis of U-I co-authorship in patents or co-inventions are used to measure the direct influence of public research (Hung, 2012; Narin & Hamilton, 1997; Zheng et al., 2011), and also the indirect and general influence of open science through citations to publications from patents (Meyer, 2000; Tijssen, 2001). Although most of these studies focus on collaboration at institutional and organisational levels, publications and patents have also been used to identify 'author-inventors' or 'star researchers', who participate in co-publication and/or co-invention and are expected to act as knowledge brokers between both communities (Breschi & Catalini, 2010).

The multiple interaction channels embedded in these U-I links have also led to quantitative and qualitative studies of alliances (Gay & Dousset, 2005), formal or informal interactions, and a combination of the different types of links (Beise & Stahl, 1999; Breschi & Catalini, 2010; Mansfield, 1991; Tijssen & Korevaar, 1997). In general, the evidence indicates that the industrial sector's dependence on public science has located the university at the core of intersectorial research production (Godin & Gingras, 2000), leading to more successful companies in terms of the number of products developed and commercialised (Zucker, Darby, & Armstrong, 2002) in research-intensive industries. In the pharmaceutical industry, for example, academia is perceived as a key source of information (Arundel & Geuna, 2004; Cohen, et al., 2002; HEFCE, 2010; Malo, 2009). On the other hand, the academic community has industry as its most important external partner and at the strategic level considers research collaboration and technology transfer as the second and fourth most important activities, respectively. This growing partnership, however, needs to be examined through publications since the majority of academics still want their work on supporting third-stream activities to gain visibility in the research community (HEFCE, 2010a). For a detailed review of the wide empirical nature of studies on U-I interaction, see Teixeira and Mota (2012).

## 2.3 Innovation Support Infrastructures

R&D and socio-economic conditions determine economic development strategies and the mechanism established to achieve a sustainable knowledge-based regional economic development. As stated above, innovation support infrastructures are intended to correct any local conditions and elements that are hindering the innovation process. They are platforms of interaction, which differ in their capabilities and functions, and may integrate organisations and elements of other innovation organizations, emerging from a process of complementation and hybridization among the entities and institutional spheres involved. The efforts and policies to create the right conditions and improve the process of translating the outputs and impact of the UK science base into economic value (Dyson, 2010; Hauser, 2010; Lambert, 2003) are not new. Since the 1980s, an innovation structure has systematically been developed of science parks, research parks and technology parks across the country. Now commonly known as science parks, these are a policy tool to nurture academia-industry collaboration to boost technological innovation and to generate socio-economic growth (A. N. Link & Scott, 2007; Vedovello, 1997). The umbrella term SP is used for research-based infrastructures with the following general characteristics: formal and operational linkages with HEIs or public research institutes (RIs); supporting the formation and growth of knowledge-intensive commercial businesses; active engagement in the transfer of science-based technologies and business skills (UKSPA, 2003). Nevertheless, commercial-based infrastructures and industrial infrastructures are not SPs as they do not necessarily have operational links with HEIs or RIs.

SPs are basically real-estate developments and are ideally located adjacent to a university. Most SPs are the result of partnerships between research-intensive universities, public authorities and private investors, and take advantage of their strong ties with these three sectors to bring together heterogeneous actors, such as universities, research centres, consulting organisations, technology transfer offices, investors, incubators, local and regional government agencies, intermediaries, and firms (Geenhuizen & Soetanto, 2008; Suvinen, Konttinen, & Nieminen, 2010). The services and support provided for tenants may include equipped offices, reception and secretarial services, conference rooms, and other facilities at below market-rates. Other services are management consulting functions such as planning, accounting, legal matters, technology transfer, intellectual property, recruitment, marketing, access to loan and venture capital, as well as networking and partnering opportunities. These services are often provided by the SP itself or by a network of external regional companies, experts and research institutions (Barrow, 2001).

In such a dynamic environment the combination of research excellence, entrepreneurial activity and public support strategies may enable academia to gain external resources and promote employment through the commercialisation of research in the form of licence agreements, consultancy services, patents, collaboration projects, and university spin-off companies. The private sector can also take advantage of this supportive physical infrastructure to establish or relocate start-up companies, spin-in companies, and R&D units to tap into innovative ideas, market science-related services to potential customers, and engage in ventures and investments with reduced risk and high growth potential. At the same time, governments benefit from these agglomerations to promote partnerships and establish intermediary organisations that facilitate the allocation of capital (funds, research grants, or seed capital) to support promising projects and new ventures, with the likelihood of generating employment opportunities and economic growth in the region (Etzkowitz, 2008; Howells, 2006; Minguillo & Thelwall, 2011b).

SPs can also be defined in terms of semi-public intermediaries or 'switchers', meaning "social actors of different kind who are defined by the context in which specific networks have to be connected for specific purposes" (Castells:51). Such intermediaries or switchers operate in the interface of these networks and have an important role in the process of reconfiguring the values, interests and protocols of communication which induce synergy and limit conflicts between the actors. However, the nature of the global network of production and application of science, technology, and knowledge management makes the SP goal of setting up connections to form a (sub)-network integrating academic-industry-government spheres more important and complex due to the multilayered context (Castells, 2009). From this perspective, innovation support infrastructures can also be seen as articulated multidimensional spaces and social networks where the economical, political, and academic networks occur and interact to commercialize academic research and create an innovative environment. These environments in turn agglomerate a range of other intermediaries and initiatives to be able to form a re-programmed sub-network based on shared goals, values and interests which support the interconnectivity and stability for a particular configuration of overlapping networks.

In summary, innovation support infrastructures are environments where the agglomeration of research and technology is complemented with an emerging network of dynamic and heterogeneous actors, and a range of incentives and services. These services are often access to academic research, lower rents, venture capital and funding, intellectual property advice, networking opportunities, and incubator facilities intended to support the creation and expansion of early-stage and growing firms. They aim to provide strategic and organizational support to overcome transaction costs for heterogeneous collaborations. This is assumed to generate regional employment, new ventures, a higher level of technological diversification and sophistication, and reactivate local and regional economies (DS Siegel et al., 2003).

### 2.3.1.    Science parks: an umbrella term

One of the main problems regarding the assessment of science parks is the lack of a classification scheme or taxonomy (A. Link & Link, 2003; A. N. Link & Scott, 2006; Löfsten & Lindelöf, 2002; Quintas et al., 1992). As Etzkowitz states, "different activities may occur under the rubric of 'science park'. Thus, it is necessary to investigate what is happening in a park rather than making an assumption on the basis of a name" (2006: 316). The multipurpose nature of SPs makes it an umbrella term. The terms used differ across different countries and the names adopted are often terms that reflect the different stakeholders, regional conditions and trajectories which have developed the SPs. Often these terms are also used to make parks more attractive without representing substantial development differences (Castells & Hall, 1994; A. Link & Link, 2003; A. N. Link & Scott, 2007; Saublens, 2007). This issue can be partially resolved by undertaking systematic studies at the national level that can help to identify particular patterns in relation to the names of the various infrastructures across different countries.

The few attempts to define these heterogeneous intermediary organisations usually define them as follows.

- *Science parks*: research-based infrastructures with formal and operational linkages with HEIs or RIs. They strive to support the formation and growth of knowledge-intensive commercial businesses, and the active engagement in the transfer of science-based technologies and business skills (UKSPA, 2003).
- *Research park or research campus:* terms often used in the United States and are broadly defined similarly to science parks in the sense that both host a majority of tenants which heavily engage in basic and applied research and have formal associations with HEIs (A. Link & Link, 2003; A. N. Link & Scott, 2007).
- *Technology park:* "a zone of economic activity composed of universities, research centres, industrial and tertiary units, which realise their activities based on research and technological development", and maintains strong links to large firms and the public research infrastructure at both national and international levels (Saublens, 2007:56).
- *Incubator:* helps young and newly founded innovative firms to establish cooperative relationships with a broad range of economic actors and focuses on compensating for the resource deficit to ensure entrepreneurial stability, sustainable economic growth and long-term business survival (M Schwartz & Hornych, 2008).
- *Business park and industrial park:* a development which provides high quality accommodation to tenants with a wide variety of activities that add value to R&D-based products through assembly or packaging, rather than doing R&D (A. N. Link & Scott, 2003), and like a *science & innovation park or centre* it does not necessarily have operational links with a higher education institution (Saublens, 2007).

### 2.4 Science parks: bridges for U-I interaction

From a regionalisation perspective, SPs form part of the regional knowledge capabilities which facilitate commercial exploitation of public research through the integration of firms in a process of open innovation "to overcome intra-firm knowledge asymmetries by tapping in to the regional knowledge capabilities and systematic innovation strengths of accomplished regional and local clusters" (Cooke, 2005;1147). Thus, the sustainability of socio-economic development among developed countries increasingly depends on their capacity to foster dynamic and strong research-based industries. In this regard, European and national policies highlight the potential role of the university as a major source of research, technology and innovation, and promote closer links with industry (Dyson, 2010; Hauser, 2010; Lambert, 2003). However, this university-industry (U-I) collaboration is not always a straightforward process as the academic and private communities belong to systems that differ in their identity and mission, bringing about transaction costs associated with the efforts employed to bridge the gap between them (Abramo, et al., 2009; Arvanitis, Kubli, & Woerter, 2008). In fact, this interaction barrier has led to an constellation of actors oriented to encourage and facilitate the multidimensional and complex process of capitalisation and transference of academic knowledge (Suvinen et al., 2010).

Some of the most important and long-standing members of this support constellation are intermediary infrastructures: incubators, science parks, and research and technology parks. The pivotal role of SPs in the commercialisation of academic research and technology (R&D) obviously has a significant impact on the goals and functions of universities, as one of the stakeholders, and in turn on part of the scientific community (Xu, Mingyuan, & Zhi'ang, 2011).

Academic assessment of SPs have mainly focused on finding out to what extent links with universities are able to stimulate growth in cutting-edge industries and a competitive advantage for businesses located on SPs in comparison to their off-park counterparts (Quintas, et al., 1992; Rothaermel & Thursby, 2005; Schwartz & Hornych, 2010; Siegel, et al., 2003; Westhead & Storey, 1995). Despite the pivotal role of SPs in the commercialisation of academic research and technology, SPs' impact on HEIs' research and technology output and on the goals and functions of universities has not been extensively investigated. Therefore, the interest in studying factors that may strengthen U-I interaction and encourage a stronger research-orientation in industry has led to suggestions that the use of a scientometric approach may give a fuller understanding of the impact of SPs on the synergy between industry and academia (Bigliardi, et al., 2006; Fukugawa, 2006; Link & Scott, 2003; Siegel et al., 2003).

### 2.4.1.    University-Industry links within SPs

The potential of SPs for more sustainable socio-economic development has generated interest and investments from both the public, academic and private sector and motivated the establishment of a broad spectrum of physical infrastructures across the UK and elsewhere in the last three decades. The mutual U-I dependence in these infrastructures is significant as about 30% of SP tenants have R&D as their main activity and more than 10% have their origins in Higher Education Institutions (UKSPA, 2012a). Moreover, SPs are now one of the main driving forces for capitalising the research capacity in the UK (Dyson, 2010; Hauser, 2010; Lambert, 2003). However, the high early expectations of policy makers for SPs seem to have been undermined by negative evidence about the ability of SPs to support the growth and development of high technology and innovative activity (Bakouros et al., 2002; K. Chan & Lau, 2005; Dee, Livesey, Gill, & Minshall, 2011; Quintas et al., 1992; Radosevic & Myrzakhmet, 2009; Donald Siegel, Westhead, et al., 2003; Westhead & Storey, 1995).

The intermediary role of SPs has been explored by different quantitative and qualitative studies to assess the performance of this interaction as a key indicator of operational success. These comparative studies have assessed whether on-park firms perform better than other firms as a result of stronger links with academia and measure the wide range of forms in which research and technology transfer can occur. Quintas and colleagues (1992) analysed a wide range of interactions, from informal contacts, sponsored research, access to facilities and resources, and the employment of academics or graduates. They argued that there are few academic spin-off firms and that research links and personnel flows between the host institution and on-park firms are no different from those with comparable off-park firms, concluding that SPs are partially constrained by an inappropriate linear model of technology transfer. Another study assessed whether U-I interactions led to a higher survival rate for on-park firms (Westhead & Storey, 1995). This study conducted interviews with comparable on- and off-park firms, finding that most links are informal and both samples linked to the same extent (86%) to a local HEI. The same sample was also used to quantitatively compare the levels of input and output between on- and off-park tenants, considering R&D spending, R&D spending as a proportion of sales revenue and allocated to conduct radical new research, patents, copyrights, new products & services. SP firms did not directly invest more in R&D than average firms and did not have significantly higher levels of technology diffusion (Westhead, 1997). A survey-based study of Surrey Research Park used a small sample of team managers, tenants, and academic researchers, and found researchers to be more likely to establish formal links (contracts, joint research, and consultancy) with off-park firms (Vedovello, 1997). The study conducted by Siegel and his colleagues (2003) applying patents, copyrights, and products also suggests that UK SPs stimulate slightly higher R&D activities among on-park firms than equivalent off-park firms. Thus, the general conclusion from UK SPs is that there is no statistically significant difference between on- and off-park firms as on-park U-I interactions or success are less than anticipated. This fact suggests that physical proximity is not a sufficient condition for facilitating interactive learning and innovation, while other conditions like cognitive proximity, defined as the capacity for sharing the similar knowledge base and expertise, is a more important factor for U-I interaction to take place (Boschma, 2005).

Similar studies have also been undertaken in other developed countries. Schwartz and Hornych (2010) report that academic links are more associated with the features of the various industrial sectors rather than to the conditions nurtured by intermediary organisations in Germany.

Swedish SPs have been systematically assessed with the help of surveys, finding that on-park firms had a substantially higher rate of job creation, sales and new products, and slightly stronger links with local universities, but were not able to generate greater R&D outputs in terms of patents, licences, and new products (Ferguson & Olofsson, 2004; Lindelöf & Löfsten, 2004; Löfsten & Lindelöf, 2002, 2005). In the case of Japan and United States, the level of technology transfer is low but seems to have a positive impact on universities, on-park firms' survival and performance so that Japanese and US SPs are essentially successful (Fukugawa, 2006; A. N. Link & Scott, 2003; Rothaermel & Thursby, 2005). In the case of European countries with less developed national innovation systems and developing countries, despite the sparse U-I links and poor SP performance, the situation around SPs may be even worse and so they seem to still represent an attractive instrument for promoting synergy and innovation (Bakouros et al., 2002; K. Chan & Lau, 2005; K.-Y. A. Chan, Oerlemans, & Pretorius, 2010; Colombo & Delmastro, 2002; Huang, Yu, & Seetoo, 2012; H. Kim & Jung, 2010; Malairaja & Zawdie, 2008; Motohashi, 2011; Radosevic & Myrzakhmet, 2009; Ratinho & Henriques, 2010).

### 2.3.2.    Technology transfer within SPs

Despite many arguments for the fundamental role of research and technology transfer as a source of competitive advantage and revenue for both U-I partners, the literature focuses on the firm level, and partially on the SP level, without properly considering the impact of SPs on HEIs. Link and Scott (2003) pointed out this lack of empirical evidence and carried out a survey to identify flows of technology from on-park organisations to universities and how these interactions may lead to a more contextualized and socially robust research output and a more entrepreneurial mission. They found that U-I involvement led to slightly increased research output and external funding, but had little measurable impact on the academic mission and patents. Some researchers argue that research production and quality, obtained from research publications, should be included as a quantitative measure of U-I interaction in any analytical framework that assesses SP performance (Bigliardi, et al., 2006; Fukugawa, 2006; Link & Scott, 2003; Siegel et al., 2003). This is because scientific publications are the main instrument in research assessment exercises to evaluate the performance of the scientific community, and thus are suited for monitoring knowledge creation when academia is involved (HEFCE, 2010a; Loet Leydesdorff & Meyer, 2010; PACEC, 2009; Whitley & Gläser, 2007).

This makes it necessary to apply a scientometric approach to examine to what extent SPs encourage U-I collaboration and entrepreneurial culture among HEIs. That should provide evidence about the degree of research activity within SPs, the impact of these quasi-academic infrastructures on HEIs, the degree of research involvement between HEIs and SPs, the structure of the collaborative networks formed by linkages between HEIs and on-park tenants, and the role of HEIs in the configuration of these networks. Therefore, it is necessary with studies that mainly focus on the academic sphere and aims to quantitatively explore whether HEIs with formal ties with SPs have a greater chance to establish U-I R&T-based links. The exploration of this networking activity is fundamental to gain a better understanding of the intermediary role of SPs between the academic and private sector, and to expand the theoretical framework that assess the complex and multidimensional networking function of SPs.

Scientometric indicators have recently been applied to the study of SPs. One study assessed, for example, how the underlying knowledge creation and diffusion in the Hsinchu region benefits the innovation capability and success of Hsinchu SP in Taiwan, employing on-park firms' patenting and patent citations (Hu, 2011). A similar study examined the use of public science in on-park firms based on Hsinchu SP, including (non-)patent citation, and public-private co-authorship of publications and patents, to show that U-I collaboration has constantly increased in terms of publications, while the patterns related to the patenting activity is stable or even declining (Hung, 2012). All this suggests that a scientometric approach is useful to study the central R&D dimension of knowledge based environments such as SPs. This could shed new light on the intermediary role of SPs regarding R&D activities, and also provide empirical evidence for the literature regarding U-I collaboration in general (Teixeira & Mota, 2012), and guide more effective U-I collaboration processes in developed countries.

## 2.5 Link analysis: a webometric method to uncover digital networks

Today's networked society is characterized by the prominent role of the global digital network of communication and how it is reshaping and expanding interactions among people. This growing re-definition of the institutional spheres and social structure has the internet and specially the web at its core. The internet enables "the articulation of all forms of communication into a composite, interactive, digital hypertext that includes, mixes, and recombines *in their diversity* the whole range of cultural expressions conveyed by human interaction." (Castells, 2009: 55). The web, therefore, needs to be studied as one of the driving forces which shape the contemporary networked society (Castells, 2000; 2009). This makes link analysis techniques interesting to study the configuration and dynamics of SPs, particularly as intermediaries and connectors of different spheres.

Hyperlinks are studied in fields such as physics, computer science and information science. In information science, the analogy between citation networks and collections of hyperlinked documents emphasises the behavioural foundations of hyperlinks (Ackland, 2009; Thelwall, 2006). Information scientists tend to acquire a social science perspective, viewing the web as a complex and multi-layered relational space that enables various forms of social, economic, and political behaviour. Here the focus is on studying the underlying structure and value of networks formed by people and organisations that might reveal offline phenomena (Bar-Ilan, 2005; Park & Thelwall, 2003; Thelwall, 2004; Thelwall, et al., 2005).

The analogy between the relational structures formed by references and citations in scientometrics with the outlinks and inlinks in webometrics has also attracted scientometricians to this field and led to an over-representation of studies based on inlink patterns, whilst other equally interesting patterns such as peer interlinking and the outlinking dimension have been left aside (Thelwall, 2003). This has also led to the extensive application of the indirect co-inlinking measure, despite the fact that direct links are more reliable indicators of similarity on the web (Thelwall, Harries, & Wilkinson, 2003). In addition, the codified and static nature of formal scholarly communication differs from the multifaceted and mutable nature of the web (Gonzalez-Bailon, 2009), and both incoming and outgoing hyperlinks are potentially social ties with meaning that can vary in relation to the contexts in which a set of actors interact on the web. Hence, considering the different types of offline relations that can be tracked on the web, the range and types of online ties may be extensive.

Hyperlinks are employed as indicators of performance through link counts and structural positions using the similarity of direct or indirect ties. Encouraging results in the study of academic institutions (Aguillo, et al., 2006; Thelwall, 2001) have led to a wide range of application contexts related to business performance and have shown that R&D investments and revenues may related to web visibility as long as the organisations are homogeneous (Vaughan & Wu, 2004; Vaughan, 2004) and their sizes are not considered (Martínez-Ruiz & Thelwall, 2010). Indirect links have also been used to map businesses market positions (Vaughan & You, 2006), and linking pages have been analysed to find out motivations for link creation (Vaughan, Kipp, & Gao, 2007). Similar approaches have been applied to identify the links of the international banking industry (Vaughan & Romero-Frias, 2010), political communication (H.W. Park & Thelwall, 2008), and the structural dynamics between organisations in a region (Faba-Pérez, et al., 2004). The relationship between the physical proximity and the strength of links connecting academic institutions has also resulted in studies that focus on uncovering stronger interactions among geographically closer public libraries (Kawamura, Otake, & Suzuki, 2009), and also among local and national government bodies that are physically closer (Holmberg & Thelwall, 2009; Holmberg, 2010; Petricek, et al., 2006).

Webometric methods have also been used to investigate scholarly communication. For example, a study that examines whether a new mode of production, like Mode 2, has strengthened the interactions with non-academic actors, supported dissemination of diverse outputs, and increased the use and exchange of digital data. This study revealed that although there are few research fields and researchers performing Mode 2 research, this new mode of production cannot be identified in the communication patterns across all fields with the knowledge production still being based on journal communication (Heimeriks, Van den Besselaar, & Frenken, 2008). Park (2010) mapped the semantic variation, disciplinary scope, and institutional structure of e-science technologies and research in South Korea, finding that the significant web presence of e-science terms differs from the underrepresentation and scarce connectivity of governmental agencies responsible for e-science facilities, and from the weak interaction between the academic and public sector. A similar study also used a relational and semantic approach to identify enterprise strategies in emerging technologies (Arora, et al., 2013). On the other hand, an explorative methodology introduced to examine scientific interdisciplinarity through relationships among researchers and their research topics also showed the complexity and limitations related to the use of unstructured data to assess academic activities (Sayama & Akaishi, 2012). Overall, despite the fact that link analysis and link counts are indicators that are not robust and cannot be used for research assessment purposes (Thelwall, et al., 2010), and the future of webometrics as an emergent research area is uncertain (Thelwall, 2010a), the main value of webometric methods is the capacity to shed light on a broader overview of cross-sector patterns of the network in which industrial, academic and government actors collaborate and operate (Thelwall et al., 2010). Thus, the application of scientometric and webometric methods provides the opportunity to see the research and networking activity expected in the SP movements.

Other studies that consider the web as an emerging inter-organisational communication tool have also used hyperlinks to study different social-political phenomena. They focus on studying ideological relationships between political parties across Europe (Romero-Frías & Vaughan, 2010), terrorist activities (H. Chen et al., 2008), social movements like anti-war and peace activists (Gillan, 2009), and the relationships, popularity and political agenda of Democratic and Republican senators from the United States Senate (J. Kim, Barnett, & Park, 2010). Link analysis has also been applied to uncover the larger inter-organisational networks established by non-government organisations located in the global North in comparison with those located in the South, confirming that the historical global divide between North and South is likely to remain despite technological innovations (Shumate & Dewitt, 2008).

## 2.6 Manifestations of R&D networks on the web

Although the re-definition of the institutional spheres and social structure is caused by global structural dynamics created by a new organizational and digital context of interaction (Castells, 2009), few studies have focused on studying the web to uncover the interactions among actors operating in R&D networks. At this early stage, most studies aimed to find new web-based standards to evaluate how the knowledge-based economy is re-articulating the social structures. For example, the first attempt examined the feasibility of webometrics with regards to exploring virtual social structures and measuring academia-government-industry dynamics (Boudourides, et al., 1999). Later, Leydesdorff and Curran (2000) investigated institutional communication between industry, university and academia. The authors used word co-occurrences and hypertext links to conclude that the relationship between the university and the other two sectors is stronger at international level while national economies promote stronger industry-government linkages. Similarly, Khan used relational and semantic techniques to map trilateral relations over the web. By using different social media tools like search engines, social network sites, and portals to collect a longitudinal dataset, he found that there is a likely association between governmental policies and the trilateral linkages in Korea (Khan & Park, 2011). The study of Stuart and Thelwall (2006) analysed the intra- and inter-sector collaboration in the automobile industry in a UK region, finding that the lack of web connectivity only partially reflects the offline collaboration. Due to the difficulty in tracing the TH connections, the authors recommend text analysis instead of link analysis. Thus, the use of terms along with the self-organization and decentralization in link formation (Flake, et al., 2002) might represent a better solution in the identification of interactions within particular social communities, like the research collaboration within the scholarly community (Kenekayoro, Buckley, & Thelwall, 2013).

An investigation into the web-based relationships of the UK's pharmaceutical industry with the help of hyperlinks, publications and patents showed that the identification of web sites of all the important organisations in the pharmaceutical industry might be an impossible task, implying that a more narrowly focused investigation is necessary (Stuart, 2008). Similarly, a study that traced the national communicative change of the Triple Helix in Spain by means of outlinks and co-outlinks, using the clustering of the ten different sectors chosen at random and which in the authors' opinion should belong to the TH model, did not find a clear triple helix structure (Garcia-Santiago & de Moya-Anegon, 2009). On the other hand, a report about the regional innovation system of the Swedish region Skåne suggests that the identified web-based structure is similar to the collaboration and innovation structures found previously, but it could not find a transparent innovation process across organizational boundaries (Daal, 2009).

Overall, the main weakness of these studies might be first, the broad context of study in which the trilateral relationships have empirically been investigated. Second, the asymmetry in the analysis of heterogeneous set of websites in terms of resources and public recognition (Gonzalez-Bailon, 2009), and third, the subjective criteria to select the different organisations to be involved in an innovation system, especially in a not codified context such as the Web. Only Ortega's study (2003) was able to show potential cross-sector interactions. He examined science and technology interactions in a specialized environment, using the outlinks to represent the relationships of research centres affiliated to two bio-related institutes in a Vector Space Model. The study showed that the different centres, according to their activities, occupied a closer position to the government and academia, whilst only a couple of research centres were able to establish trilateral relationships.

Due to the low number of studies and questionable results, web-based findings are still merely suggestive of hypotheses that need further investigation, which makes it necessary to develop a set of methods for studying the online nature of dynamic and innovative environments. This might make it possible to take advantage of the vast information available on the web to provide a broad overview of the complex configuration of R&D networks, which could then be further analysed and disclosed by other approaches.

## 2.7 The intermediary role of science parks on the web

SPs can be seen as micro-communities of cross-sectoral and interdependent organisations united by an identity, a mission, a set of routines and a strategic core, which interact with the external environment as unified entities that change over time and that should be studied from a social network perspective (Bøllingtoft & Ulhøi, 2005; Phan, Siegel, & Wright, 2005). The R&D networks associated with SPs have an institutional intermediary role that gives them an inherent multidimensional and synergetic nature, encompassing different kinds of organisations and sectors which can be studied at different levels of analysis (Tijssen, 1998).

However, despite the importance of the networking function of SPs in technology transfer, firm-formation and social development, they have so far been investigated without considering them as intermediary organizations with multiple levels of analysis from a social network perspective. This might be because there is currently no data available nor a systematic framework to study the diversity of links established within and across industry, higher education, and government agents involved in SPs (Phan et al., 2005). Few studies have empirically investigated SPs' ability to foster networking capabilities. These small scale studies primarily use interviews and surveys that investigate bilateral relationships and provide evidence that is difficult to generalise (Bøllingtoft & Ulhøi, 2005; Bøllingtoft, 2012; K. C. Chen, et al., 2011; Hansen, et al., 2010; McAdam & McAdam, 2006; Motohashi, 2011; Sá & Lee, 2012; Soetanto & Jack, 2011), while other studies take a more theoretical perspective (Ahmad & Ingle, 2011; Hongli & Zhigao, 2010).

This makes it difficult to understand and assess the relational and intermediary capacity of SPs. It is therefore necessary to develop a multidimensional approach to deal with the relational structure generated by the agglomeration of heterogeneous groups of actors. This could shed new light on the network structure formed by the formal and informal ties, and information flows within and across industry, university, and public agents embedded in a specific geographic location, economic system, and political and social context. According to Castells, cyberspace enables the analysis of different institutional networks that coexist in the digital dimension and become interconnected through hyperlinks, building global digital networks of interactions which transcend territorial and institutional boundaries (2009:4, 24). This highlights the important role of the internet and websites for these agglomerations (Steinfield & Scupola, 2008) and as an important social interface for communication (Heimeriks et al., 2008). The study of web-based interactions could help tackle the complexity, scalability, and self-configuration of these multi-layered and diverse R&D networks and make it possible to deal with the relational and self-configured structure generated by heterogeneous groups of organisations, which otherwise would need to be studied through a wide set of indicators.

# Chapter 3: Global research questions and methods

## 3.1 Introduction

This chapter is divided into two main sections. The first section lists the research questions that arise from the overall aim and objectives of the thesis. The second section describes the main methods adopted in the different empirical studies to tackle the research questions.

## 3.2 Global research questions

The main aim of this thesis is to determine the potential application of a scientometric and webometric approach to the study and assessment of UK innovation support infrastructures. This aim is guided by two main research questions:

(1) What insights can scientometric methods give into the role of public science and Higher Education Institutions (HEIs) in R&D networks associated with SPs?

(2) Can web-based patterns reflect the configuration of R&D support infrastructures associated with SPs?

These two broad research questions are split into specific questions that are empirically answered in order to achieve the main aim of this thesis.

### R&D activity in the UK SP movement

SPs are prime candidates for collecting empirical evidence on the emergence and development of knowledge-intensive industries, and the use of bibliographic data has been suggested for this (Bigliardi et al., 2006; Fukugawa, 2006; A. N. Link & Scott, 2003; Donald Siegel, Westhead, et al., 2003). It can be used in quantitative methods to determine the levels of R&D activities and U-I interactions that innovation infrastructures in general, and science parks in particular, are expected to promote to support the development of cutting-edge industries. However, scientometric approaches that use public-private co-authorship of publications and patents have only recently been used to study particular parks (Hu, 2011; Hung, 2012).

This lack of evidence makes it necessary to first determine the information that can be drawn from this data source regarding the R&D activities produced by on-park organisations and fostered by SPs.

This is addressed by the following broad research question:

- Can scientific publications help to shed light on the R&D activities fostered by the different support infrastructures?

Before addressing this question it is important to know what types of innovation infrastructures are established across the country. The information that scientific publications may give could reveal which physical and organisational infrastructures are the most research-intensive; where these are established and which type of organisations engage in these environments; how onsite-organisations collaborate, the collaboration between on-park firms with knowledge producers, and how these linkages extend across the country and beyond; what scientific disciplines underpin these industries as well as the quality and impact of the research produced. Scientometric indicators may thus add to the existing battery of indicators and make it possible to systematically monitor support infrastructures on a large-scale. This is important because besides the UKSPA evaluation (2003) most previous UK studies have only focused on one or a few SPs, making it difficult to draw general conclusions because SPs are unique physical infrastructures and social entities that are also strongly influenced by external conditions (A. N. Link & Scott, 2007). Thus, scientometric indicators could provide complementary statistical information to that provided by socio-economic indicators and surveys to give a better picture of the heterogeneous SP movement, which is necessary to better understand its role and how it needs to be evaluated (Geenhuizen & Soetanto, 2008). (Dataset & Methods: Section 3.3.2.; Results & Discussion: Section 4.2)

**Role of public science and HEIs in the UK SP movement**

While cutting-edge science and technology is becoming increasingly important for regional competitiveness and economic growth, relationships between producers and consumers of research-based knowledge and skills are a central issue for the development of policies and regional innovation strategies to increase the application of scientific research (Cooke, 2002; Etzkowitz, 2008). For instance, the EU's Competitiveness and Innovation Framework Programme (CIP) aims to foster successful engagements between both communities to enhance competitiveness and integration (European Commission, 2012). Thus, policy makers expect SPs to provide the desired academia-industry synergies and economic added value. HEIs are an important source of economic advantage as the main producers of public science and can provide a vibrant technology base, skills, support and advice local businesses, attracting inward investments (Witty, 2013). SPs are expected to bring academia closer to industry to exploit the research strength of universities and foster the creation of university spin-offs, strategic alliances and to attract new technology-based firms and the R&D units of existing companies that desire collaboration with academic researchers (Etzkowitz, 2008; Vedovello, 1997).

Despite their critical roles in modern economies, the intermediary role of SPs in the interaction between on-park tenants and HEIs has only partially been studied, producing contradictory and fragmented evidence. This lack of evidence makes it necessary to analyse the collaborations between academia and industry to analyse the role of academia in the configuration of the collaborative networks embedded within, or triggered by, SPs. Therefore, this thesis also focuses on examining the university-industry partnership as a bidirectional interaction and with mutual benefits and impacts (Hansson, Husted, & Vestergaard, 2005; Phillimore, 1999), and assesses the extent to which SPs support firms to evolve from a passive knowledge-consumer role to actively engage in the creation of new knowledge and research.

This is addressed by the following research questions:

- What are the main structural properties of the inter-institutional collaboration network of on-park organisations?
- What are the main structural properties and role of HEIs in the inter-institutional collaboration network of on-park and off-park organisations?
- What are the main structural properties and role of HEIs in the firms, HEIs, and RIs inter-institutional collaboration network?

The answers to these questions will shed light on the internal and external structures of collaborations established by on-park organisations. This relational dimension is essential in the assessment of SP performance and success as these collaborative links are important for highly-innovative firms (Soetanto & Jack, 2011). In addition, it might help to deepen understanding of the complex and multidimensional networking function of SPs. (Dataset: Section 3.3.2.; Methods: Section 3.3.5. & 5.2; Results & Discussion Section: 5.3-4)

**Impact of SPs and third mission activities on HEI performance**

SPs may help to overcome shortcomings in the exploitation of academic research as their mission is to reduce the gap between academia and industry. They are expected to provide the support, expertise and resources to fill the lack of business acumen among academics and to help introduce new products into markets (Dahlstrand, 1997), create spin-offs and new technology-based businesses (NTFBS) (Michael Schwartz & Hornych, 2010a), and attract external investments (Shane & Stuart, 2002). However, it is not clear whether SPs encourage academic research and entrepreneurship.

Studies of the influence of SPs on hosting universities and academic R&D activities are surprisingly limited (Huggins, Johnston, & Steffenson, 2008; A. N. Link & Scott, 2003; Donald Siegel, Waldman, & Link, 2003). Quantitative studies are therefore needed to determine the effect of SPs on the 'third mission' of universities (i.e. knowledge transfer) as well as to what extent SPs help HEIs to provide knowledge for industry and act as the central actor in innovation systems. In this context, it is particularly relevant to examine the characteristics of this key mutual relationship with the help of R&D publications, as important proxies to assess a research community's performance.

This is addressed by the following research questions:

- How long after SPs are established do they start to promote research creation and cross-fertilization with academia?
- Which factors help R&D production and U-I collaboration in SPs?
- Do HEIs (and RIs) with a formal relationship – hosting, partnering, both or none - with SPs have a greater capacity to produce R&D and collaborate with the private sector or do other factors have a greater influence on U-I research interactions?

In addressing these questions, the analysis will mainly focus on revealing whether SPs may help to strengthen the university-industry (U-I) interaction and the extent to which SPs may benefit HEIs as the main stakeholder and source of knowledge to be exploited. This is important since HEIs are expected to be a key driving force in the growth of the knowledge economy. (Dataset: Section 3.3.2-3.; Methods: Section 3.3.4; Results & Discussion Section: 6.2-3)

## Web-based structure of the R&D support infrastructure associated with science parks

Currently there is neither data available nor a systematic framework to study the diverse links established within and across the industry, higher education, and government agencies involved in SPs (Phan et al., 2005). This makes it difficult to understand and assess the relational and intermediary capacity of these organisational innovation structures, which aim to strengthen the cross-sectoral connections necessary to generate highly innovative environments. An interesting approach to tackle the complexity of these R&D networks is to take into account to what extent these interactions could be manifested in the digital dimension (Stuart, 2008). Castells argues that cyberspace enables the analysis of different institutional networks that coexist in the digital dimension and become interconnected through hyperlinks, thus building global digital networks of interactions, which transcend territorial and institutional boundaries (2009:4, 24). In addition, the key role of ICTs and the Web for enterprises and organizations in knowledge-intensive agglomerations (Steinfield & Scupola, 2008) suggest that Web hyperlinks could be an important data source for empirical studies. Consequently, it is useful to assess whether a webometric approach can tackle the complexity of these diverse R&D networks and uncover the relational structure generated by heterogeneous groups of organisations, which otherwise would need to be studied through a wide range of alternative indicators.

This is addressed by the following research questions:

- Can webometric methods identify interactions between institutional sectors and interactions between various types of organisations associated with SPs?
- Are the main features of SP web interlinking networks in line with the findings of official reports and surveys which evaluate the UK R&D infrastructure?
- Can the organisations and their operational interactions associated with R&D infrastructures be identified in a hyperlink network generated by the web sites of organisations associated with a SP?
- Do the links between organisations in a SP hyperlink network reflect the potential behaviour of the different types of actors expected to engage in R&D infrastructures?

In addressing these questions, the viability of link analysis to identify a SP's intermediary roles is examined. It is also important to determine whether web-based and offline characteristics of SPs reflect similar patterns in order to investigate whether the general picture provided by the digital dimension may shed new light on the structures established in these dynamic and innovative environments. (Dataset: Section 3.3.6.; Methods: Section 3.3.7-8. & 8.2; Results & Discussion Section: 7.3-4 & 8.4)

## 3.3 Methods

The project applies quantitative methods. Bibliometric methods are used to study the R&D activities associated to the UK SP movement and U-I interactions. Webometric methods, especially link analysis, are used to reduce the complexity of mapping the wide range of relationships established among the different organizations that engage in the science parks and thus provide a broad overview of their off-line interactions. These informetric methods are used along with visualization and social network analysis (SNA) techniques as well as descriptive statistics to analyse the data. The theoretical basis is underpinned by different knowledge domains, including regional innovation systems, economic geography, research and development policy, research and technology transfer, as well as research into the entrepreneurial university and the university's third mission.

### 3.3.1.    UK Science park movement as the population of analysis

#### a.   Science park movement: Scientometric approach

Science parks (SPs) are policy tools that seek to be a location for different institutional sectors to converge and collaborate to develop a R&D support infrastructure that facilitates the exploitation of research to foster the growth of research-intensive industries. This is expected to create regional competitiveness and sustainable socio-economic development. According to the United Kingdom Science Park Association (UKSPA) "a Science Park is a business support and technology transfer initiative that: (1) encourages and supports the start up and incubation of innovation-led, high-growth, knowledge-based businesses; (2) provides an environment where larger and international businesses can develop specific and close interactions with a particular centre of knowledge creation for their mutual benefit; (3) has formal and operational links with centres of knowledge creation such as universities, higher education institutes and research organisations" (UKSPA, 2012b).

Since the 1980s an innovation infrastructure of science parks, research parks and technology parks across the country has been systematically developed. Nowadays the UKSPA has over 100 full members, including support infrastructures called Science Parks, (bio-) Incubators, Innovation Centres, Research Parks, Technology Centres, Technology Parks, and Technopoles. The objects of analysis of this thesis are all the full members of the UKSPA and other infrastructures with similar names across the UK. Bibliographic records were the main data source used in this thesis to identify research-active infrastructures in the UK, as described in detail below. The information drawn from this structured dataset, such as quantity, impact, quality, and collaboration, was applied to find the main patterns of research activity of UK science parks. (Sample applied in Chapter 4-6)

### b. Yorkshire and the Humber: Webometric approach

Inter-link analysis and hyperlinks were the main method and data source, respectively, employed for the two webometric studies of the thesis (chapters 7 & 8). The objects of analysis of these exploratory studies were SPs indexed as full members by the UKSPA in the region of Yorkshire and the Humber, and which had their own website and provided a list of their tenants with their respective URLs. This region was selected since three of its four SPs indexed by the UKSPA were considered adequate for the exploratory study, but most importantly because it is an interesting region that has invested in a growing innovative infrastructure as a tool for industrial restructuring (Huggins & Johnston, 2009). In addition, the SPs in this region host a wide range of company sizes; from small and medium-sized enterprises and spin-offs to large manufacturing groups, and types; from specialised knowledge and high technology to advanced manufacturing. This heterogeneity was important in order to gain good insights into the different organisations involved in this context, as well as the degree of interaction between the different organisations and sectors. (Sample applied in Chapter 7-8)

### 3.3.2. Structured dataset: bibliographic records

#### a. Data collection

Publications associated with UK SPS were retrieved from Elsevier's Scopus database covering a period of 35 years (1975-2010). We used two different approaches to retrieve the records of the research publications produced by any organisation located within a SP in the UK. First, with the help of the SP list provided by the UKSPA and the electronic version of the *Atlas of Innovation* created by the *World Alliance for Innovation* (Wainova) we identified the names of 82 full UKSPA members across the country. This allowed the creation of Scopus queries with the specific names of the different SPs (e.g., *AFFIL ("norwich research park") AND (LIMIT-TO(AFFILCOUNTRY, "United Kingdom"))*). Second, to extend the first search and identify potential non-members of the UKSPA and track down the high diversity of the SP movement, we used truncated queries with terms that are broadly used to name research-based infrastructures in the country, such as science-, technology-, innovation park, incubator, etc., as well as terms of commercial-based infrastructures, such as business-, industrial-, enterprise park, and business centre (i.e. *AFFIL("sci* park") AND (LIMIT-TO(AFFILCOUNTRY, "United Kingdom")*)). Both specific and truncated queries were restricted to the year 2010 covering journals, book series, and conference proceedings, while excluding editorials, erratum, letters, and notes. This selection of document types is based on their relevance as public communication channels for industry research outputs (Cohen et al., 2002). The search yielded 10,920 records. A similar search strategy was used on the Web of Science (WoS) database (Thomson Reuters) but approximately two thousand fewer records were retrieved using this method. Note that not all onsite organisations mention the SPs where they are located as part of their affiliation addresses in research publications, so this search approach may not take all the relevant publications into account. This is the main dataset for the bibliometric studies carried out in chapters 4, 5 and 6.

#### b. Data processing and cleaning

The bibliographic data of these publications, such as titles, authors, affiliations, abstracts, and sources, were exported to an ad-doc relational database created for this study. Data cleaning and standardisation was used to identify all publications listing at least one author address referring to a UK SP, and the author address was checked for assignment to the organisation stated by the author. The research produced by departments, sub-units, or company groups was assigned to the parent entity, and only research centres associated with HEIs were treated independently in order to get more fine-grained results. In the case of firms, name changes, mergers, or acquisitions were taken into account where possible. But in most cases organisations with different physical locations are treated separately to quantify the impact of SPs on the immediate environment.

Most hospitals in SPs are teaching hospitals and are classified as HEIs, as recommended in the *Frascati Manual* (2002). For conciseness, multiple health centres or hospitals from the same region or city were collapsed into one organisational entity. Similarly, non-UK organisations were broadly classified according to country and grouped together into different types of organisation (higher education, industry, government, and on-park organisation), while UK-based organisations were clustered into six groups (higher education, industry, government, on-park organisation, non-profit organisation, and research institute). The 'foreign on-park organisation' group includes organisations located in a SP outside the UK. In addition to this typology, they are also grouped according to four other main attributes (type of organisation, location, type of location, and district). This process gave 9,771 publications produced by at least one onsite-organisation. This is the main dataset for the bibliometric studies carried out in chapters 4, 5 and 6.

### c. Research subject areas

The research subject areas and categories were taken from the Scopus journal classification scheme. The publications placed in journals indexed in more than one subject area were counted in each one. These areas were also used to identify the degree of participation of the private and academic sectors, of the regions, and of the U-I collaboration (Chapters 4). The research subject categories were also used to broadly identify the thematic areas in which the collaborations associated with the SP movement could roughly be based on (Chapters 5).

### d. Quality and impact

Reputation, in the form of citations given by the research community, was used to determine the popularity and impact of the research. Prestige was determined in two ways. First, quality was approximated by the number of citations received by the journals of the publications. This is quantified by the two citation based indicators; Scimago Journal Ranking (SJR) and Source Normalised Impact per Paper (SNIP), as both are designed to evaluate the prestige and visibility of journals in relation to the particular characteristics of a research area. Second, impact was approximated by the number of citations received by each individual publication. In chapter 4 these two indicators were applied along with the Wilcoxon signed-rank test, which is the non-parametric equivalent of the t-test, to assess if there is a significant difference between the observed and expected quality and impact of the research across subject areas.

### e. Inter-organisational co-authorship

Inter-organisational co-authorship is a form of joint research which is widely used as a proxy to measure collaborative R&D activities and a formal indicator of R&D network links (Tijssen &

Korevaar, 1997; Vedovello, 1997). This approach is in line with studies that analyse the inter-relationship between scientific and technological (publications and patents) competence as the underlying conditions to support and raise the emergence of cutting edge technology industries (Van Looy, Callaert, & Debackere, 2006). Despite U-I co-authorship being a partial indicator of research collaboration (Katz & Martin, 1997; Tijssen et al., 2009), the analysis of U-I creation of knowledge in the form of research publications represents an interesting indicator of knowledge transfer between public science and industry. It includes important factors such as sharing problems, skills, expertise and resources to resolve problems and improve the ways of doing things through research and consultancy agreements, research support, cooperative research, and knowledge and technology transfer (Santoro & Chakrabarti, 2002). Inter-organisation collaboration was mainly used in chapter 7, but it was also included in chapters 4, 5 and 6.

### 3.3.3. Structured dataset: patents and secondary data sources

#### a. Patents

The 78 HEIs and RIs with five or more publications co-authored with on-park organisations were selected to further gather the number of patents produced by these institutions until 2010. The UPSTO database was used because it provides the best coverage. The patents produced by the 78 institutions were identified because patents are widely used to monitor U-I interactions (Huang et al., 2012; Donald Siegel, Westhead, et al., 2003). However, patenting is not a core function of universities, and is not widely included in the reward system because it is expensive in comparison with the expected rewards (Loet Leydesdorff & Meyer, 2010). Consequently, academic patent applications are decreasing as result of insignificant economic returns (Geuna & Muscio, 2009), and U-I patenting is declining despite increased SP capabilities to collaborate and produce research (Hung, 2012). Furthermore, patents are not a good proxy to measure companies' commercial success (Zheng et al., 2011), being considered by firms as one of the least effective sources of knowledge, compared to scientific publications, conferences and informal interactions with academic researchers (Agrawal & Henderson, 2002). Citations from patents are also an invalid indicator of science dependence (Meyer, 2000; Tijssen, 2002). Moreover, SMEs make up the vast majority of tenants (93%) on UK SPs (UKSPA, 2012a) but SMEs rarely patent. The patent data along with the bibliographic data were used in the chapter 6 study that analyses whether HEI connections with SPs associate with better R&D outputs.

#### b. Secondary data sources: quantitative & qualitative

The bibliographic and patent data sets were complemented with data extracted from different sources. The UKSPA and the electronic version of the WAINNOVA *Atlas of Innovation*, SPs' websites and managers were used to identify when each park was established. The classification of the Department for Business, Innovation & Skills (BIS) (PACEC, 2009) which classifies 108 HEIs and PROs into four different groups according to their research spending, academic research staff, 2001 RAE average quality score, and research intensity, was used to estimate their research quality.

The *Higher education – business and community interaction survey 07-08* (HEFCE, 2009) was used to obtain qualitative and quantitative data regarding how HEIs associate with SPs and incubators, as well as qualitative and quantitative indicators which measure the intensity of a HEI's research and technology transfer activity. The qualitative variables are: technology transfer, supporting SMEs; research collaboration, spin-off activity, entrepreneurship training, seed corn capital, and venture capital. The quantitative variables are: income from collaborative research, contract research, consultancy, facility and equipment service, intellectual property, spin-off activities with some HEI ownership, spin-off actives not HEI owned, and active patents.

The bias introduced by the size of the different HEIs was reduced by normalising all the variables by dividing them by the total number of academic staff. The staff data was obtained from the Higher Education Statistics Agency (HESA) database. These data sets were used in the studies conducted in the 5 and 6.

In the study presented in chapter 4, the UK Competitive Index (UKCI) (Huggins & Thompson, 2010) were also used to analyse whether the research activity across the UK SP movement is related to the regional competitiveness indexes.

### 3.3.4. Statistical methods

Non-parametric statistical techniques were used because the main dataset employed in this thesis is not normally distributed.

#### a. Kaplan-Meier and Cox regression

Kaplan-Meier and Cox regression are two non-parametric survival analysis techniques employed in chapter 5 and 6. These were chosen for robustness and the ability to analyse incomplete data using censoring techniques to calculate the probability and conditional probability that an event occurs at a certain point in time without parametric assumptions. These techniques are popular in clinical studies and other research areas. The Kruskal-Wallis rank test evaluates whether two or more independent groups come from the same population. The null hypothesis is that the samples are drawn from identically-distributed populations. The main limitation of this test, like its parametric counterpart one-way ANOVA, is that when it identifies that there is a difference it does not identify the specific group or groups of the samples that differ, which makes it necessary to perform Mann-Whitney tests for all pairs of groups to identify where the differences lie. These are also applied for estimating failure rates in engineering, epidemiology, economics and sociology. Publication analysis was used to evaluate the potential bias involved in the delayed of completing, submitting and publishing clinical trials (Haidich & Ioannidis, 2001; Ioannidis, 1998). In the SP literature this technique was also applied to investigate patents (Squicciarini, 2009), the survival rates between on-park and a control group of off-park firms, (Ferguson & Olofsson, 2004; Westhead & Storey, 1995), and the study of post-graduation survival rates and exit predictors after graduation in Germany (Michael Schwartz, 2009).

The Kaplan-Meier method was used to determine the median time between the SP year of creation and the year when its tenants start to produce research (Chapter 6). As publications can be the result of research carried out by a firm, RIs, or collaboration between firms and RIs within or outside of the parks, these four different distributions were also estimated. In addition, the long-rank test was used along with the Kaplan-Meier estimations to find statistical differences between the groups, where SPs were classified according to their type[1].

A multivariate semi-parametric Cox Proportional Hazard regression was also used to detect the extent to which the start of the research output was influenced by a number of covariates (Chapter 6). These covariates are properties of the different infrastructures, including: *infrastructure typology, age, establishment decade, UKSPA full-membership,* and *region*. The censored observations identify SPs with low research production by the end of the studied period (right-censored), and also those with research production before their establishment (left-censored). These statistical analyses were run with SPSS software, version 19, and all p values are two-tailed.

### b. *Kruskal-Wallis test*

The Kruskal-Wallis test was used to determine if the institutions that have a relationship with a *SP*, *On-campus incubator*, or *Incubator in the locality*, and that are based on *hosting*, *partnering*, *both*, or *none* show any statistically significant difference in relation to the number of patents, publications and number of publications produced in collaboration with on- and off-park firms. This test was carried out to explore if formal partnerships between HEIs and SPs encourage more entrepreneurship among HEIs (Chapter 6). As this test only reports if a distribution differs across the groups but not which groups differ, Mann-Whitney tests were also performed. Spearman correlations were also used to measure whether other factors are more likely to strengthen U-I interactions. The factors analysed are widely used to assess the R&D transfer capabilities, and were grouped into qualitative variables related to HEIs' strategies (*research collaboration, technology transfer, support to SMEs, spin-off activity*) and infrastructure (*business advice, entrepreneurship training, seed corn investment, venture capital*), and also quantitative variables related to HEI incomes from commercial activities (*collaborative research, contract research, facility & equipment service, intellectual property (IP),* and *spin-off activities with some HEI ownership, of spin-off actives not HEI owned, and active patents*).

---

[1] The characteristics of the different types of infrastructures have been described in detail previously (A. N. Link & Scott, 2003, 2006, 2007; Saublens, 2007) (see section 2.4.1.).

### c. Mann-Whitney

The Mann-Whitney test is also a rank-sum test and is the non-parametric analogue of the independent samples *t-test*. It is used to test the null hypothesis that two samples come from the same population, having the same median or whether the observations in one of two groups have larger values in the sense a higher sum of ranks (Field, 2009).

Apart from these statistical techniques, the thesis also includes other standard methods: the Wilcoxon signed-rank test (Chapter 4), the Gini coefficient (Chapter 7), the Pearson correlation coefficient, and Spearman's rank correlation coefficient.

### 3.3.5.    Social Network Analysis (SNA): Stochastic techniques

Stochastic techniques were used to identify the modular structure of the inter-organisational collaboration network associated with the SP movement at three different levels (Chapter 5). Other common SNA measures were also used, including local, structural (Chapter 7) and centrality measures (Chapter 5).

The structural analysis used cohesion measures to estimate the degree of integration of the websites, the size of the network, the cohesion of the websites and the level of mutual interaction between them. The measures applied were: inclusiveness, that describes the number of websites that are integrated into the network and is the total number of nodes minus the isolated nodes (Scott, 2000); connectivity gap, that measures the total number of in- or outlinks established in the in- and out-data sets minus the number of in- or outlinks in each data set to give the proportion of links which the data set needs to reach the maximum number of links obtained from the data sets; density, that measures the proportion of all possible connections that are actually present; and reciprocity, that indicates the proportion of relations (links) that are reciprocal (S Wasserman & Faust, 1998). The local analysis used centrality measures like (in- and out-) degree and betweenness, to estimate the degree of interaction between particular websites and group of organisations. The centrality measures mainly applied in chapter 5 were: degree, that measures the level of activity through the number of direct ties to other actors in the network; betweenness, that measures the level of influence and control through the shortest paths between any two node passing through a certain node; closeness, that measures the minimum distance through the average shortest path from a node to all other nodes; and also other more advanced centrality measures (Scott, 2000; Stanley Wasserman & Faust, 1994).

### a. On-park collaboration network

To analyse the inter-institutional collaboration network among on-park organisations, a simulated annealing method was applied to identify the different groups or modular structure of the network and functional role of its actors based on the link type frequency. The simulated annealing method is "a stochastic optimization technique that enables one to find 'low cost' configurations without getting trapped in 'high cost' local minima." (Guimera & Amaral, 2005: 2). This stochastic approach goes beyond approaches that focus on degree and global properties alone, and is most accurate for small networks (Guimera & Amaral, 2005). The algorithm "assesses the significance of the modular structure of each network by comparing it with a randomization of the same network" (Guimera, et al., 2007: 63). The role of each node is then determined by two properties; (1) "relative within-module degree z, which quantifies how well connected a node is to other nodes in their module, and (2) the participation coefficient P, which quantifies to what extent the node connects to different modules" (Guimera et al., 2007: 63). It basically classifies nodes into seven universal roles, according to their pattern of intra- and inter-module connections, defined by their within-module degree and their participation coefficient. These roles are divided into four non-hubs (within-module degree < 2.5):

- (R1) ultra-peripheral nodes, that is, nodes with all their links within their module;
- (R2) peripheral nodes, that is, nodes with most links within their module;
- (R3) satellite connector nodes, that is, nodes with a high fraction of their links to other modules;
- (R4) kinless nodes, that is, nodes with links homogenously distributed among all modules.

The hubs (within-module degree ≥ 2.5) are:

- (R5) provincial hubs, that is, hubs with the vast majority of links within their module;
- (R6) connector hubs, that is, hubs with many links to most of the other modules;
- (R7) global hubs, that is, hubs with links homogenously distributed among all modules (Guimera, et. al., 2007: 63-4).

### b. SP and off-park collaboration networks

To map the role of HEIs, as the main external partners of the SP movement and the degree of reliance of the SP movement on external organisations, inter-institutional collaboration between on- and off-park organisations was used. This asymmetric relationship was represented as a bipartite network with two sets of nodes, SPs (teams) and off-park organisations (actors), where a link indicates collaboration between the off-park actor and one of the on-park organisations. This bipartite network was analysed with the help of the model proposed by Guimera and his colleges (2007; Sawardecker, et. al., 2009) based on modularity maximization through simulated annealing (described above) and identifies "groups (modules) of actors that are strongly connected to each other through co-participation in many teams". This method's accuracy outperforms spectral decomposition, bipartite recursively, and (bi)clique percolation (Sawardecker et al., 2009).

The bipartite affiliation network was then projected onto a SP-SP (column-column or team-team) network to identify the structure of the SP movement based on the collaborative patterns the SPs share with off-park organisations. In the transformation the values were normalised through the different similarity measures (calculated in UCINET 6), weighting rows (off-park organisations) inversely by row size (Wasserman & Faust, 1994: 312-14, 338). Additionally, the method Pathfinder network was applied to reduce the density of the network caused by the high concentration of a few HEIs as main SPs' partners. This uncovered the main network structure as only the shortest path between two nodes remained (Quirin, et. al., 2008).

### c. Firms-HEIs-RIs collaboration network

In the third network, formed by collaborations of at least 7 publications among HEIs, firms, and RIs, the method Pathfinder network was also applied to simplify the high density of a symmetric network of 300 edges representing 4,085 collaborations. This threshold was primarily established to facilitate the visual representation of the network as the full network is then used for further analyses and representation. It helped to reduce the links to 107 (36%) and represents 1,913 (47%) collaborations. In this inter-institutional collaboration network the nodes represent three different types of organisations that are linked by bidirectional links, edges, representing co-authorship between two organisations. The social boundary of the networks was imposed by the joint creation of research in partnership with any organisation located in any support infrastructure in the UK.

### 3.3.6.    Unstructured dataset: hyperlinks

To obtain data on the SP link networks, the websites of the *Advanced Manufacturing Park* (www.attheamp.com), Leeds Innovation Centre (www.leedsinnovationcentre.com), and York

Science Park (www.yorksciencepark.co.uk) were crawled with SocSciBot in May 2010. The web crawler identified site outlinks to websites with potentially a formal or informal relationship with the SPs. The crawler collected 215 site outlinks which were then manually checked to identify the type of relationship with the SPs (see Table 3.1). Each targeted website was classified by sector (Industry, Academia, Government), following similar criteria used by Ortega (2003), and according to nine different categories that may be relevant within R&D networks (Etzkowitz, 2008; Howells, 2006; OECD, 2002). Websites that were deemed to be irrelevant were excluded (e.g. maps.google.co.uk; twitter.com; nationalexpress.com; adobe.com; youtube.com; news.bbc.co.uk). The number of relevant websites identified was 190, but after reducing the URLs to their respective domains (e.g., wlv.ac.uk) or sub-domains (e.g., cybermetrics.wlv.ac.uk) to assign each (sub-)domain to an organization and minimise the impact of multiple levels of websites, the number of organizations was reduced to 183 (see Table 3.1). Therefore, the total number of websites analysed (including the SPs) in the last study was 186 (Chapter 7). The problem of websites with multiple domain names were resolved through the selection of the domain name with the highest number of pages, inlinks, outlinks, and the longest time of activity according to Internet Archive.

**Table 3.1** Number of websites linked to by the SPs and their manually classified type of relationship with the SPs.

|  | Adv. Man. Park | Leeds Inn. | Yorkshire SP | *Total* |
|---|---|---|---|---|
| External links | 33 | 54 | 123 | 210 |
| Organisations | 32 | 48 | 103 | 183 |
| **Type of relationship** | | | | |
| Tenant | 26 (81%) | 34 (71%) | 81 (79%) | 141 (77%) |
| Information | | 3 (6%) | 18 (17%) | 21 (11%) |
| Support | 3 (9%) | 4 (8%) | | 7 (4%) |
| Partnership | | 5 (10%) | 2 (2%) | 7 (4%) |
| Other | 2 (6%) | | 1 (1%) | 3 (2%) |
| Membership | | 2 (4%) | 1 (1%) | 3 (2%) |
| Incubator | 1 (3%) | | | 1 (1%) |

### 3.3.7. Interlinking-analysis: Combining three data sources (SocSciBot, Yahoo! and Bing)

Once the 186 organisations (websites) to be analysed were identified, in order to develop an effective and reliable method to map and analyse the cross-sectoral interactions and relationships established by means of hyperlinks between the websites, the datasets for the analysis were collected from three different sources. The complexity of the web and the low overlap in coverage by search engines makes it necessary to combine various tools and sources to obtain the most reliable results (Thelwall, 2008a). For this reason, the web crawler SocSciBot and the commercial search engines Yahoo! and Bing were all used to gather the data. Interlink analysis was applied to reveal the existing interconnections between the set of websites and make it possible to study the bi-directionality of each data set (in/outlinks). In addition, the analysis of inter-links in closed environments enables us to explain the meaning of the links between the set of co-members, because the origin and destination of the links belong to the community studied. Due to the lack of studies which use different data sets to compare the links between peers, it was first necessary to compare the results from the networks based on the in- and outlinks to observe if similar structures are obtained, as well as the differences and similarities to identify which data might provide the best results.

The set of site inlinks (links pointing to a web domain or sub-domain from another domain or sub-domain) were retrieved from the commercial search engine Yahoo! with the help of the free Webometric Analyst software, which is a social science web analysis tool developed by Thelwall (2004) and can be found at [http://lexiurl.wlv.ac.uk](http://lexiurl.wlv.ac.uk). This software, also automatically split queries with results exceeding the maximum of 1,000 hits permitted when using Yahoo! for this purpose (Thelwall, 2008b). This gave up to 19,619 inlinks per query.

The set of site outlinks (links pointing from the domain or sub-domain to another domain or sub-domain) were collected using both the web crawler SocSciBot and the search engine Bing. Combining the web crawler and search engine data was designed to minimise the bias and limitations of each tool and achieve the most complete list of outlinks possible. Web crawlers automatically download a web page, extract hyperlinks from that page, and then download these web pages to extract more hyperlinks in a continuous process until there are no more hyperlinks to follow. Without ethical principles, the use of these fast and powerful tools to extract information over the Internet can cause network problems in terms of cost, denial of service and privacy, and can also to financial penalties to the owner of the crawled web sites (Cothey, 2004; Thelwall & Stuart, 2006). Due to the time and resources required by the web crawler, each website was crawled to a search depth of two levels only. The crawler made it possible to collect updated information and include sub-domains in the second data set since Bing is not able to retrieve the outlinks of sub-domains. On the other hand, Bing's coverage complemented the relatively superficial crawling when large websites were analysed. The outlinks were collected from Bing through Webometric Analyst and the capability to split the queries with more than 1,000 results was used as well. The comparison and combination of the results was important to improve the quality and quantity of the data set, as illustrated by the low average (4%) overlap among site outlinks.

The number of inlinks collected through Yahoo! was 337,911 and the number of outlinks gathered through SocSciBot and Bing were 6,597 and 104,890 respectively. After the links were reduced to (sub-)domains, the duplicates of each website were eliminated and the outlinks were combined, 183,006 inlinks and 80,588 outlinks remained. This reduction of 46% for the inlinks and 28% for the outlinks suggests that a high proportion of inlinks tend to come from the same group of websites while the outlinks tend to be spread around a wide range of websites. In addition, Yahoo! having 183,006 inlinks and Bing having only 77,657 outlinks suggests that the websites studied do not link to the same extent as they are linked to. Even taking into account that Bing does not allow retrieving outlinks of sub-domains, which was relevant to only 7 (3.7%) websites of the sample with 11,594 (6%) inlinks, the in-data set through Yahoo! still returned 90,824 (113%) more links than the out-data set through Bing. Assuming that the results of the two sources have similar levels of accuracy, this means that the websites receive twice as many links than the ones they give. It is important to note that this comparison reflects two different behaviours (inlinks and outlinks) and not the coverage of both search engines, although the number of in- and outlinks returned could be biased towards the coverage of both search engines, and the type of organisations studied.

It is important to point out that the advanced link search queries used to collect the datasets are no longer available and should be replaced by the different alternatives and data sources proposed (Thelwall, Sud, & Wilkinson, 2012; Thelwall, 2011; Vaughan & Yang, 2012).

### 3.3.8. Inter-linking analysis: matrix

The inlinks and outlinks collected were exported to a relational database constructed for the webometric study, and the different bi-directional adjacency matrix (Actors x Actors) were constructed for each SP and dimension. The frequencies with a value greater than or equal to 1 were dichotomised to measure the number of websites that were interconnected rather than the intensity of the connections, and the diagonal entries (self-links) were eliminated. Dichotomization is recommended because the frequencies of the direct links established between a pair of heterogeneous websites may not be representative of offline relationships, leading to misunderstandings. The differences in the link behaviour among various types of websites makes it necessary to only take connections involving a large number of direct links into account to identify patterns of the strength and direction of a relationship in the network. Finally, the networks were represented and analysed with the help of Social Network Analysis (SNA) and visualisation techniques. Different cohesion and centrality measures were calculated using UCINET and represented with NetDraw.

In summary, the web-based data used in this exploratory study was divided in two different unidirectional datasets, one based on inlinks collected with Yahoo! and another based on outlinks collected with SocSciBot & Bing, which together contain all the links pointing to and from the websites. The scheme to obtain the interlinking networks, based on either the in-data set or the out-data set, then consists in identifying only the links between the websites of the sample within each data set (see Figure 3.1). These interconnections between peers should form the same network from either the inlinks or outlinks. However, due to the differences in the gathering of the datasets and the inherent shortcomings and capabilities of the different sources, both datasets were compared in order to see the features of each one and determine how they could be used (see Chapter 7).

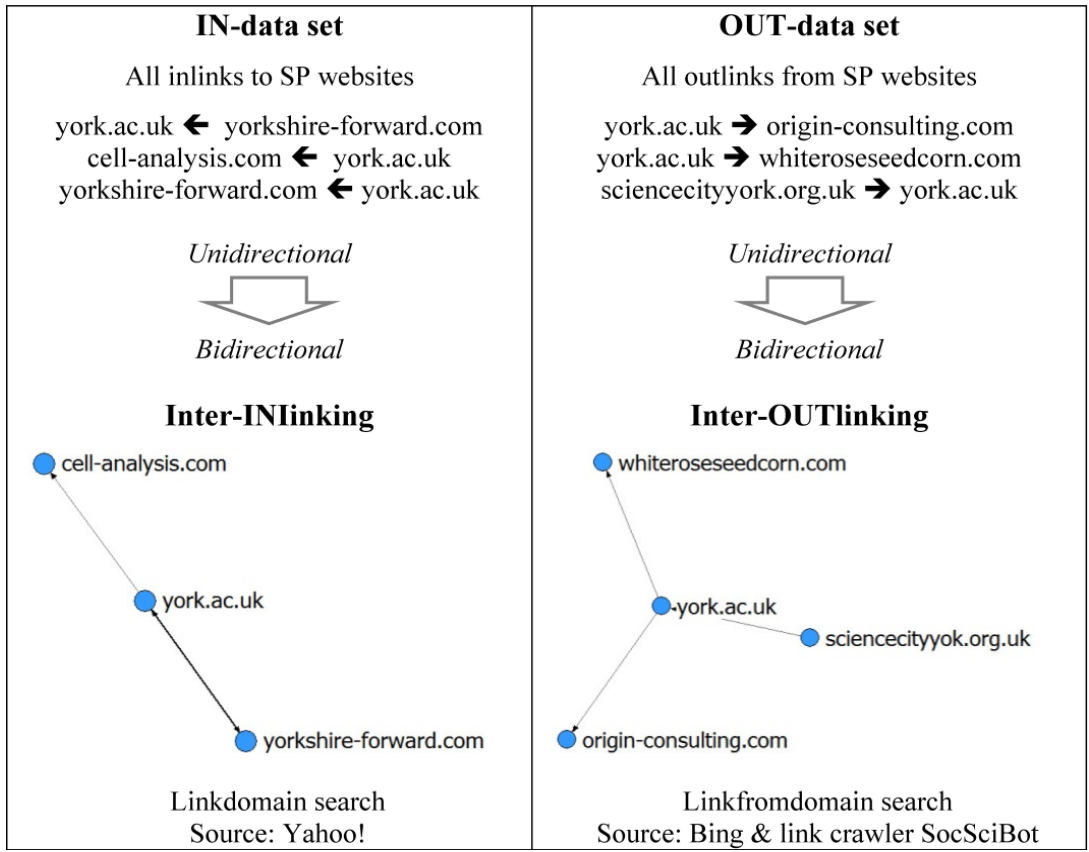

**Fig. 3.1** Example of how unidirectional hyperlinks (in-links or out-links) of a set of websites form bidirectional connections (interlinking networks) that allow the use and combination of different data sources.

# Chapter 4: R&D activity in the UK SP movement

## 4.1 Introduction and goals

This first empirical chapter investigates whether scientific publications can give plausible suggestions about whether UK science parks and similar support infrastructures successfully foster scientific cooperation and innovation. For this, research publications associated with UK SPs (Section 3.3.2) are analysed by region, infrastructure type and organisation type. This study analyses the capacity of the UK SP movement to encourage and generate R&D. This is a first step to examine whether scientific publications can be used as a valid indicator for monitoring the R&D ability fostered by SPs, and then gain a fuller understanding of their impact. This objective is summarised in the following research question:

- Can scientific publications help to shed light on the R&D activities fostered by SPs and other similar support infrastructures?

In order to address this question, evidence about the following are extracted from a bibliometric analysis of scientific research publications:

(1) what types of SPs and support infrastructures are established across the country, and which infrastructures are the most research-intensive;
(2) where these are established and which type of organisations engage in these environments;
(3) how do onsite-organisations collaborate, including collaboration between on-park firms and knowledge producers, and how do these links extend across the country and beyond;
(4) which subject areas attract most on-park research and what is the contribution of the geographic regions and what is the level of U-I collaboration across different subject areas;
(5) does research production associated with SPs have a higher quality and impact than the average research across the different subject areas.

These aspects provide an insight into the R&D activities and U-I links that are expected to be fostered by the different support infrastructures, and the extent to which on-park research is integrated into the wider scientific community.

## 4.2 Results and Discussion

As background information, the data set extracted from *Scopus* outperforms the *Web of Science* in terms of representing the heterogeneous publication output of the mainly private oriented research community associated with the SP movement (see Figure 4.1). The coverage of WoS and Scopus seems to be very similar until the mid-1990s, after which Scopus exhibits an exponential growth compared to the flat and even decreasing WoS coverage. No bias that would account for the difference could be identified by the publication sources or type of sources indexed by Scopus, as demonstrated by the normal distribution of the top 30% largest journals in Scopus (see Figure 4.1). The WoS output trend confirms previous findings indicating that WoS-indexed research produced by industry is steadily declining (Tijssen, 2004). These findings suggest that the publication output of the SP movement is underrepresented in WoS. Scopus' broad coverage policy, with about 70% more sources than WoS (López-Illescas, Moya-Anegón, & Moed, 2008), offers more comprehensive coverage of industrial research. This is especially true when conference proceedings are important (Meho & Rogers, 2008). The likely underrepresentation of private research in WoS represents a significant limitation for U-I studies, as any conclusions drawn are limited by the properties of the bibliographical database used.

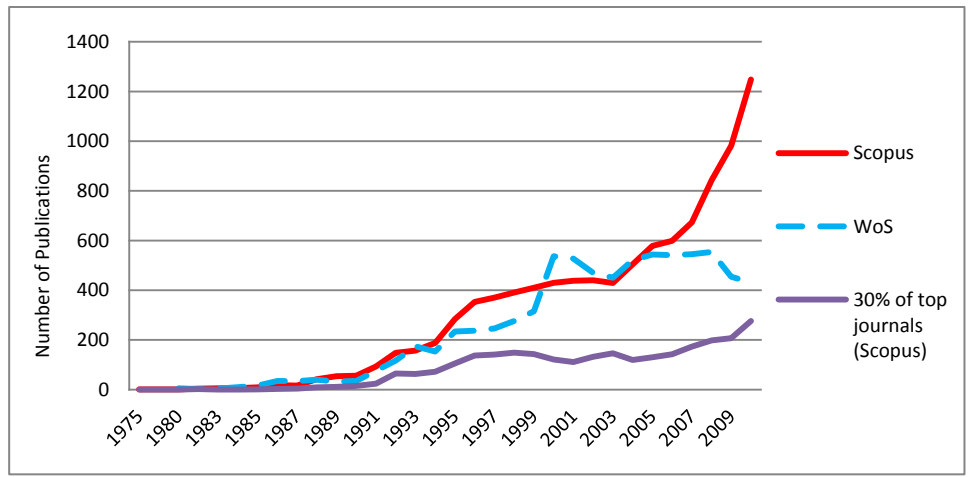

**Fig. 4.1** Publications from the UK SP movement from 1975 to 2010 extracted from Scopus (n=10,920) and WoS (n=8,057).

### 4.2.1. Historical development of the SP movement in the UK

A trend analysis of research publications sheds light on the historical development of the SP movement, and the degree of research activity relating to the various types of SP infrastructures during the last three decades. The increase in the number of SPs that are involved in research (Figure 4.2) coincides with the constant growth in the number of publications (see Figure 4.3). Before the 1990s there were, on average, 4.5 research-publishing SPs every year. During the 1990s this increased to 24.5 and in 2010 61 SPs published research. Similarly, the output trend started to become substantial at the beginning of the 1990s, reaching over 400 publications in 2000, with a further three-fold increase by 2010.

The first research papers in the UK SP movement appear to belong to the first university spin-off in the country (*Edinburgh Instruments*) and the first two infrastructures, which were set up in 1971, *Heriot-Watt Research Park* and *Cambridge SP*. Then came *Birmingham Aston SP* (1981), *Wilton Centre* (1977), and then the research production of a wave of SPs established in the mid-80s. This development was not formally guided by government policy but 62% of the investments were provided by the public sector (Gower & Harris, 1994), which at the same time reduced public funding for universities to (1) encourage HEIs to collaborate with manufacturing industry to gain access to external resources, and (2) encourage industry to gain knowledge and resources to enhance its competitiveness (Guy, 1996; Vedovello, 1997; Westhead & Batstone, 1998; Westhead & Storey, 1995). The trend of the research-oriented infrastructures shows that at the beginning only the oldest and probably most dynamic infrastructures were capable of promoting technology transfer. There is a gap between the trend of UKSPA members and that of research-active infrastructures. The latter needed nearly ten years to mature to a level where most seem to be regularly research active, as reflected by the high indices of occupancy, job- and firm-creation of the movement by 2000 (UKSPA, 2012a).

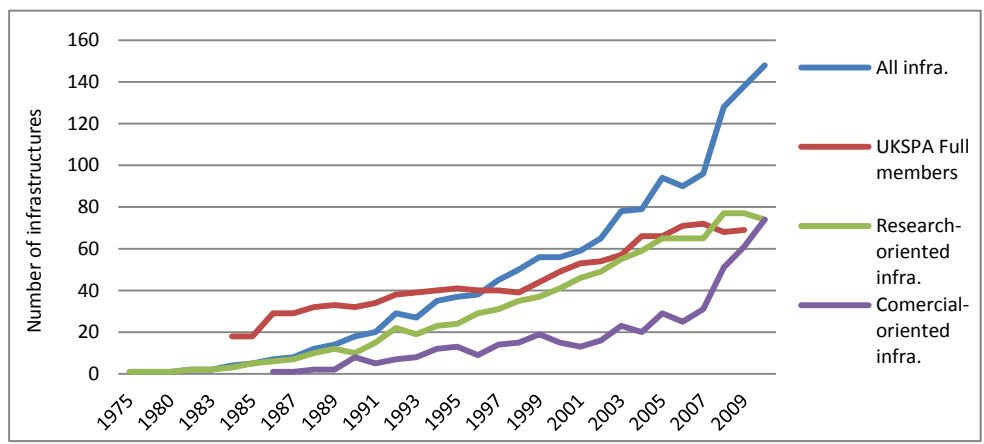

**Fig. 4.2** Comparison between the number of research- and commercial infrastructures producing research publications in each year with the number of UKSPA full members from 1975 to 2010.

In contrast to the first wave of SPs, the second wave had active commitments from HEIs to support the regional economy, with management goals targeted at technology transfer (Hansson et al., 2005). Active involvement from the private sector and dedicated policy strategies at regional, national and European levels in the late 1990s (Storey & Tether, 1998; Wynarczyk & Raine, 2005) seemed to make considerable progress toward improving technology transfer (HM Treasury & DIUS, 2008). However, according to Wainova, the UKSPA covers about 80% of SPs across the country and only in the last three years have there been more research-based infrastructures[2] than UKSPA members, suggesting that not all UK infrastructures that should foster research transfer host R&D-active organisations. Interestingly, the total number of 319 infrastructures identified in this study (i.e., Science-, Technology-, Innovation-, and Research Parks, Incubators, Research Campuses, Science & Innovation Centres, Industrial- and Business Parks) is similar to the estimate of the UK Business Incubation (UKBI) professional organisation, that the broad SP movement is formed by approximately 300 infrastructures in the UK. Furthermore, according to the number of on-park companies located on the premises of the UKSPA infrastructure members (UKSPA, 2012a) this study covers 31% of the tenants, similar to the proportion of on-park organisations that pursue a competitive strategy focused on radical research (Westhead, 1997).

Commercial-based infrastructures[3] started to become research active from the 1990s and have grown rapidly in number during recent years, whereas the number of UKSPA full members and research-oriented infrastructures has levelled off, indicating stagnation in the creation of new SPs. The high proportion of commercially-oriented infrastructures is the result of less resources and effort being needed in comparison with the research-oriented ones, but their research capacity seems to be low. Despite Business Parks (BPs) being the most visible commercially-oriented infrastructures, they are the fourth most important subcategory of infrastructure in terms of research publications (with only 584 publications spread across 174 different BPs, representing only 3.4 publications for each BP) (see Figure 4.3), while the subcategories science- and research parks produced 107 and 315 publications each on average respectively. This significant difference reveals the sporadic production of the commercially-oriented infrastructures. It is not clear whether BPs are actively supporting R&D activities as a way to tackle the lack of public funding and investments, of knowledge-based companies, and in general to become more attractive and improve their reputation (Gower & Harris, 1994). The growing interest from commercially-oriented business parks to promote R&D activities as a means to add value to the products and services of their tenants gives new opportunities for further expansion of the SP movement, as it has been able to redefine itself to attempt to nurture greater research production in the last two decades.

---

[2] Research-based infrastructures host a majority of tenants which heavily engage in basic and applied research and have formal association with research producer, hosting or being physically close to HEIs and RIs units. Under this category can be found; Science parks, Technology parks, Research parks or campuses, Incubators.

[3] Commercial-based infrastructures do not necessarily have operational links with research producers and provide high quality accommodation to tenants which engage in a wide variety of activities that are not necessarily based on R&D activities.

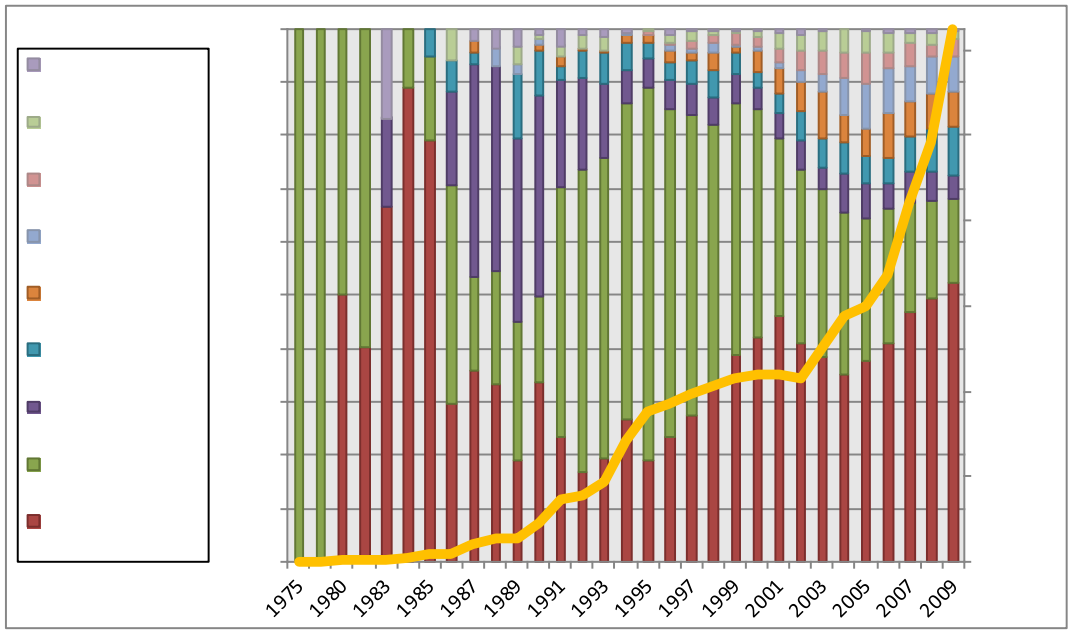

**Fig. 4.3** Infrastructures'[4] share of publications produced in each year (left) in relation with the total number of publications (dotted line) produced in each year (right). (Source: Scopus 1975-2010).

During the first five years discussed only two Research and Science parks appeared, located in Scotland and the East of England. In 1984-87 the first Innovation centres, Industrial- and Business parks emerged, and the SP movement spread across nine regions. In the next five years there was a total increase in the number of infrastructures, from 22 in 1988 to 53 in 1992, and

---

[4] This classification is primarily based on the official names of the support infrastructures as there is no taxonomy or classification scheme of support infrastructures (see Section 2.4.1.).

Research Campuses and Technology Parks were set up. The following period had a considerable increase in Research and Science Parks, reaching 27 (34%) different infrastructures in 1995 and sharing 46% and 26% of the total publications respectively, while Science & Innovation Centres produced 18% of the publications. This led to steady publication growth in the following years. In the period 1996-99 the development continued with Incubators as the newcomers, and during the early years of this century the first Bioincubator, BioCentre, and Science Centres appeared, while Business Centres and Enterprise Parks started to get involved in R&D activities later. In the last fifteen years the publications was still predominantly from Science- (40%) and Research Parks (32%) despite the high variety and maturity of the infrastructures, while other infrastructures such as Business Parks (6%) and Technology Parks (5%) only played a peripheral role in spite of their recent increase in publications.

### 4.2.2.    Regions and types of organisations involved in the SP movement

Traditionally, the SP movement has been driven by the assumption that physical proximity between industry and sources of knowledge should trigger innovation (A. N. Link & Scott, 2003; Vedovello, 1997). Since these policy instruments are oriented to support research-intensive industries as a means to generate growth and employment by adding synergy and dynamism to a socio-economic context, SPs are also essential for the development of research intensive clusters (Saublens, 2007). Therefore, the research activity and types of organisations of the movement across regions are identified in this section in order to observe how it is spread across regions, if it forms part of larger clusters and if it could be related to regional innovative performance and which organisations are the major driving forces for research visibility in the SP movement.

As Table 4.1 reports, 52% of the SP Scopus publications are from the East of England followed by the South East (14%) and Scotland (12%), and these regions also have the highest number of onsite organisations. In contrast, Northern Ireland, London and Wales have the lowest publication rates. In terms of established infrastructures, the largest agglomerations are in the South East (20%), the East of England (13%), and Scotland (13%). The high visibility of these areas reflects the fact that knowledge creation and absorptive capacity have a direct impact on the industrial competitiveness. This is shown by these particular three regions, excluding inner London, being the major centres of biotechnology in the UK according to the knowledge-space dynamics produced by the interaction and synergy of capabilities embedded in these regions (biotech firms, public science basis and funding, patents and publications, biotech organisations, and formal alliances) (Birch, 2009; Cooke, 2001b; Kitson, Howells, Braham, & Westlake, 2009).

A closer look at the patterns of the two groups of infrastructures reveals the research or commercial orientation of the different regions (see Table 4.1). Those with the most infrastructures, such as the East of England and Scotland, have a balance between research- and commercial oriented ones and their main research publishing is based on the former. The South East, despite being a research-active region, also provides the largest research-publishing commercial structure across the country, having a high visibility in both areas due to the activity and number of industrial and scientific organisations. All this agrees with previous findings (Birch, 2009; Cooke, 2001b) that the South East is more market-driven whilst the East and Scotland are university-based clusters. Other regions with a probable commercial identity are the South West and the North West, as shown by their large commercial structures. However, their limited research capacity suggests a similar situation to the other regions. Porter (2003) also highlights the lack of university-company interaction outside the life sciences and the university-based cluster around Oxford and Cambridge. In general, the SP movement depicts a gap between the South and North of the country, with the exception of Scotland and the North West.

A similar regional divide is found in SP funding and establishment policy, with the South being largely privately funded and the rest of the country being publicly supported (Quintas et al., 1992), with infrastructures located in prosperous environments in the South becoming relatively more developed, and those established in depressed and disadvantaged areas in the North aiming to encourage industrial regeneration (Westhead & Batstone, 1998). Finally, the low position of London is clearly influenced by a scarce SP infrastructure, which makes it impossible to uncover its science and technology-based industry or leading HEIs. A similar example is the North West, where the Daresbury S&I Campus was set up in 2005 to maximise the process of technological commercialisation around the four decades old Daresbury Laboratory (Kitson et al., 2009). Figure 4.4 displays maps of the agglomerations of infrastructures and outputs across the UK.

The two southern regions occupy top positions in the UK Competitive Index (UKCI), a composite index that benchmarks regions and localities based on a set of factors that reflect the link between macro-economic performance and innovative business behaviour (Huggins, 2003). Furthermore, there are significant correlations between the number of infrastructures (0.63), output (0.70), and onsite-organisations (0.80) with the UKCI input factors (see Table 4.2). This suggests that the SP movement plays a significant role in the design of regional innovation strategies and to some extent might reflect the degree of competitiveness where they are located. However, it is not clear whether SPs are adequate mechanisms to regenerate declining industries and less competitive areas, rather than intermediaries that support and maximise the exploitation of already existing dynamism and learning capabilities embedded in innovative areas, as shown by the infrastructures within biotech clusters. Also, to what extent could policies be successful in promoting SPs in areas that lack the support capabilities and research basis of dynamic and innovative areas? Besides this, the relationship found between the SP movement and biotech concentrations, as well as the concentration of the movement's research production in life science and bio-related fields (see Figure 5), indicates that, at least among research-active parks, the SP movement provides or exploits the right conditions for the expansion of high technology industries.

**Table 4.2** Pearson correlations between the research activity (in terms of number of infrastructures, publications, on-park organisations) across the UK SP movement and regional competitiveness indexes (input[5] & output[6]).

|  | Competitiveness Index - Input | Competitiveness Index - Output |
|---|---|---|
| Infrastructures (without London) | 0.63 (0.81) | 0.36 (0.67) |
| Output (without London) | 0.70 (0.80) | 0.44 (0.64) |
| On-park org. (without London) | 0.80 (0.92) | 0.56 (0.80) |

Regarding the type of organisations that produce research and technology publications, industry (48%) and RIs (44%) are the main onsite producers. Figure 4.5 shows how the activity trend of the private sector has grown exponentially since the 1990s and experienced a striking upward growth after 2005. RIs, on the other hand, were the major research producers during the 1990s and although they have kept growing, businesses now lead output. This suggests that industry's passive role as a traditional knowledge consumer might have changed to a learning-by-doing process to actively producing specific knowledge and remaining plugged into the scientific network, particularly in high tech sectors where there is high dependence of high quality research (Marston, 2011). However, this distribution varies in relation to the type of infrastructure. Figure 4.6 shows that, although this categorisation is subjective in the sense that it is based on the names of the parks, the R&D activities found in the groups of infrastructures are in line with the research intensity expected according to their definitions. Thus, the output of public science producers tends to be based in research-based environments, such as research parks and campuses, which have more than three quarters of the output, while a more diverse

---

[5] Input factors: R&D Expenditure; Economic Activity Rates; Business Start-up Rates per 1,000 inhabitants; Number of Business per 1,000 inhabitants; GCSE Results - 5 or more grades A* to C; Proportion of Working Age Population with NVQ Level 4 or Higher; Proportion of Knowledge-Based Business (Huggins & Thompson, 2010).

[6] Output factors: Gross Value Added per Head at Current Basic Prices; Exports per Head of Population; Imports per Head of Population; Proportion of Exporting Companies; Productivity - Output per Hour Worked; Employment Rates (Huggins & Thompson, 2010).

**Table 4.1** Research intensity of the SP movement by region. (a) Number of infrastructures, publications, on-park organisations and the association with the competitiveness index. (b) Number of infrastructures and publication of research- and commercial-oriented infrastructures. (c) Proportion of publications and organisations across five types of organisations.

| Region | SP Movement (a) | | | | | Infrastructures (b) | | | | Organisations (c) | | | | | | | | | |
|---|---|---|---|---|---|---|---|---|---|---|---|---|---|---|---|---|---|---|---|
| | Infras. | Output | Onsite organisations | Competitiveness Index - Input | Competitiveness Index - Output | Research-oriented infra. | | Commercial-oriented infra. | | Organisations' output share | | | | | Organisations' presence share | | | | |
| | | | | | | Infras. | Output | Infras. | Output | Firms | RIs | HEIs | Gov | NPO | Firms | RIs | HEIs | Gov | NPO |
| East Midlands | 20 *6.3* | 205 *1.978* | 66 *6.203* | 92.1 | 90.2 | 10 *50* | 185 *90.244* | 10 *50* | 20 *9.7561* | 0.03 | 0.00 | 0.02 | 0.00 | 0.15 | 0.06 | 0.05 | 0.08 | 0.00 | 0.13 |
| East of England | 41 *12.9* | 5335 *51.33* | 240 *22.56* | 119.4 | 107.6 | 20 *48.78* | 5261 *98.614* | 21 *51.22* | 74 *1.3858* | 0.40 | 0.68 | 0.33 | 0.04 | 0.28 | 0.23 | 0.21 | 0.12 | 0.08 | 0.25 |
| London | 11 *3.4268* | 120 *1.158* | 48 *4.511* | 107.7 | 115 | 6 *54.545* | 111 *92.5* | 5 *45.455* | 9 *7.5* | 0.02 | 0.00 | 0.03 | 0.00 | 0.02 | 0.05 | 0.00 | 0.04 | 0.00 | 0.06 |
| North East England | 10 *3.1* | 493 *4.743* | 32 *3.008* | 82.0 | 82.7 | 4 *33.333* | 477 *96.781* | 6 *50* | 16 *3.2193* | 0.09 | 0.00 | 0.02 | 0.09 | 0.00 | 0.03 | 0.00 | 0.04 | 0.08 | 0.00 |
| North West England | 35 *11* | 506 *4.868* | 86 *8.083* | 97.6 | 86.4 | 11 *31.429* | 383 *75.692* | 24 *68.571* | 123 *24.308* | 0.06 | 0.03 | 0.07 | 0.01 | 0.11 | 0.08 | 0.08 | 0.08 | 0.04 | 0.06 |
| Northern Ireland | 5 *1.5576* | 90 *0.868* | 8 *0.752* | 84.7 | 82.5 | 3 *60* | 85 *94.444* | 2 *40* | 5 *5.5556* | 0.00 | 0.00 | 0.11 | 0.02 | 0.00 | 0.01 | 0.00 | 0.04 | 0.04 | 0.00 |
| Scotland | 39 *12.2* | 1290 *12.41* | 140 *13.16* | 93.2 | 77.3 | 18 *46.154* | 1235 *95.736* | 21 *53.846* | 55 *4.2636* | 0.08 | 0.18 | 0.04 | 0.34 | 0.07 | 0.13 | 0.26 | 0.12 | 0.15 | 0.13 |
| South East England | 63 *19.7* | 1415 *13.61* | 209 *19.64* | 113.7 | 114.8 | 18 *28.571* | 1275 *90.106* | 45 *71.429* | 140 *9.894* | 0.18 | 0.10 | 0.08 | 0.11 | 0.04 | 0.20 | 0.18 | 0.15 | 0.19 | 0.06 |
| South West England | 32 *9.9688* | 236 *2.277* | 55 *5.169* | 93.8 | 82.8 | 5 *15.625* | 136 *57.627* | 27 *84.375* | 100 *42.373* | 0.02 | 0.00 | 0.17 | 0.17 | 0.24 | 0.05 | 0.03 | 0.08 | 0.12 | 0.13 |
| Wales | 24 *7.4766* | 147 *1.418* | 46 *4.323* | 80.0 | 76.1 | 9 *37.5* | 103 *70.068* | 15 *62.5* | 44 *29.932* | 0.02 | 0.01 | 0.03 | 0.07 | 0.00 | 0.04 | 0.13 | 0.08 | 0.04 | 0.00 |
| West Midlands | 20 *6.3* | 385 *3.714* | 81 *7.613* | 85.7 | 88.8 | 9 *45* | 341 *88.571* | 11 *55* | 44 *11.429* | 0.06 | 0.00 | 0.06 | 0.09 | 0.04 | 0.07 | 0.03 | 0.12 | 0.15 | 0.13 |
| Yorkshire and the Humber | 19 *6* | 172 *1.659* | 53 *4.981* | 82.9 | 82.3 | 7 *36.842* | 140 *81.395* | 12 *63.158* | 32 *18.605* | 0.03 | 0.00 | 0.03 | 0.06 | 0.04 | 0.05 | 0.05 | 0.08 | 0.12 | 0.06 |
| **Total** | 319 | 10394 | 1064 | | | 120 | 9732 | 202 | 662 | | | | | | | | | | |

Competitiveness Index Source: www.cforic.org

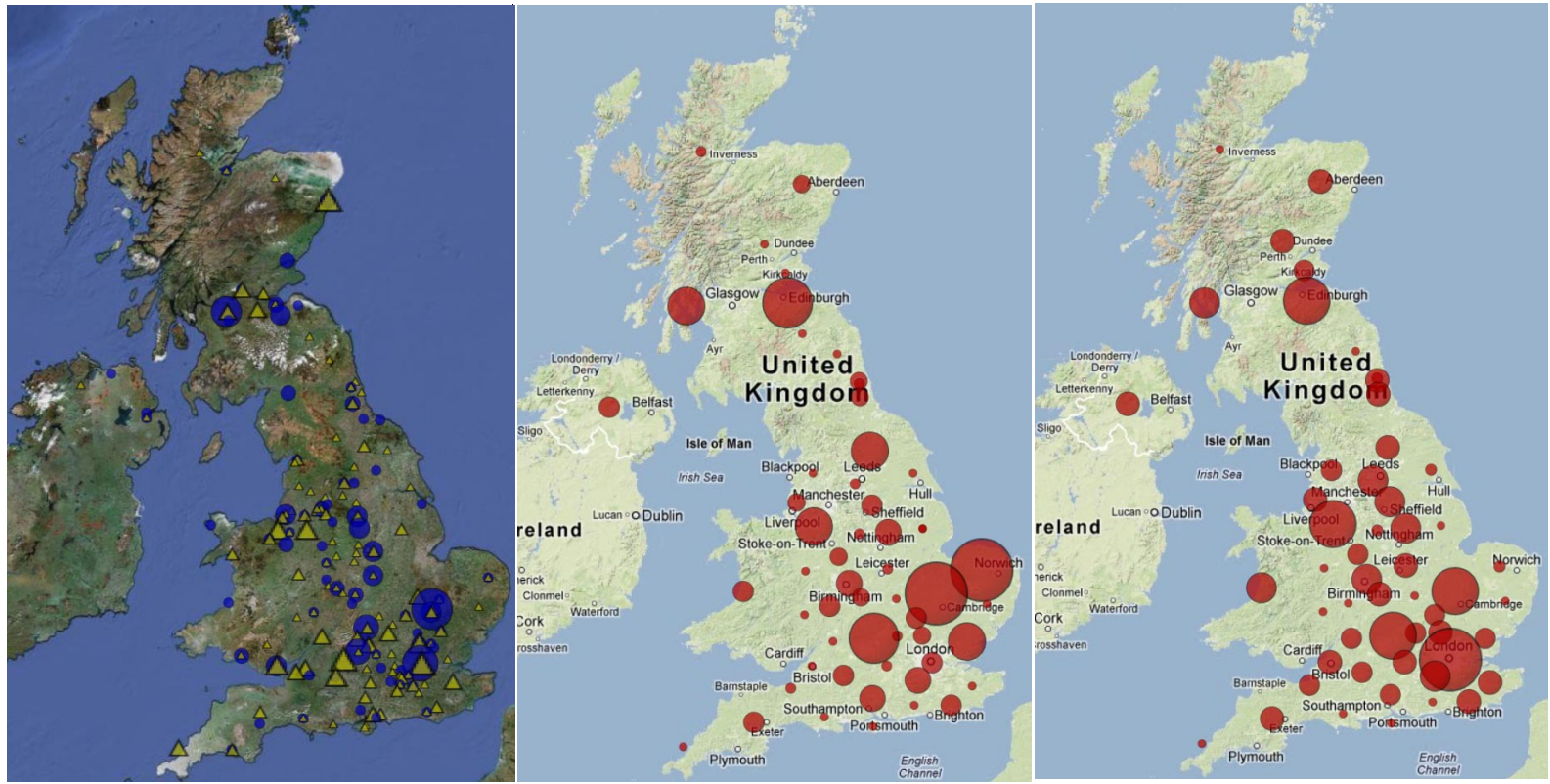

**Fig. 4.4** (Left figure) Number of research-oriented infrastructures (blue nodes) and business-oriented infrastructures (yellow nodes) agglomerated by district. Nodes size represents the number of infrastructures established in each district. [Interactive version available at: *http://home.wlv.ac.uk/~in1493/sp-movement.html*]; (Center figure) Research output of on-park organisations agglomerated at county-level; (Right figure) Research output of off-park organisations agglomerated at county-level. Red nodes size represents the number publications in each county. Source: Scopus 1975-2010

collection of research producers is primarily found in science and technology parks. This is a result of the majority of high-tech tenants heavily engaging in basic and applied research and development. Incubators with prominent private activity reflect the limited capacity of newly founded ventures to undertake research. On the other hand, the infrastructures with no formal ties with HEIs, such as innovation parks, science & innovation centres, and business & industrial parks, reveal the limited R&D capacity of their research-intensive tenants.

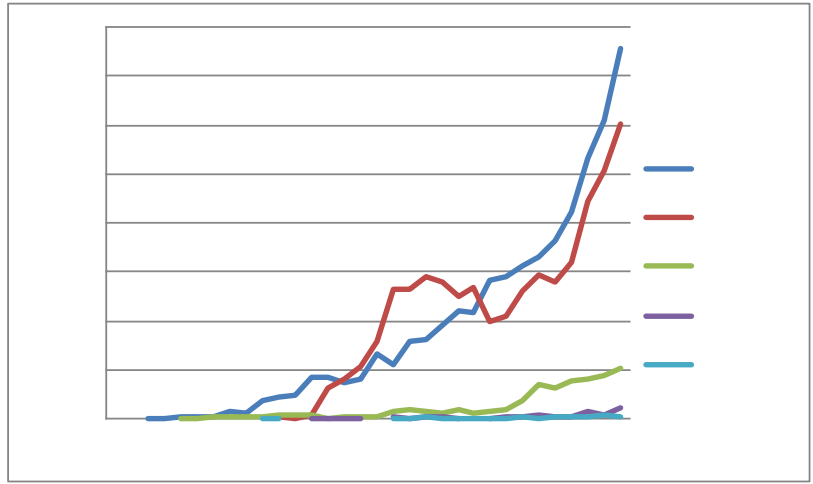

**Fig. 4.5** Chronological output of research publications in terms of on-park organisation type (Source: Scopus; 1975-2010; n= 10,920).

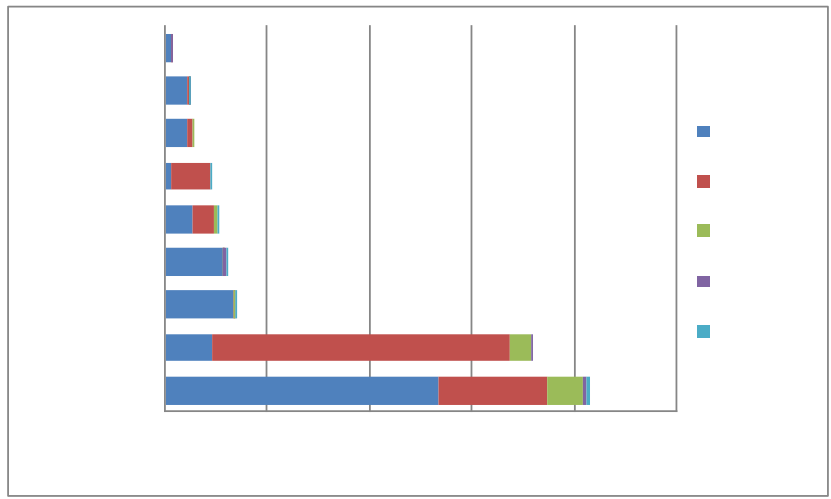

**Fig. 4.6** Number of publications of the different types of support infrastructures in terms of on-park organisation type (Source: Scopus; 1975-2010; n=10,920).

Although the aggregate data shows that the research publications are produced by organisations that are usually based on the expected infrastructures and it helps to have a more accurate distinction between infrastructures, this conceals important differences between infrastructures. For example, the fifteen most research-active infrastructures show how R&D activities differ among similar infrastructures (see Table 4.3). In regard to the region and type of infrastructure, again the East of England, the South East, and Scotland, and the science- and research parks dominate the top positions in terms of research output. Cooke (2001) considers this as an important catalyst in the development of the biotechnology industry in the three regions. However, the differences regarding the organisations driving R&D publications within each park give three main profiles:

- parks whose production is the result of an anchor research institution (Norwich RP, Pentlands SP, Babraham RC, Scottish Enterprise TP, Roslin BioCentre);
- parks whose production relies on relocated R&D units from multinational high-tech and pharmaceutical companies (New Frontiers SP, Wilton Centre, Birmingham RP);
- parks whose production relies on a significant number of small and medium new-technology based companies and academic spin-offs, as a result of apparently better support and networking activities (Cambridge SP, Granta P, St. John's IP, Surrey RP, Manchester SP, Heriot-Watt RP).

**Table 4.3** Scopus publications of the fifteen most research-publishing support infrastructures in the UK.

| Infrastructure name | Type of infra. | Region | Firms | RIs | HEIs | Gov. | NOP | Total |
|---|---|---|---|---|---|---|---|---|
| Norwich Research Park | Research Pk | East of England | 29 (0.9) | 2876 (93.7) | 165 (5.4) | | | 3070 |
| Cambridge Science Park | Science Pk | East of England | 646 (92.4) | | 46 (6.6) | | 7 (1.0) | 699 |
| Harwell Oxford | Science Pk | South East | 253 (38.6) | 402 (61.4) | | | | 655 |
| New Frontiers Science Park | Science Pk | East of England | 502 (100.0) | | | | | 502 |
| Pentlands Science Park | Science Pk | Scotland | 30 (6.1) | 459 (93.7) | | | 1 (0.2) | 490 |
| Wilton Centre | S&I Centre | North East | 424 (100.0) | | | | | 424 |
| Granta Park | Science Pk | East of England | 335 (100.0) | | | | | 335 |
| Babraham Research Campus | Research Camp | East of England | 62 (23.0) | 204 (75.6) | | | 4 (1.5) | 270 |
| Scottish Enterprise Technology Park | Technology Pk | Scotland | 10 (4.4) | 215 (95.6) | | | | 225 |
| St. John's Innovation Park | Innovation Pk | East of England | 163 (98.8) | | | | 2 (1.2) | 165 |
| Surrey Research Park | Research Pk | South East | 140 (92.7) | 1 (0.7) | 10 (6.6) | | | 151 |
| Manchester Science Park | Science Pk | North West | 94 (69.1) | | 42 (30.9) | | | 136 |
| Roslin BioCentre | Science Pk | Scotland | 24 (18.9) | 103 (81.1) | | | | 127 |
| Heriot-Watt Research Park | Research Pk | Scotland | 81 (65.3) | 11 (8.9) | 21 (16.9) | 11 (8.9) | | 124 |
| Birmingham Research Park | Research Pk | West Midlands | 97 (79.5) | 13 (10.7) | 11 (9.0) | 1 (0.8) | | 122 |

This confirms the high diversity and the difficulty generalizing the evidence based on a few cases and is in line with Castells and Hall's (1994;92-3) conclusion that parks where development focuses on attracting multinationals and research centres only succeed at one level, bringing firms and jobs, but the synergy and cross-fertilization of an innovative environment remain minimal.

**Table 4.4** The most research-active firms and RIs located in the East of England, the South East, and Scotland.

| Firms | | | | | |
|---|---|---|---|---|---|
| Rank | Name | Infrastrucutre name | Output | Field | Type of company |
| | **East of England** | | | | |
| 1 | GlaxoSmithKline [Harlow] | New Frontiers SP | 502 | Bio-pharma | R&D Lab (UK) |
| 4 | TWI Ltd | Granta Park | 138 | Technology engineering | Consultant (UK) |
| 5 | Toshiba Research Europe Ltd. | Cambridge SP | 119 | Computer Science and Engg. | R&D Lab (Japan) |
| 6 | Unilever [East] | Colworth SP | 115 | Food biotechnology | R&D Lab (UK/NL) |
| 8 | UCB Celltech-Chiroscience [Oxford GlycoSciences] | Granta Park | 107 | Bio-pharma | Spin-out (UK/BE) |
| 9 | Vernalis [Ribo targets / British Biotech] | Granta Park | 72 | Bio-pharma | Spin-out (UK) |
| 12 | Chirotech Technology Ltd | Cambridge SP | 58 | Bio-pharma | R&D Lab (India) |
| | **South East England** | | | | |
| 3 | Diamond Light Source Ltd. | Harwell Oxford | 227 | Multidisciplinary research | Synchrotron facility (UK) |
| 7 | QinetiQ Ltd. [Farnborough] | Cody Technology Pk | 107 | Defence technology | R&T company (UK) |
| 18 | Surrey Satellite Technology Ltd | Surrey Research Pk | 35 | Aerospace | Spin-out (UK/France) |
| | **Scotland** | | | | |
| 10 | Quintiles Scotland Ltd [Syntex Research Centre] | Heriot-Watt Research Pk | 62 | Bio-pharma | Consultant (USA) |
| 24 | Biomathematics and Statistics Scotland | Edinburgh Tech Transfer Cent | 25 | Bioinformatics | Consultant (UK) |
| 30 | Codivien [CardioDigital Ltd] | Elvingston Science Cent | 22 | Bio-pharma | R&D Lab (USA) |
| **Research insitutes** | | | | | |
| Rank | Name | Infrastrucutre name | Output | Field | Funder |
| | **East of England** | | | | |
| 1 | Institute of Food Research | Norwich RP | 1756 | Food biotechnology | BBSRC |
| 2 | John Innes Centre | Norwich RP | 1042 | Food biotechnology | BBSRC |
| 6 | Babraham Institute | Babraham Research Camp | 199 | Functional genomics | BBSRC |
| 7 | Sainsbury Laboratory | Norwich RP | 153 | Food biotechnology | Gatsby - BBSRC |
| 12 | Sanger Institute | Wellcome Trust Genome Camp | 52 | Functional genomics | Wellcome Trust |
| 14 | European Bioinformatics Institute | Wellcome Trust Genome Camp | 30 | bioinformatics | EC-Wellcome Trust |
| | **South East England** | | | | |
| 4 | Rutherford Appleton Laboratory | Harwell Oxford | 363 | Multidisciplinary research | STFC |
| 11 | IT Innovation Centre | Chilworth SP | 57 | IT | Univ of Southampton |
| 13 | Mammalian Genetics Unit | Harwell Oxford | 39 | Human genetics and functional genomics | MRC |
| | **Scotland** | | | | |
| 3 | Moredun Research Institute | Pentlands SP | 402 | Agro-biotech | EBAR |
| 5 | NERC Radiocarbon Facility | Scottish Enterprise Tech Pk | 215 | Earth and Environmental Science | SUERC - NERC |
| 8 | Roslin Institute | Roslin BioCentre | 103 | Agro-biotech | BBSRC |
| 10 | Veterinary Laboratories Agency | Pentlands SP | 72 | Agro-biotech | DEFRA |

Table 4.4 shows that even though most research active institutions and businesses are located in different infrastructures in the three top regions, both the basic-public sector and the applied-industrial sector produce research that shares a common interest around similar subject areas. For example, in the East of England the top industrial and public organisations are based in different parks but share the same industrial interest, namely bio-pharmacology and food biotechnology. A similar R&D interest among organisations and between both sectors was also found in the other two regions, suggesting that the agglomeration of the critical mass of knowledge and capabilities embedded in a dynamic environment could be more relevant than sharing the same roof or campus to foster fruitful knowledge exchange. The concentration of like-minded research institutions in particular regions indicates the importance of regional innovation policies to guide industrial restructuring, and the importance of the networking role played by the SP movement in building bridges between the different actors. All this leads to the consideration that the external environment is a pre-condition in order to set up a SP, and that the design of a regional innovation strategy is essential for SPs' performance and success.

### 4.2.3. The main partners of the SP movement

The main function of SPs' networking function is to promote relationships between industry and university to gain access to mutually exchange of resources as a result of joint R&D projects as well as public and private research (Westhead & Storey, 1995). Thus, publication patterns represent a rich source of data about the nature and extent of technology transfer between both sectors (Cockburn & Henderson, 1998b; Lee & Win, 2004). As the analysis of onsite organisations only uncovers part of the SP movement, and there is a need to get a better understanding of the networking function inherent in these innovation infrastructures. This section describes the collaborative patterns of the movement to identify which types of actors are the most collaboratively oriented, which types of infrastructure foster collaboration, which regions are the best connected, and the role of academia and off-park organisations as external sources of knowledge.

For this analysis, only publications co-authored by members of two or more organisations were selected. Inter-institutional collaboration represents 70% (6,825) of the total UK SP movement Scopus publications. Inter-institutional collaboration can also be the result of single authors who represent two or more organisations. Here, these types of researchers primarily work for both industry and HEIs, and only account for 50 (0.005) publications. There was an increase in the average number of organisations co-authoring the publications since 2003; from 1.6 and 1.7 during the 1980s and 90s to 2.4 in the 21st century, reaching 3.4 in 2010.

**Table 4.5** Distribution of collaborations between on- and off-park organisations[7].

| Type of collaboration | National | International | Total |
|------------------------|----------|--------------|-------|
| On-park org - On-park org | 1101 *(0.13)* | 64 *(0.01)* | 1165 *(0.11)* |
| On-park org - Off-park org | 2110 *(0.24)* | 1626 *(0.16)* | 3736 *(0.35)* |
| Off-park org - On-park org | 5488 *(0.63)* | 257 *(0.13)* | 5745 *(0.54)* |
| Total | 8699 | 1947 | 10646 |

The collaboration patterns among and between on- and off-park organisations (see Table 4.5), suggests a limited knowledge exchange within the movement in comparison to the high impact of the off-park organisations in the transfer of knowledge to support on-park research-intensive industry. The collaborative capacity of on-park organisations to exchange knowledge with on- and off-park organisations is basically limited to the national territory (1,101) and national off-park organisations (2,110). This depends on on-site organisations being dominated by industry (55%) and, to a lesser extent, RIs (35%). On the other hand, the national off-park organisations, which are dominated by academia, have established 5,488 collaborations, making HEIs the main partners of on-park organisations for R&D activities. Overall, the dynamism of on-park organisations to get involved in inter-organisational collaboration (46%) is similar to the extent of collaborative ties built by external organisations with the SP movement (54%). Overall, the dynamism of on-park organisation to get involved in inter-organisational collaboration only constitutes a small part (20%) in relation to the extensive collaborative ties built by external organisations with the SP movement (80%). Collaboration between off-park firms and RIs with on-park organisations is rare. Another interesting fact is that 83% of onsite firms generated collaborative research in the last ten years, resulting in a growing interest by onsite firms in developing their absorptive capacity to effectively take advantage of public research produced by HEIs and RIs, which in turn act as partners and connect industry to the global research network. In addition, 91% of all these partnerships were established in the last ten years, and the increasing transfer of knowledge between off-park organisation at both national and international levels is one of the main drivers of this growth.

---

[7] Appendix 1 illustrates the types of collaboration based on the physical location of on- and off-park organisations.

**Table 4.6** Regional ranking based on collaborations between (1) on-park and on-/off-park organisations, and between (2) off-park and on-/off-park organisations.

| | On-park org - On-/Off-park org | | | | Off-park org - On-/Off-park org | | | | | |
|---|---|---|---|---|---|---|---|---|---|---|
| | Rank | National | Rank | International | Rank | National | Rank | International | Rank | Total |
| East of England | 1 | 1370 | 1 | 1039 | 4 | 1146 | 4 | 1025 | 1 | 4580 |
| South East England | 3 | 473 | 2 | 230 | 1 | 1965 | 1 | 1607 | 2 | 4275 |
| Scotland | 2 | 545 | 3 | 144 | 2 | 1615 | 3 | 1375 | 3 | 3679 |
| London | 8 | 70 | 8 | 22 | 3 | 1331 | 2 | 1515 | 4 | 2938 |
| North West England | 4 | 255 | 4 | 83 | 6 | 961 | 5 | 932 | 5 | 2231 |
| Yorkshire and the Humber | 9 | 67 | 10 | 20 | 5 | 1012 | 6 | 717 | 6 | 1816 |
| West Midlands | 5 | 139 | 5 | 47 | 9 | 520 | 7 | 507 | 7 | 1213 |
| South West England | 6 | 105 | 7 | 24 | 8 | 618 | 8 | 463 | 8 | 1210 |
| East Midlands | 7 | 79 | 6 | 34 | 7 | 696 | 9 | 277 | 9 | 1086 |
| North East England | 11 | 55 | 12 | 12 | 10 | 280 | 10 | 64 | 10 | 411 |
| Wales | 10 | 34 | 9 | 22 | 11 | 219 | 11 | 47 | 11 | 322 |
| Northern Ireland | 12 | 19 | 11 | 13 | 12 | 100 | 12 | 30 | 12 | 162 |
| Total | | 3211 | | 1690 | | 10463 | | 8559 | | 23923 |

To identify the degree of collaboration within and outside the SP movement, Table 4.6 illustrates to what extent the national on- and off-park organisations collaborate with on- and off-park organisations at the national or international levels. First, regarding the on-park organisations, the East of England is the best connected at the national and international levels. The South East and Scotland also top the ranking as a result of their large R&D structures in terms of tenants and research engagement. Second, looking at the influence of off-park organisations, the research base in the South East, Scotland and London are the most attractive sources of knowledge for national and international organisations. The top position of London also illustrates the regional research capacity and entrepreneurship available in the region, regardless of its virtually non-existing park infrastructure (Sainsbury, 1999). Surprisingly, the remarkable institutional infrastructure developed in the East of England has little participation from offsite organisations, as their collaborative ties are weak outside the movement, and the main reason for this is the important role of RIs and their slightly low propensity to collaborate (Noyons, Moed, & Luwel, 1999). The South East has, on the other hand, more global links than other regions as result of the agglomeration of large firms. These two regions' interactions are similar to those based on the level of formal alliances in the biotech sector (Birch, 2009). The East of England concentrates a high number of local alliances, whereas the local and low connectivity patterns of London and Scotland differ from the evidence of the high inter-organisational knowledge exchange provided by the HEIs in these regions. However, all in all the high concentration of alliances (70%) in the East of England, the South East, and London, as found by the same author, is somewhat similar to the 50% of knowledge exchange and interactions concentrated in these regions, providing the mass of research and technology necessary to innovate.

**Table 4.7** Distribution of publications by on-park organisations co-authored with national and international organisations according to the type of support infrastructures where they are located (Source: Scopus; 1975-2010; n=10,920).

| | Firms | | HEIs | | RIs | | Gov. | | NOP | | On-park org | Total | | Total |
|---|---|---|---|---|---|---|---|---|---|---|---|---|---|---|
| | Nat | Inter | Nat | Inter | Nat | Inter | Nat | Inter | Nat | Inter | Inter | Nat | Inter | Total |
| Science Pk | 487 | 413 | 529 | 163 | 200 | 3 | 62 | 4 | 12 | 1 | 13 | 1290 | 597 | 1887 |
| Research Pk | 219 | 250 | 241 | 110 | 255 | 4 | 28 | | 10 | | 3 | 753 | 367 | 1120 |
| Research Camps | 86 | 77 | 199 | 348 | 92 | | 3 | 1 | 13 | 1 | 25 | 393 | 452 | 845 |
| Technology Pk | 70 | 91 | 96 | 7 | 24 | | 47 | | 9 | | | 246 | 98 | 344 |
| Business Pk | 97 | 49 | 71 | 14 | 12 | | 21 | | 5 | | 3 | 206 | 66 | 272 |
| Science & Innovation Cent | 59 | 28 | 60 | 3 | 22 | 1 | 8 | | 2 | | 3 | 151 | 35 | 186 |
| Innovation Pk | 35 | 34 | 27 | 3 | 3 | | 4 | | 4 | | 2 | 73 | 39 | 112 |
| Incubator | 15 | 21 | 47 | 6 | 13 | | 1 | | | | | 76 | 27 | 103 |
| Industrial Pk | 12 | 8 | 8 | | 1 | | 2 | | | | 1 | 23 | 9 | 32 |
| Total | 1080 | 971 | 1278 | 654 | 622 | 8 | 176 | 5 | 55 | 2 | 50 | 3211 | 1690 | 4901 |
| Total (%) | 2051 (0.42) | | 1932 (0.39) | | 630 (0.13) | | 181 (0.04) | | 57 (0.01) | | 50 (0.01) | | | |

The collaborative patterns between infrastructures and organisations help to identify which infrastructures best promote collaboration across organisations and which on- and offsite organisations are the favoured partners (see Table 4.7). Infrastructures with a large public scientific base, such as, science- and research parks, and research campuses, seem to be the most connected, hosting organisations that equally collaborate with national and international businesses. Nonetheless, their connections with HEIs and RIs remain primarily within the UK, which coincides with previous findings (BIS, 2009). On the other hand, firms and HEIs concentrate 81% of the collaborations and are the most frequent partners for the SP movement members. Indeed, firms primarily collaborate with other firms (1,209; 54%) and HEIs (774; 34%), while RIs' partners include HEIs (938; 44%), industry (737; 34%), and RIs (338; 16%). All this underlines the central role of HEIs (Etzkowitz, 2008; Godin & Gingras, 2000), and illustrates how academia-industry collaboration adds financial value and enriches the knowledge base while at the same time facilitates the R&D links and technology transfer needed for supporting research-based industrial and innovative activities (A. N. Link & Scott, 2003).

Despite the low proportion of collaborations fostered by the SP movement in general, the degree of interaction and synergy among firms, HEIs, and RIs, is likely to be a sign of the successful promotion of academia-industry links found previously (UKSPA, 2003), and although this study does not directly compare the proportion of collaboration among off-park firms, on-park research publication exhibits added-value as the research output of off-park firms reports lower rates of research generated in the form of inter-organisational collaboration and a weak integration among firms, HEIs and RIs (BIS, 2009; Cockburn & Henderson, 1998b). It therefore contrasts with previous findings where these interactions were found to be insignificant (Vedovello, 1997), informal and with little difference to off-park firms (Quintas et al., 1992; Westhead & Storey, 1995). Conversely, it might coincide with studies that argue that SPs facilitate links with HEIs and RIs (Soetanto & Jack, 2011), and stimulate research productivity among firms (Donald Siegel, Westhead, et al., 2003). However, this study cannot conclude that the innovation infrastructures are successful policy tools as the research and technology outputs and collaboration are phenomena well concentrated and limited to only three regions in the UK.

Since industry-academia interaction plays a central role, it is necessary to examine from which region the demand for knowledge comes from and from which regions this knowledge is supplied. Thus, stemming from the assumption that research collaborations between industry and HEIs might represent a relationship between knowledge consumers and knowledge suppliers (Abramovsky & Simpson, 2011), the links established by onsite firms with on- and offsite firms, HEIs, and RIs are seen as a sign of demand for collaboration and exchange of knowledge, while the links established by offsite HEIs, firms, and RIs with onsite firms are interpreted as a sign of supply of collaboration and exchange of knowledge (see Table 4.8). These two relationships developed in each region are analysed in terms of: intra-regional level, inter-regional level, European level (collaborations with countries across Europe), and global level (collaborations with countries beyond Europe), national dependency (difference between the inter-regional and intra-regional supply or demand across the country, which illustrates their degree of dependence on inputs beyond the region), and off-park influence (difference between the supply and demand at national or international level, which illustrates the degree of influence of off-park inputs).

**Table 4.8** Regional demand and supply of collaboration. 'Demand' is considered to be collaboration between on-park organisations and off-park organisations; conversely 'supply' is considered to be collaborations between off-park organisations and on-park organisations. These two types of knowledge exchange are reported by: intra-regional level, inter-regional level, European level, and global level. National dependency illustrates the degree of dependence on inputs beyond the region, measuring the difference between the inter-regional and intra-regional supply or demand across the country. Off-park influence illustrates the degree of influence of off-park inputs, measuring the difference between supply and demand at national or international levels.

| | Demand by On-park firms to on- and off-park firms, HEIs, RIs | | | | | | | | | Supply by Off-park HEIs, Firms, RIs to On-park firms | | | | | | | | | Off-park influence | |
| | National | | International | | Total | | Total | Nat. dependecy | Collab. per consumer | National | | International | | Total | | Total | Nat. dependecy | Collab. per consumer | influence | |
| | Intra-reg | Inter-reg | European | Global | Nat | Inter | | | | Intra-reg | Inter-reg | European | Global | Nat | Inter | | | | Nat | Inter |
|---|---|---|---|---|---|---|---|---|---|---|---|---|---|---|---|---|---|---|---|---|
| East Midlands | 18 | 45 | 26 | 8 | 63 | 34 | 97 | 27 | 2.9 | 48 | 174 | 42 | 11 | 222 | 53 | 275 | -126 | 4.4 | 159 | 19 |
| East of England | 239 | 311 | 159 | 132 | 550 | 291 | 841 | 72 | 8.4 | 303 | 108 | 326 | 344 | 411 | 670 | 1081 | 195 | 5.0 | -139 | 379 |
| London | 21 | 21 | 12 | 6 | 42 | 18 | 60 | 0 | 2.9 | 33 | 318 | 19 | 29 | 351 | 48 | 399 | -285 | 6.2 | 309 | 30 |
| North East England | 29 | 18 | 6 | 6 | 47 | 12 | 59 | -11 | 3.5 | 43 | 45 | 96 | 63 | 88 | 159 | 247 | -2 | 2.0 | 41 | 147 |
| North West England | 25 | 80 | 30 | 27 | 105 | 57 | 162 | 55 | 5.1 | 87 | 233 | 79 | 77 | 320 | 156 | 476 | -146 | 4.0 | 215 | 99 |
| Northern Ireland | 8 | 2 | 6 | 6 | 10 | 12 | 22 | -6 | 5.5 | 5 | 24 | 1 | 0 | 29 | 1 | 30 | -19 | 5.0 | 19 | -11 |
| Scotland | 42 | 85 | 30 | 27 | 127 | 57 | 184 | 43 | 3.7 | 155 | 222 | 71 | 67 | 377 | 138 | 515 | -67 | 4.6 | 250 | 81 |
| South East England | 67 | 208 | 66 | 40 | 275 | 106 | 381 | 141 | 4.3 | 192 | 294 | 321 | 218 | 486 | 539 | 1025 | -102 | 6.1 | 211 | 433 |
| South West England | 7 | 27 | 7 | 4 | 34 | 11 | 45 | 20 | 3.0 | 20 | 126 | 20 | 14 | 146 | 34 | 180 | -106 | 3.4 | 112 | 23 |
| Wales | 11 | 17 | 13 | 9 | 28 | 22 | 50 | 6 | 3.8 | 17 | 19 | 8 | 18 | 36 | 26 | 62 | -2 | 1.8 | 8 | 4 |
| West Midlands | 29 | 97 | 25 | 19 | 126 | 44 | 170 | 68 | 4.7 | 81 | 127 | 43 | 118 | 208 | 161 | 369 | -46 | 4.4 | 82 | 117 |
| Yorkshire and the Humber | 19 | 36 | 11 | 5 | 55 | 16 | 71 | 17 | 5.1 | 52 | 235 | 11 | 27 | 287 | 38 | 325 | -183 | 5.2 | 232 | 22 |
| Total | 515 | 947 | 391 | 289 | 1462 | 680 | 2142 | | | 1036 | 1925 | 1037 | 986 | 2961 | 2023 | 4984 | | | | |
| (%) | (0.24) | (0.44) | (0.18) | (0.13) | (0.68) | (0.32) | (1.00) | | | (0.21) | (0.39) | (0.21) | (0.20) | (0.59) | (0.41) | (1.00) | | | | |

The demand for collaboration from onsite firms is tightly related to the size of the SP infrastructure developed in each region and thus the top regions (East of England, the South East, and Scotland) also have the highest number of on-park firms that develop R&D activities in collaboration. Hence, the East of England clearly has the highest level of ties within the region (239), across the country (311) and abroad (291). Interestingly, the overall demand for collaboration is greater at the inter-regional level than the intra-regional level, suggesting that geographical proximity is not an important factor when onsite firms establish ties with on- and off-park firms, HEIs, and RIs. This seemingly contradicts the fact that 90% of the on-park firms that have a link with HEIs or RIs connect to local institutions (UKSPA, 2003). The regions in which national dependence is highest are the South East, the East of England, and the West Midlands, and the overall national links are more important than the international, as they represent 68% of the total links, while most international organisations are European. Finally, the share of collaborations per onsite firm indicates that on average they establish 4.4 collaborations and the East of England and the East Midlands are the most and least collaborative regions respectively.

Regarding the supply of knowledge by offsite HEIs, firms, and RIs to onsite firms at the national and international levels, the East of England (1,081) and the South East (1,025) have the most external inputs. Only the East of England reports intra-regional ties that are three times stronger than the inter-regional ties, having access to a research base capable of supporting the regional innovative infrastructure, while the most interconnected on-park firms with off-park organisation across the UK are located in London. At the international level, the contribution of Europe and other countries are similar, and the East of England and the South East are again the most connected regions and the only two whose international ties are stronger than the national ones. This clearly indicates that the public research base developed in the country represents a more relevant source of knowledge and technology than the research produced abroad and, in particular, within the same region. Furthermore, only the regions with higher number of infrastructures establish more international links.

External suppliers established, on average, 4.3 collaborations and their inputs were the most intense in the South East and London. On the other hand, the level of influence of external inputs across regions determined by the difference between the demand from onsite firms and the supply of external knowledge shows that London (309), Scotland (250), and Yorkshire and the Humber (232) are highly dependent on external UK based organisations, while the East of England is the only region that could be considered as independent. However, this autonomy is only limited to the national sphere because it is the region second most connected to external organisations outside the UK, only surpassed by the market-oriented South East. This may be an indication of the integration of these two regions in global life science flows, the excellence of the regional research infrastructures, and experience in technology transfer and commercialisation (Kasabov & Delbridge, 2008). All this indicates that geographical proximity does not determine university-industry interactions, which contradicts one of the main assumptions behind the development of SPs. Similarly, Fukugawa (2006), who studies the propensity of on-park firms to engage in joint research with academia in Japan, has also obtained similar results. In addition, there is evidence based on off-park firms in the UK that suggests that firms with high levels of absorptive capacity appear to favour quality over geographical proximity when it comes to collaboration with academia, and the collaborative arrangements only remain geographical if there are top universities nearby (Laursen, Reichstein, & Salter, 2011; Moodysson & Jonsson, 2007). This could also help to explain the high proportion of local interconnection in regions with a high quality research base, such as the East of England.

### 4.2.4. Research subject areas, collaborative efforts of the SP movement

Scholarly journals are the main venues for formal interaction and communication for different scientific communities, making it possible to identify the intellectual and social aspects shared. These two aspects provide the framework that forms each knowledge domain, and the distance between domains can be determined by the degree of similarity between their cognitive and reputational systems, which in turn shapes the structure of science as a whole (Minguillo, 2010). Hence, the output of the SP movement can help, among other things, to shed light on their degree of intellectual and social integration into the wider scientific community. To do this, the research areas with the largest number of publications were identified based on the journals where the research is frequently disseminated.

The most frequent Scopus-indexed type of source for the research generated by SPs is journals (91%), in comparison to conference proceedings (7%), serials (1%), and generic (1%). The low rate of conference proceedings is somewhat surprising because conferences are considered as potential venues of interaction for industry and academia (D'Este & Patel, 2007; Lee & Win, 2004), and indeed, in the last ten years there has been an increasing trend for participating in conferences, as shown by the fact that 83% of all conference publications were published between 2005-2010, representing 12% of all publications over the last five years. This growth is the result of the intensification of R&D activities in technology areas, such as Engineering, Physics and Astronomy, and Materials Science. Overall, the SP movement prefers to publish in journals and the expansion of technology fields has recently increased the use of conference proceedings as source of communication.

Regarding the most important research fields, the chronological development of the top nine subject areas, covering 80% of the total output, shows that *Biochemistry, Genetics and Molecular Biology* is the largest research area with 18% of the total output (Figure 4.7). It started in the mid-1980s and had its first breakthrough in the mid-1990s due to the establishment of RIs (e.g. Institute of Food Research, and the John Innes Centre), the parallel relocation of the pharmaceutical industry (e.g. GlaxoSmithKline) and the emergence of new spin-offs. In 2005 it again had exponential growth partially caused by the diversification and maturity of the industry and new emerging RIs (e.g., Babraham Institute). This trend differs from the relative decline suffered by *Chemistry*, and *Agricultural Biological Sciences* between 2000 and 2007. The other top subject areas have followed a constant growth and have similarly achieved a remarkable upward increase since 2005. Three related subject areas have experienced recent exponential growth, *Physics and Astronomy*, *Materials Science*, and *Engineering*, and this is partially caused by the RIs *Rutherford Appleton Laboratory* and the private sector (e.g. AkzoNobel R&D, Diamond Light Source, TWI). On one hand, these two sets of fields represent the emerging physics and materials engineering industrial sector and, on the other hand, the partially weakening health and life science industrial sector, consisting of three subject areas: *Biochemistry, Genetics and Molecular Biology*, *Chemistry*, and *Agricultural Biological Sciences*. Both groups also differ in terms of research and technology producers as the first is slightly dominated by firms (64%) and the second by public RIs & HEIs (72%) (see Table 4.9), suggesting the maturity of new research-based industrial sectors, mostly produced by the private sector, that coexists with the well-established and publicly backed bio-tech industry within the SP movement.

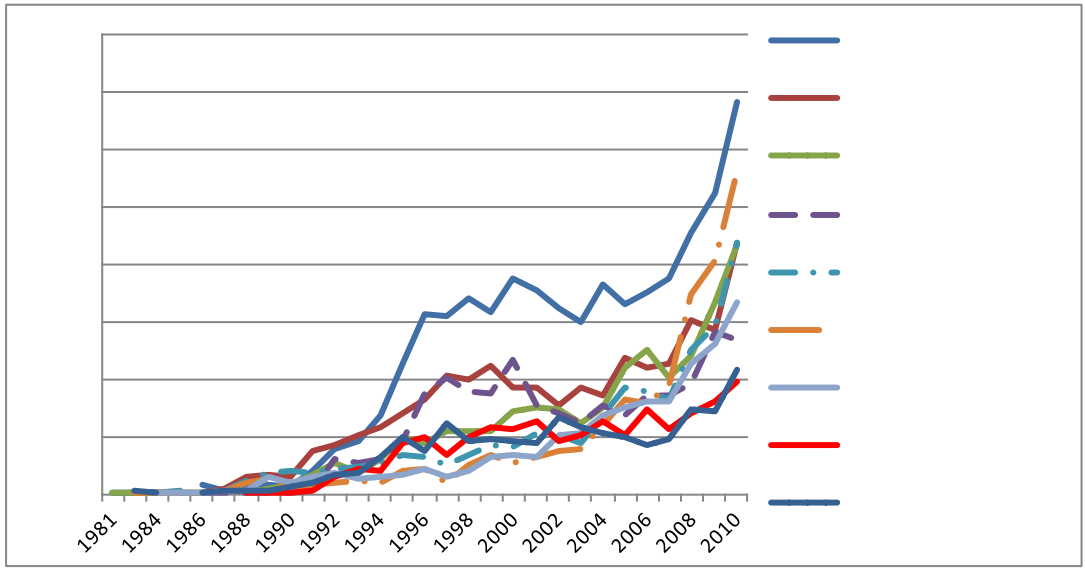

**Fig. 4.7** Chronological development of the top nine subject areas for the SP movement (Source: Scopus; 1975-2010; n=10,920).

The ranking of the top 15 subject areas in terms of Scopus publications (Table 4.9) illustrates characteristics of the research associated with the SP movement, the research profile of the three regions with the greatest research-intensive innovation structures, and the collaboration between on-park organisations (firms or HEIs/RIs) with on- or off-park organisations (firms or HEIs/RIs). At the regional level, the most productive is the East of England with the top subject area *Biochemistry, Genetics and Molecular Biology*. This depends upon the high concentration of small and large biotech firms (Birch, 2009), that in turn are highly dependent upon public RIs, as shown by the low share of private research (38%). This region also produces significant research in *Agricultural Biological Sciences* and *Chemistry*, and despite generating considerable research in other research fields, the region seems to be public science-based and specialised in food-biotechnology and bio-pharmacology. The research and technology from the South East is framed within four important areas *Physics and Astronomy*, *Materials Science*, *Engineering* and *Chemistry*, and even though there are public RIs that support the two first research areas, the role of industry as a research producer is significant (63%). Another region with a similar profile is the North East. The South East region seems to rely on private research to develop an industrial sector around physics and materials engineering. Finally, Scotland, with a reduced private research capacity (35%), relies on public research (e.g. *Moredun RI, Roslin Institute, Veterinary Laboratories Agency*) to concentrate research related to *Immunology*, *Medicine*, *Veterinary* and *Biochemistry, Genetics and Molecular Biology*, which in turn is exploited by the agro-biotech industry, confirming previous findings (Cooke, 2001b). On the other hand, the subject areas with the highest rate of private participation are *Pharmacology* (81%), *Materials Science* (67%), and *Engineering* (66%); conversely the highest academic contribution is found in *Agricultural and Biological Sciences* (85%) and *Immunology* (80%).

**Table 4.9** Distribution of the top subject areas according to private and academic publications, regions (East of England, South East, Scotland), and inter-institutional collaborative efforts (Source: Scopus; 1975-2010; n=10,920).

| # | Research area | Output | | | | | | Three main regions' Output | | | | | | Collaboration | | | |
|---|---|---|---|---|---|---|---|---|---|---|---|---|---|---|---|---|---|
| | | $n = $ 17,341 | % | # | Industry n (45%) | # | HEIs/Ris n (52%) | # | a | # | b | # | c | # | All n (25%) | # | U-I n (56%) |
| (1) | Biochemistry, Genetics & Molecular Biology* | 3182 | 18% | (10) | 36% | (5) | 62% | (1) | 26% | (5) | 9% | (4) | 10% | (11) | 18% | (9) | 46% |
| (2) | Chemistry* | 2009 | 12% | (6) | 58% | (9) | 41% | (3) | 12% | (4) | 12% | (12) | 2% | (6) | 34% | (5) | 67% |
| (3) | Medicine*** | 1572 | 9% | (9) | 39% | (7) | 55% | (4) | 8% | (7) | 5% | (2) | 15% | (12) | 15% | (12) | 44% |
| (4) | Agricultural and Biological Sciences* | 1535 | 9% | (15) | 12% | (1) | 85% | (2) | 13% | (13) | 2% | (5) | 8% | (13) | 14% | (14) | 32% |
| (5) | Physics and Astronomy** | 1334 | 8% | (7) | 58% | (11) | 39% | (7) | 4% | (2) | 13% | (11) | 4% | (7) | 33% | (2) | 73% |
| (6) | Materials Science** | 1300 | 7% | (2) | 67% | (13) | 32% | (8) | 4% | (1) | 20% | (10) | 4% | (1) | 52% | (1) | 74% |
| (7) | Engineering** | 1097 | 6% | (3) | 66% | (14) | 31% | (9) | 4% | (3) | 12% | (9) | 5% | (5) | 35% | (3) | 71% |
| (8) | Immunology and Microbiology*** | 1015 | 6% | (13) | 18% | (2) | 80% | (6) | 7% | (16) | 1% | (1) | 15% | (14) | 14% | (13) | 33% |
| (9) | Pharmacology, Toxicology & Pharmaceutics | 1006 | 6% | (1) | 81% | (15) | 17% | (5) | 7% | (9) | 4% | (8) | 5% | (8) | 20% | (7) | 65% |
| (10) | Chemical Engineering | 551 | 3% | (5) | 58% | (10) | 41% | (10) | 3% | (10) | 3% | (14) | 1% | (3) | 38% | (6) | 65% |
| (11) | Environmental Science | 484 | 3% | (11) | 34% | (6) | 62% | (11) | 2% | (12) | 2% | (7) | 6% | (9) | 20% | (10) | 45% |
| (12) | Computer Science | 391 | 2% | (4) | 64% | (12) | 33% | (14) | 1% | (6) | 5% | (13) | 1% | (4) | 36% | (4) | 68% |
| (13) | Mathematics | 294 | 2% | (8) | 55% | (8) | 44% | (16) | 1% | (8) | 4% | (15) | 1% | (2) | 39% | (8) | 63% |
| (14) | Veterinary*** | 287 | 2% | (14) | 16% | (3) | 80% | | | | | (3) | 12% | (15) | 10% | (15) | 25% |
| (15) | Earth and Planetary Sciences | 285 | 2% | (12) | 33% | (4) | 63% | | | (11) | 2% | (6) | 8% | (10) | 18% | (11) | 44% |

*a* East of England (n=54%; I=38%); *b* South East (n=14%; I=63%); *c* Scotland (n=12%; I=35%)
* Food-biotechnology and Bio-pharmacology; ** Physics and Material engineering; *** Agro-biotechnology

Regarding inter-institutional collaboration, only 25% of all the Scopus research published has been co-authored by two or more different institutions, with *Materials Science* being the area with the highest collaborative effort. From these collaborations, more than half (56%) are U-I, and there is a strong relationship ($r_s$=0.86) between the ranking of private output and U-I collaboration across the research areas. This shows that the research-intensive industries within the SP movement are, to some extent, able to capitalise on academic knowledge. Interestingly, the comparison between research areas in terms of U-I collaboration shows that the three top areas belong to the physics and materials engineering industry, implying that the South Eastern agglomeration is the most successful in fostering U-I interaction. On the other hand, the low ranking of the other two main industrial agglomerations, food-biotechnology and bio-pharmacology, and agro-biotechnology − mainly based in East of England and Scotland respectively - is affected by the central role of the public research and especially RIs. Although most RIs are meant to closely support and cooperate with local businesses, they are industry-related and the outcome of the cooperation with private sector may not necessarily lead to the publication of research articles.

Quantitatively speaking, the interdisciplinary field of *Biochemistry, Genetics and Molecular Biology* is the main research field of the movement, and the East of England possesses the main private and public agglomeration across the country, which in turn is related food-biotechnology and bio-pharmacology, in line with other findings (Birch, 2009). Despite two closely related areas (*Chemistry*, and *Agricultural Biological Sciences*) to the food-biotech and bio-pharma sector suffering a slight decline between 2000 and 2007, the research output of this important sector is underpinned by the convergence of recognised centres of research excellence that form an important public science base, along with a considerable group of international companies and spin-offs. The high visibility of this sector is also partially the result of the heavy publishing activity of bio-related companies (Cockburn & Henderson, 1998b). The other two sets of top agglomerations are related to the South East or Scotland; the first is configured by an emerging private and multidisciplinary research base that is exploited by the physics and materials engineering sector, while the latter has a considerable public research base focused on agro-biotechnology. The characteristics of both agglomerations also have been highlighted by Cooke (2001), while the slight decline in research of areas considered within food-biotechnology and bio-pharmacology may reflect the important weakening of the pharmaceutical industry in the UK and Europe (Rafols et al., 2012). The chronological trend followed by, at least, these three main agglomerations illustrates the potential influence of public strategy in the establishment of research units and partnerships within SPs as a way to support the emergence of new industries. Link and Scott (2003) also show how the historical development of SPs in the United States is influenced by public policies, promoting an early emergence of medical centres and aerospace technology that are then replaced by a biotechnology and biomedical industry. This policy-driven development may also be the reason for the difference between the subject areas distribution of the SP movement with those found among patenting off-park firms where *physics*, *engineering*, *clinical medicine*, *chemistry*, and *biomedical science* are the most popular fields, for example (Godin, 1996).

In terms of collaborative efforts, only a quarter of the output is the result of an inter-institutional collaboration, of which more than half is between HEIs/RIs and industry. This national rate of U-I collaboration is low in comparison with the 34% found on the Hsinchu science park, for example (Hung, 2012). The significant involvement of the private sector in publications related to physics and materials engineering, makes this domain the most successful in bridging the U-I gap and represents an attractive market niche for the commercialisation of academic R&D. The explanation for the active participation of industry in R&D activities in this domain is that industry needs to develop its own expertise in physics, while the life science sector relies more on external research (Godin, 1996). However, the central role of the public research infrastructure, mostly RIs, in the high visibility of the other two main domains (Food-biotechnology and Bio-pharmacology, and Agro-biotechnology), seems to generate an unexpectedly low rate of U-I collaboration. Most RIs tend to have a lower publication average in comparison with Universities, as factors such as human resources, value to publishing, and rewarding system differ between HEIs and RIs (Hayati & Ebrahimy, 2009; Noyons et al., 1999). In fact, the top position for the areas related to *Physics and Materials engineering*, in terms of U-I collaboration, coincides with the study of Abramo and his colleagues (2009) who found that U-I collaboration in Italy is chiefly established in *Electronic and engineering*, outperforming other domains, such as *Chemistry* and *Agro-biotechnology*. The authors' explanation is the low level of development of the Italian industry, however this finding suggests that this domain is more likely to encourage a closer interaction between both sectors.

### 4.2.5.    Quality and Impact of the SP movement

Quality is defined here as the capacity to place publications in journals that attract many citations. The quality of the output was obtained through comparing the expected quality (the average value of the SJR and SNIP given to each subject area in 2010) with the observed quality (the average value of the 2010 SJR and SNIP of the journals where on-park organisations publish). If the observed quality is higher than the expected quality then this is evidence that the research of on-park organisations is good enough to be disseminated among the more prestigious journals in the area. On the other hand, the impact of the output, defined by the number of citations that each publication receives, is obtained through comparing the expected impact (the average number of citations received by the publications in each subject area), with the observed impact (the average number of citations received by on-park organisations' publications). If the observed impact is higher than the expected one it is assumed that the on-park research is relevant and attracts the attention of the research community.

Table 4.10 illustrates that the SP movement as a whole is capable of publishing in the more influential Scopus journals and these publications have a higher impact than the national average. Based on the SNIP indicator, the difference between the observed and expected quality suggests that the areas with highest quality are *Earth and Planetary Sciences*, *Chemical Engineering*, and *Agricultural and Biological Sciences*, while those with lower quality are *Immunology and Microbiology* and *Physics and Astronomy*. The comparison based on the SJR supports the high quality of on-park research, with the areas of highest quality being *Earth and Planetary Sciences* and *Computer Science*. In terms of impact of the output, between the period 1996 and 2010, 79% of the publications have been cited and the observed impact is higher (18.44) than the expected one (16). However, only five areas seem to have higher impact than expected, the highest being; *Chemical Engineering*, *Agricultural and Biological Sciences*, and *Veterinary*. On the other hand, the areas with the lowest relative impact are: *Earth and Planetary Sciences* and *Physics and Astronomy*.

**Table 4.10** Quality and impact of research publications published in the top 15 subject areas (Source: Scopus; 1975-2010; n=10,920).

| | Quality | | | | Impact (1996-2010) | | |
| --- | --- | --- | --- | --- | --- | --- | --- |
| | SNIP | | SJR | | Observed | | Expected |
| | Observed | Expected | Observed | Expected | n=18.44 | St dev | n=16 |
| Biochemistry, Genetics and Molecular Biology* | 1.42 | 0.78 | 0.68 | 0.42 | 25.12 | 40.66 | 28.46 |
| Chemistry* | 1.35 | 0.88 | 0.23 | 0.15 | 16.50 | 27.45 | 18.76 |
| Medicine*** | 1.26 | 0.77 | 0.41 | 0.13 | 18.75 | 33.88 | 17.86 |
| Agricultural and Biological Sciences* | 1.33 | 0.64 | 0.25 | 0.10 | 23.21 | 36.97 | 18.51 |
| Materials Science** | 1.15 | 0.91 | 0.14 | 0.10 | 11.06 | 23.35 | 11.57 |
| Physics and Astronomy** | 1.12 | 1.14 | 0.13 | 0.11 | 8.01 | 21.48 | 15.18 |
| Engineering** | 1.34 | 0.80 | 0.12 | 0.06 | 7.52 | 21.35 | 8.12 |
| Immunology and Microbiology*** | 1.39 | 1.45 | 0.63 | 0.40 | 21.45 | 28.98 | 24.01 |
| Pharmacology, Toxicology and Pharmaceutics | 1.03 | 0.49 | 0.29 | 0.15 | 18.87 | 30.52 | 17.72 |
| Chemical Engineering | 1.42 | 0.63 | 0.28 | 0.09 | 15.43 | 29.65 | 10.7 |
| Environmental Science | 1.37 | 0.67 | 0.13 | 0.08 | 15.03 | 27.04 | 18.55 |
| Computer Science | 1.62 | 1.49 | 0.70 | 0.06 | 6.56 | 43.30 | 10.23 |
| Mathematics | 1.20 | 1.01 | 0.07 | 0.05 | 6.36 | 49.56 | 9.95 |
| Veterinary*** | 1.02 | 0.56 | 0.10 | 0.06 | 13.12 | 24.71 | 9.23 |
| Earth and Planetary Sciences | 1.40 | 0.51 | 1.10 | 0.07 | 10.13 | 17.09 | 17.96 |

* Food-biotechnology and bio-pharmacology; ** Physics and material engineering; *** Agro-biotechnology
SNIP Source: www.journalindicators.com
SJR Source: www.scimagojr.com

The Wilcoxon signed-rank test compares the expected and observed values, confirming that the quality measured by the SNIP ($z$=-3.238, $p$<.05) and SJR ($z$=-3.409, $p$<.05) of the journals within the different subject areas is significantly higher than the expected. On the other hand, the level of impact obtained by the publications is only slightly higher than expected with a difference that is not statistically significant ($z$=-.966, $p$>.05). This reveals that the organisations associated with the SP movement are able to publish in high-quality journals, although the impact of these publications on the scientific community varies across areas and tends to be only slightly greater than the average.

Different factors may lead areas with high quality to have low impact and vice versa. When the top quality research areas are compared based on the three main regional agglomerations (not shown), the observed quality reveals that research in food-biotechnology and bio-pharmacology industries in the East of England has a much higher value (2.63) than the agro-biotech industry in Scotland (2.40), and the physics and materials engineering industry primarily located in the South East (2.0). The citations, however, show that only the agro-biotech sector has a positive impact (0.74), whereas the impact of food-biotechnology and bio-pharmacology (-0.3) and physics and materials engineering sectors (-2.75) are below the expected values. The main reason for this could be the nature of the research. As Godin (1996) claims, basic research produced by industry in biotechnology and chemistry is more useful for the research community and thus more cited than the applied research produced by industry in physics. The applied nature of the research generated in physics and materials engineering is reflected in the greater dissemination of research in the form of conference proceedings, for example. Another reason could be that the private-oriented sectors have only experienced a strong increase over the last ten or five years, and thus have had less time to be cited.

To find the reason for the inconsistency between the quality and impact of the output the characteristics of the impact across regions, infrastructures, and types of organisations were examined. First, Table 4.11 reports the citation rates of the on- and off-park organisations. Interestingly, at the national level the evidence indicates that on-park research production, chiefly conducted by the private sector, had a slightly lower impact (19.2) than the off-park production (22.1) which is chiefly conducted by HEIs. At the regional level, the low impact of the private research base in the South East, which occupies the tenth position, differs from the top positions of the primarily public research generated in the East of England and Scotland. The impact of the off-park organisations shows that the exchange of research with off-park organisations located in the North East, London, and the East of England attracted the interest of the research community, increasing its impact.

**Table 4.11** Citation rates of regions, infrastructures, and organisations in terms of on-park and off-park organisations.

| Citation per publication | | | | | | | | | |
|---|---|---|---|---|---|---|---|---|---|
| **Region** | **on-park** | | **off-park** | | **Infrastructure** | **on-park** | **Organisation** | **on-park** | **off-park** |
| | # | n=19.2 | # | n=22.1 | | n=19.7 | | n=19.7 | n=21.2 |
| East of England | 1 | 26.9 | 1 | 30.2 | Research Camp | 48.6 | Research Institutes | 25.7 | 25.6 |
| North West England | 2 | 16.0 | 2 | 29.4 | Research Pk | 27.8 | Firms | 15.2 | 17.4 |
| Scotland | 3 | 13.3 | 3 | 28.2 | Incubator | 16.0 | HEIs | 14.3 | 19.9 |
| North East England | 4 | 13.3 | 4 | 27.1 | Science Pk | 14.8 | Government | 6.3 | 10.2 |
| South West England | 5 | 12.4 | 5 | 19.9 | Innovation Pk | 13.8 | Non-profit organisations | 5.8 | 182.0 |
| East Midlands | 6 | 11.7 | 6 | 19.1 | Science & Innovation Cent | 12.6 | **% of uncited publications** | **on-park** | **off-park** |
| London | 7 | 10.4 | 7 | 17.7 | Industrial Pk | 8.9 | Organisation | n=0.21 | n=0.21 |
| West Midlands | 8 | 9.7 | 8 | 16.8 | Business Pk | 8.6 | Research Institutes | 0.13 | 0.15 |
| Yorkshire and the Humber | 9 | 8.5 | 9 | 15.5 | Technology Pk | 8.3 | Firms | 0.27 | 0.25 |
| South East England | 10 | 7.9 | 10 | 15.3 | | | HEIs | 0.29 | 0.21 |
| Wales | 11 | 7.3 | 11 | 12.3 | | | Government | 0.40 | 0.23 |
| Northern Ireland | 12 | 3.4 | 12 | 11.0 | | | Non-profit organisations | 0.43 | 0.26 |

Similarly, the level of impact of the infrastructures and organisations (see Table 4.11), clearly shows that the closer the research production is to public RIs the greater the research impact. Infrastructures with a greater part of their output generated by RIs, research campuses (48.6) and research parks (27.8), and, to a lesser extent, incubators (16), and science parks (14.8), have a greater impact than the business-oriented infrastructures, namely industrial- (8.9) and business- parks (8.6). Most of these RIs are recognised centres of excellence and the research produced by RIs, regardless of being on (25.7) or off park (25.6), leads to the highest impact for the on-park research community. On the other hand, it is difficult to argue that the research produced with the participation of either firms or HEIs could receive more citations due to the high level of collaboration between both.

Overall, the publications of the SP movement have the quality to appear in leading journals and have similar impact to the national average, being consistent with the higher quality (Cockburn & Henderson, 1998b) and impact (Marston, 2011) of private research in biomedicine, for example. Thus, the observed quality and impact on the different fields do not seem to be related to each other, even though a journal's prestige is the most important factor for future impact in some science and technology areas (Bornmann & Daniel, 2007). The evidence suggests that the degree of impact is determined by the public or private origin of the research. Hence, the regions with a greater public research base, such as the East of England and Scotland, have a higher impact on the research community, while those with a higher rate of private research, such as the South East, have less impact. In support of this, publications related to research oriented infrastructures and organisations (e.g., research campuses, research parks, and RIs) draws greater interest from the scientific community. This difference is also apparently linked to the applied nature of the research conducted by the private sector, and which has less scientific impact (Godin, 1996). This finding also reflects the distance between basic and applied research, as it is widely considered as one of the main interaction barriers between the public and private sectors (Bruneel, D'Este, & Salter, 2010). Thus, despite the private sector tending to establish collaborations with research leaders, they tend not to be able to publish their publication in top quality journals (Abramo et al., 2009), however this fact is partially contradicted as the on-park research in general have a significant higher quality. For this reason, the use of citations as a proxy to assess the quality of private research may not be suitable, as the diverse objectives of both communities from research differ in terms of intellectual and reputational goals, undermining to some extent the interest of private research in the actions of the scientific community.

### 4.3 Conclusions

This study draws on scientific and technology publications produced by on-park organisations in the UK to examine whether scientific publications can give plausible results when used as an indicator for monitoring R&D fostered by SPs and similar support infrastructures. To tackle this goal, the focus has been on the analysis of four different aspects that illustrate to what extent this indicator can help to complement other methods, and also to expand the knowledge of the SP movement as a whole.

The results show that, first, the development of the UK SP movement is characterised by a constant increase in publications from the 1990s with exponential growth since 2000. Scotland and the East of England have a high R&D intensity and are the regions where the first infrastructures were created. The first type of infrastructures established across the country with the highest proportion of research output was Science Parks, Research Parks, and Science & Innovation Centres. Although the number of commercially-oriented infrastructures (e.g., business parks and industrial parks) is higher due to the reduced need for resources in comparison with research-oriented infrastructures (e.g., science parks), they produce less publications. However, the recent increase in the publication output of commercial infrastructures represents new opportunities for R&D activities and partnerships with knowledge producers, while reflecting that despite the heterogeneity of the SP movement, the intensity of R&D activities is highly concentrated.

Second, regarding R&D intensity at the regional level, the East of England, the South East and Scotland have the highest proportion of infrastructures and publications. The biotechnology industry plays a central role in the publications of the SP movement, and the nature of these agglomerations differs, with the East of England and Scotland being driven by public-research and the South East being driven by commercial interests. However, this high concentration of knowledge-active on-park organisations and the relation between the level of R&D intensity with the competitiveness of the infrastructure developed across the regions suggest that SPs may only be able to exploit dynamism and competitiveness that already exists in a region – the so-called spontaneous clusters (Chiesa & Chiaroni, 2005). SPs do not seem to be the most adequate policy tools to revitalise less-favoured regions nor to reduce the uneven development and the unequal distribution of innovative firms in the country, having a limited impact on policy driven clusters (Chiesa & Chiaroni, 2005).  Hence, this study is also in opposition to the rationale underlying the development of SPs (Quintas et al., 1992; Siegel, et al., 2003). Regarding the most research-active organisations, due to an increase in private R&D activities since 2000, industry has overtaken research institutions as the main research and technology publication producer, and both are together responsible for 92% of the overall on-park output. In SPs the different types of organisations (Industry, RIs, HEIs, Government, and Non-profit organisation) involved produce knowledge, while the R&D activity of the other infrastructures fundamentally depends upon either the private (Science & Innovation centres, Innovation parks, Incubators, and Business and Industrial parks) or the public research basis (Research parks and Research campuses, and Technology parks). The level of R&D activities among the different support infrastructures could also be useful in the formulation of a better classification scheme of the different infrastructures. Finally, based on the organisations driving the R&D activities within the parks, it is also possible to identify those who either rely on anchor public RIs, on R&D units of large companies, or on new technology-based firms.

Third, regarding inter-organisational collaboration, interconnections at the national level and with off-park organisations are the most popular, while the high number of collaborations established by off-park organisations, mostly HEIs, confirms their central role in the development of an external research network that supports on-park firms and RIs. The collaborations established by on- and off-park organisations at the regional level shows that the regions with the highest number of interactions are, as expected, those with the largest SP research agglomerations, while the regions where off-park organisations are highly associated with on-park organisations are the South East, Scotland and London. Regarding the infrastructures that best promote inter-organisational collaborations, Science- and Research Parks are the most successful, and industry and HEIs are the most frequent partners for SP movement members, with the private sector primarily connected with itself and academia. SPs increase the density of academia-industry links and overall institutional involvement. However, this apparent effectiveness is only limited to the structures developed in three regions and cannot be extrapolated to the whole SP movement. Finally, regarding industry-academia interactions, the public research base developed in the country represents a more relevant source of knowledge and technology than those located abroad and in particular within the same region because on-park firms tend to collaborate with partners beyond their local region. Hence, one of the main assumptions behind SPs is once again questioned by this study (Fukugawa, 2006; Vedovello, 1997). The reason for this could be the lack of relevant and top quality universities nearby (Laursen et al., 2011). Furthermore, only the regions with great agglomerations have access to many international links, and the R&D activities involving on-site industry and academia do not follow physical proximity, questioning one of the main reasons behind the popularity of SPs as policy tools. However, this unnecessary physical proximity to establish collaborations for the SP movement is to some extent expected as the interactive learning and knowledge creation implied in the co-creation of research has the cognitive proximity as main condition. It implies interactions between actors with similar knowledge base, skills and expertise. Here, spatial proximity may rather play a complementary role in building and strengthening cognitive proximity but the potential to access to novelty and new knowledge is more important than other proximity constrains, like, physical, institutional, organisational or social (Boschma, 2005).

Fourth, regarding the research areas that produce most of the on-park publications and the contribution of the geographic regions and U-I collaboration across the different areas, the findings reveal that the R&D publications are frequently generated in four subject areas:

*Biochemistry, Genetics and Molecular Biology*, *Chemistry*, *Medicine*, and *Agricultural and Biological Sciences*, and the mass of research accumulated in the three top regions is characterised by; (1) public science-based research specialised in food-biotechnology and bio-pharmacology in the East of England, (2) private science-based research specialised in physics and materials engineering in the South East, and (3) public science-based research specialised in the agro-biotech sector in Scotland (Glasson, Chadwick, & Smith, 2006; Kasabov & Delbridge, 2008; Leibovitz, 2004; Sainsbury, 1999). The reason for the significant private research in the South East is that its industrial sector needs to develop their own expertise, while the life science sector relies more on public scientific expertise (Audretsch, 2001; Godin, 1996). *Pharmacology*, *Engineering*, and *Materials Science* are the areas with the highest rate of private participation. There are also other interesting aspects around the movement, which suggest that it has also apparently been able to redefine itself to nurture stronger university-industry links in the last two decades. However, this positive image of the movement is not completely true because research production is clearly concentrated around only three top regions, where these infrastructures take advantage of highly dynamic environments. The synergy expected within SPs is again questioned here as inter-institutional collaboration is only limited to a quarter of the publications, of which more than half is U-I collaboration.

Fifth, in terms of the quality and impact of the Scopus publications produced by SPs, on-park organisations tend to publish in significantly higher quality journals, and the on-park research has similar impact to the national average. The relationship between quality and impact varies for the same subject area, especially among the set of areas related to the three top domains and regions. A closer look at the impact produced by the regions, infrastructures, and organisations reveals that the closer the output is to HEIs and RIs the greater the impact, while the closer the output is to firms the lower the impact. This is a sign of the interaction barriers between the public and private sectors that are usually caused by the focus on either basic or applied research, which is also illustrated by the limited impact of the private research on the scientific community. The domain with the highest U-I interaction is private research-oriented physics and materials engineering.

In conclusion, this study provides evidence that research impact is likely to be associated with the nature of the organisation producing the research rather than its relation to a physical intermediary infrastructure. The low level of interest in private research from the scientific community suggests that citation-based indicators may not be the best tools to assess the private research community and academic research organisations (e.g., schools, departments and RIs) that have built up strong links with industry. Furthermore, important aspects, such as geographically high concentrations of on-park research activities, low U-I collaboration rates, and limited integration into the research community, question the idea of SPs catalysing knowledge-transfer across regions, and of SPs being policy tools intended to support the transition from declining to innovative industries as a way of reducing the unequal distribution of research-intensive industry across the UK. Thus, this evidence is helpful for policy makers in assessing the actual impact of policies and in guiding the directions of a more effective and realistic transfer policy for SPs and U-I collaboration in general. There are also other interesting aspects of on-park research output which suggest that the development of the UK SP movement is characterised by a constant increase in the research production from the 1990s with exponential growth since 2000. The central role of most RIs in supporting local industries makes it necessary to map their research performance and links with the private sector. On the other hand, the coverage gap found in the WoS database suggests that the sources where industry in general is able to publish and interact with the wide scientific community might be less likely to be indexed in the WoS. It is therefore necessary to empirically examine the bias of this database against private research.

Overall, all this suggests that publications represent an interesting proxy to investigate the R&D dimension of SPs, answering the research question. However, the results here are only indicative because although the main goal of SPs is to facilitate knowledge and technology transfer, formal research dissemination only uncovers part of this transference, and not all U-I interactions result in (co-authored) articles (Katz & Martin, 1997). Another important limitation is that this study might not cover all of the research generated within the SP movement because not all on-park organisations mention the name of the infrastructures where they are based as part of their affiliation address. The rapid increase of the output over recent years can also generate bias against part of the publications as they have less time to be cited. In addition, the results could also favour the visibility of high-research intensive industrial sectors where publications play an important role, while the R&D intensity of other sectors in the movement could be underrepresented.

Apart from these main limitations, a scientific publication is a formal and fairly reliable proxy for knowledge creation and exchange, and can help to complement the battery of indicators already used to study the SP movement. Therefore, this study introduces a new insight as it expands knowledge about innovative infrastructures through mapping the research capacity of the UK SP movement as a whole. It represents a first step towards a better quantitative understanding of the R&D activities of the movement and discloses interesting insights into the real impact of support infrastructures on effective knowledge transfer. It can facilitate collaboration among parks, especially by promoting connectivity between entrepreneurial research institutions and businesses, since they will be able to easily identify attractive locations to be established or re-located and get information about the research profile of relevant partners or competitors at regional, national and international levels.

Stakeholders and policy makers could also track highly intensive research areas, the development of new parks, parks which might develop into potential clusters, and that in turn may lead to stimulate and intensify the transference of knowledge and technology through a better funding allocation for entrepreneurial research institutions and research-oriented firms. Hence, this approach can be further exploited to obtain a deeper understanding of the inter-organisational collaboration within the SP movement, and to find the answers to important questions such as: Do SPs facilitate the collaboration between hosting universities and on-park firms? Are SPs the main intermediaries between hosting universities and industry? How long after SPs are established do they usually start to promote research creation and cross-fertilization? What are the reasons behind the frequent collaboration among on-site firms? Is there a difference between fields in terms of the collaborative patterns at regional, local or national levels? And which field requires more collaborative efforts? In addition, it can also be applied to different countries and geographical areas. Mapping the R&D activities of the SP movement across Europe, for example, would be essential to promote a better consolidation of a well-interconnected European innovation infrastructure, which is one of the main goals of the EU's Competitiveness and Innovation Framework Programme (CIP).

# Chapter 5: The structure of the R&D networks associated with the UK SP movement

## 5.1 Introduction and goals

This chapter analyses the collaborative links established by the organisations located on science parks and similar support infrastructures to map the extent to which SPs rely on public science (i.e., universities) to get involved in R&D activities in three different ways (on-park organisations; SP and off-park organisations; firms-HEIs-RIs). This objective is supported by goals to identify (1) the role played by HEIs in these R&D collaboration networks, and (2) whether HEI research performance is associated with stronger U-I collaborations. Social network analysis and visualization techniques (Section 3.3.5), and descriptive statistics (Section 3.3.4) are applied to analyse the publications produced in collaboration with any on-park organisation in the UK (1975-2010) (Section 3.3.2). This study provides evidence of the relational structure of the research-based collaborations embedded within SPs.

The main goals of this study are translated into the following research questions;

- What are the main structural properties of the inter-institutional collaboration network of on-park organisations?
- What are the main structural properties and role of HEIs in the inter-institutional collaboration network of on-park and off-park organisations?
- What are the main structural properties and role of HEIs in the firms, HEIs, and RIs inter-institutional collaboration network?
- Are the most research intensive HEIs the main partners for on-park firms?

These research questions can help to shed light on the internal and external structure of collaborations established by on-park organisations. This relational dimension is essential in the assessment of SPs' performance and success as these collaborative links are associated with the performance of highly innovative firms (Soetanto & Jack, 2011). It could also help to extend the analytical framework used to evaluate these innovation infrastructures, and more importantly provide evidence that can reduce the risks involved in the development of policies oriented towards the development of a large network of SPs in the UK. This contribution can also provide new insights into U-I collaboration in general (Teixeira & Mota, 2012).

## 5.2 Methods

This study covers all research papers published in journals, book series, and conference proceedings between 1975 and 2010 indexed in the Scopus database. All papers with at least one author affiliation for an organisation located in any UK SP or similar infrastructure was selected. This gave 9,771 records of which only a quarter is the result of an inter-institutional collaboration[8] and of which (56%) is between HEIs/RIs and industry. The Scopus journal classification scheme was used to broadly identify the research subject areas and subject categories where the papers are produced. The publications placed in journals indexed in more than one subject area are counted in each one. This allows the depiction of the thematic areas in which the collaborations could roughly be based on.

First, to analyse the inter-institutional collaboration network among on-park organisations, a simulated annealing[9] method was applied to identify the different groups or modular structure of the network and functional roles of its actors based on link type frequencies. Second, to map the role of HEIs, as the main external partners of the SP movement and the degree of reliance of the SP movement on external organisations, inter-institutional collaboration between on- and off-park organisations was used. This asymmetric relationship was represented as a bipartite network with two sets of nodes (SPs (teams) and off-park organisations (actors)), where a link indicates collaboration between the off-park actor and one of the on-park organisations. This bipartite network was also analysed with the help of a model based on modularity maximization through simulated annealing to identify groups of actors that are strongly connected to each other through co-participation in many teams. This bipartite affiliation network was normalised and then projected onto a SP-SP network to identify the structure of the SP movement based on the collaborative patterns the SPs share with off-park organisations. The Pathfinder network method was applied to represent the network. Third, to analyse the firms-HEIs-RIs collaboration network only collaborations of at least 7 publications among HEIs, firms, and RIs were considered. The Pathfinder network method was also applied to represent the backbone of this dense network.

Finally, because the threshold of 7 or more publications depicted only the core of the network, a HEIs-firms-RIs network was constructed of two or more collaborations to find if there was a relationship between HEIs' quality levels and centrality levels in the collaboration network. To identify the quality level of the research institutions the classification made by BIS (HEFCE, 2009) was used, which classifies 108 HEIs and PROs into four different quality groups according to their research spending, academic research staff, 2001 RAE average quality score, and research intensity. The centrality measures applied were: degree, which measures the level of activity through the number of direct ties to other actors in the network; betweenness, which measures the level of influence and control through the shortest paths between any two nodes passing through a certain node; closeness, which measures the minimum distance through the average shortest path from a node to all other nodes; and also other more advanced centrality measures (Scott, 2000; Stanley Wasserman & Faust, 1994). This comparison was calculated with the help of the Kruskall-Wallis, Jonckheere, and Mann-Whitney non-parametric tests.

---

[8] The contribution of each organisation counts as one regardless of the number of co-authors from the same organisation or different organisations (Godin & Gingras, 2000: 274-5).

[9] Simulated annealing is "a stochastic optimization technique that enables one to find 'low cost' configurations without getting trapped in 'high cost' local minima." (Guimera & Amaral, 2005: 2)

### 5.3 Results

#### 5.3.1.    On-park collaboration network

Although this network considers on-park collaborations with two or more publications, the low rate of on-park collaboration produces a network where nodes are, on average, connected with only one other node (1.14). The high number of isolated dyads generates 19 modules, i.e. groups or sub-groups that form a connected sub-network, but only 5 modules form the main component. The modularity $M^{10}$ of the best partition using simulated annealing is 0.719, while the average modularity $<M>$ of the randomizations of the network is 0.659 (S.D. 0.006). This means that the network is significantly modular as the modularity is larger than the modularity of its randomizations. Interestingly, the modular structure of this main component shows that most on-park collaborations cluster at the regional level (see Figure 5.1). Nevertheless, physical proximity between partners is not an important feature in the overall network since collaborations between tenants from the same park (27%) and region (34%) together are not much higher than those outside the region (39%). In terms of the frequencies of these collaborations, the partnerships between two tenants from the same park are on average (12.1) three times as many collaborations than the partnerships at regional (4) and national level (4.5). However, the reason for this stronger and more productive interaction within the same park is that these partnerships are mostly between RIs rather than involving knowledge transfer. This suggests that SPs are not able to promote cross-fertilisation among tenants since cooperation within parks is rare and tends to be between RIs and HEIs. Here, most of the organisations come from the East (33%), South East (20%), and Scotland (14%), forming the main sub-networks.

---

[10] Modularity is the assessment of the significance of the different groups' modular structure, that form the network in comparison with the randomization of the same network. The modular structure of the network and functional role of its actors' nodes, is identified based on the link type frequency between the actors of the network. The role of each node is determined by two properties: (1) "relative within-module degree z, which quantifies how well connected a node is to other nodes in their module, and (2) the participation coefficient P, which quantifies to what extent the node connects to different modules" (Guimera et al., 2007: 63).

In terms of roles, this low density network only has three out of potentially seven different roles[11] with 99% of the nodes being peripheral. In fact, according to the method of Guimera and his co-workers, 75% are ultra-peripheral nodes (R1); 14% are peripheral nodes (R2); and only one node (GlaxoSmithKline) is a provincial hub (R5). This pharmaceutical company is the provincial hub of the main component and plays a central role in its Eastern module and outside. Among the R1 nodes, only 26% are HEIs or RIs while 70% are firms, of which more than a third are spin-offs. Spin-offs have the highest number of collaborations among firms. These ultra-peripheral nodes (R1) are in turn tied together to 13 peripheral nodes (R2) formed by HEIs (23%) and RIs (62%) and two large pharmaceutical companies (15%). The role and collaboration pattern of these three types of nodes show the function of the organisations in this R&D network, where the level of research intensity and size of the organisations suggest a more central role. Research-active organisations are the main partners across the different modules and contribute to the growth and development of the on-park network.

---

[11] This method classifies nodes into seven universal roles, according to their pattern of intra- and inter-module connections, defined by their within-module degree and their participation coefficient. These roles are divided into four non-hubs (within-module degree < 2.5):
- (R1) ultra-peripheral nodes: nodes with all their links within their module;
- (R2) peripheral nodes: nodes with most links within their module;
- (R3) satellite connector nodes: nodes with a high fraction of their links to other modules;
- (R4) kinless nodes: nodes with links homogenously distributed among all modules.

The hubs (within-module degree ≥ 2.5) are:
- (R5) provincial hubs: hubs with the vast majority of links within their module;
- (R6) connector hubs: hubs with many links to most of the other modules;
- (R7) global hubs: hubs with links homogenously distributed among all modules (Guimera, et. al., 2007: 63-4).

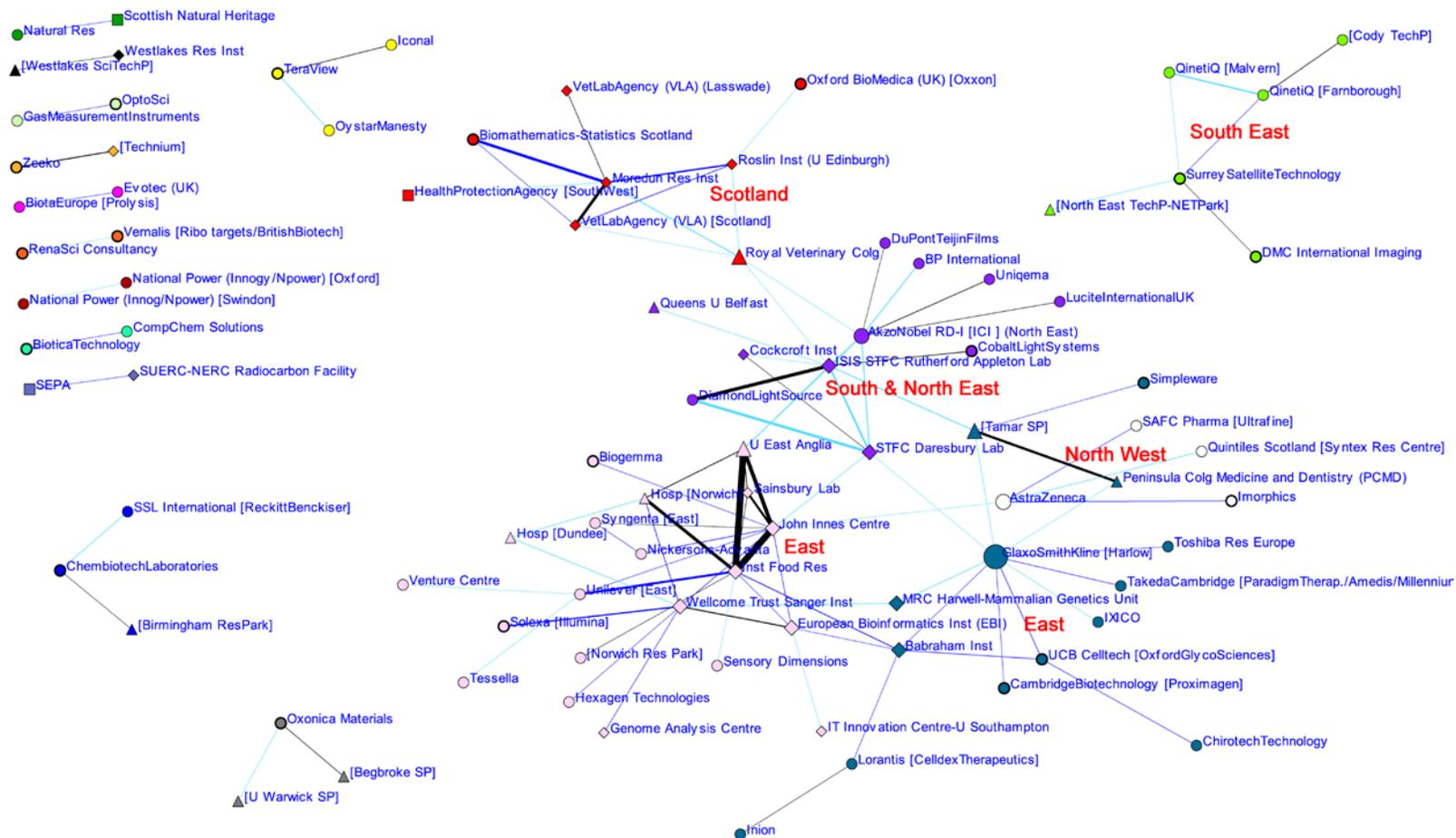

**Fig. 5.1** Collaboration network of on-park organisations. 90 Nodes – 12 HEIs (triangles), 19 RIs (diamonds), 40 Firms (circles), 19 Spin-offs (bold circles); 103 Edges (≥ 2 collaborations) representing 1,086 collaborations. 27% On-park links (black), 34% Regional links (blue), 39% National links (sky blue). The node colours and sizes represent the different modules and roles, respectively. The edge thickness is proportional to the collaborations rate between two nodes.

### 5.3.2.    SP and off-park collaboration network

In terms of the collaboration between on-park and off-park organisations represented in Scopus, HEIs are involved in 77% of the off-park interactions in the UK, while firms and RIs are only responsible for 12% and 7% respectively. The 2-mode network that represents the collaborative links between SPs and off-park actors in the UK illustrates the central role of the University for the SP movement (see Figure 5.2). In this bipartite network there are two sets of nodes, SPs (*teams*) and off-park organisations (*actors*), and a link indicates three or more collaborations between an off-park and an on-park organisation. This accounts for 69% of all these collaborations. At the core of the network there is a high concentration of a small group of HEIs that are the main partners of the SPs. Only *Norwich SP* and *Pentlands SP* are able to attract multiple actors outside the core, as the first has the highest number of private partners and the latter has many Scottish partners. HEIs' intermediary role as an off-park interconnector is quantified by having on average a much higher betweenness (.0098) than firms (0.000,15) and RIs (0.000,23), as well as higher centrality measures in general. The strong links between central HEIs with physically close SPs cannot be extrapolated to the whole network as intra-regional links only represent 28% of these links and 37% of the collaborations. These are more visible as they are slightly stronger (10.7) than the inter-regional (7.2) links, on average.

The bipartite network was projected onto one unipartite network, SP-SP, to analyse the collaborative patterns shared by the SPs (Figure 5.3). The symmetrical SP-SP network represents the similarity between a pair of SPs based on the number of equivalent off-park organisations that co-collaborate with each pair of SPs. This transformation was normalised with the help of two measures; Pearson correlation (Figure 5.3) and covariance (Figure 5.4). As expected, the normalisation with Pearson correlation gave best results in terms of intra-regional connectivity and 2-mode modular structure, since it measures the structural equivalence determined by the similar connectivity patterns of a collection of actors. The network (Figure 5.3) confirms that SPs based in the same region share to some extent collaborative ties with the same off-park organisations. This method was not used as it only focuses on similar connectivity patterns of a collection of actors and ignores the strength of particular connections (Wasserman & Faust, 1994: 368-9) and is not suitable for the Pathfinder network method as it is based on the strength of particular connections between each pair of actors. In contrast the normalisation with covariance was used as this method makes it possible to compare the level of similarity between each pair of SPs and to identify the role of important SPs (Wasserman & Faust, 1994: 539-41, 577-80). The combination of this measure along with the pruning technique gives a shortest path spanning tree of SPs defined by the degree of similarity of equivalent off-park partners. In this network (Figure 5.4) only the strongest links between pairs of SPs remain. This differs from the previous diagram (Figure 5.3) where 52% of the links are intra-regional, whereas here only 45% and only a few important SPs, based in Scotland, the South East, and the East of England, remain interconnected.

Given the position of the SPs in this Pathfinder network, the functional roles were identified with the help of the stochastic approach of Guimera and Amaral (2005). On the other hand, the modular structure identified in the original 2-mode network was used to best keep the information of the original SP—off-park organisation interconnections since the modular structure is defined on subsets (not pairs) of actors and events (Guimera et al., 2007). In terms of the 2-mode modular structure (see Figure 5.2), the high concentration of few HEIs as main external partners results in one main component of 8 modules of which only one module (green nodes) concentrates 66% of all the parks, with a high rate (72%) of intra-modular connectivity. Only the module of Scottish parks (white nodes) is well defined because they interact with a different group of off-park partners. The best represented regions are Scotland 18%, South East 15%, and East of England 15%. In terms of roles, there are 5 out of 7 potential roles (see Figure 5.4). 85% are ultra-peripheral nodes (R1); 8% are peripheral nodes (R2); the *Granta Park* is the satellite connector node (R3); the *Wellcome Trust Genome Campus* is the provincial hub (R5); and the *Harwell Oxford* and *New Frontiers SP* are connector hubs (R6). The core of the network is connected by *Harwell Oxford*, and some other South Eastern parks, as a result of its broad multidisciplinary research capacity led by physics and material engineering. This central group is connected to the right through the satellite connector, *Granta Park*, which hosts important technology engineering and bio-pharmacological tenants, the ideal interface to more bio-pharmacological parks, such as *Cambridge SP*. On this right side, the group is led by the other hub, *New Frontiers SP*, which represents the links of its important bio-pharmacological tenant *GlaxoSmithKline*. On the left side, the central group is connected through the agro-biotech group of parks from Scotland. Here the group is led by the provincial hub, *Wellcome Trust Genome Campus*, which hosts RIs specialised in functional genomics and bioinformatics (see Chapter 4). Overall, this network has a thematic structure where the research produced by these three sub-groups is mainly characterised by the following subject areas: biochemistry, pharmacology, immunology, agriculture and biology (to the left); physics and material engineering, chemistry, computer science (in the centre); biochemistry, pharmacology, chemistry, medicine (to the right).

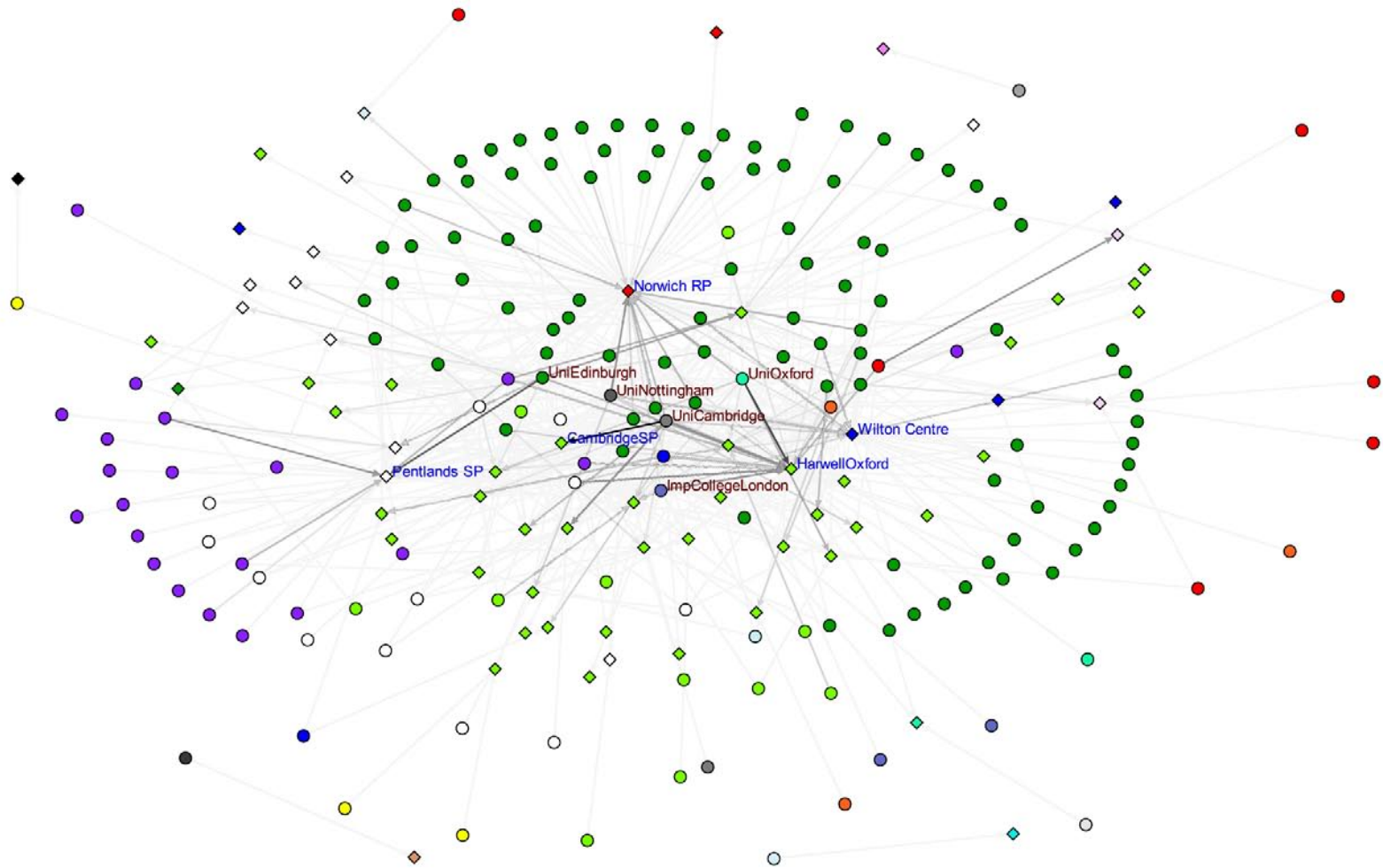

**Fig. 5.2** Bipartite network of the collaborations between SPs (diamonds) and off-park organisations (circles). 233 nodes (65 diamonds, 168 circles); 609 Arcs (representing ≥ 3 collaborations); 5,010 Collaborations. Modular structure: 32 Modules (12 diamonds, 20 circles). Link opacity is proportional to the number of collaborations. The node colours represent the different modules.

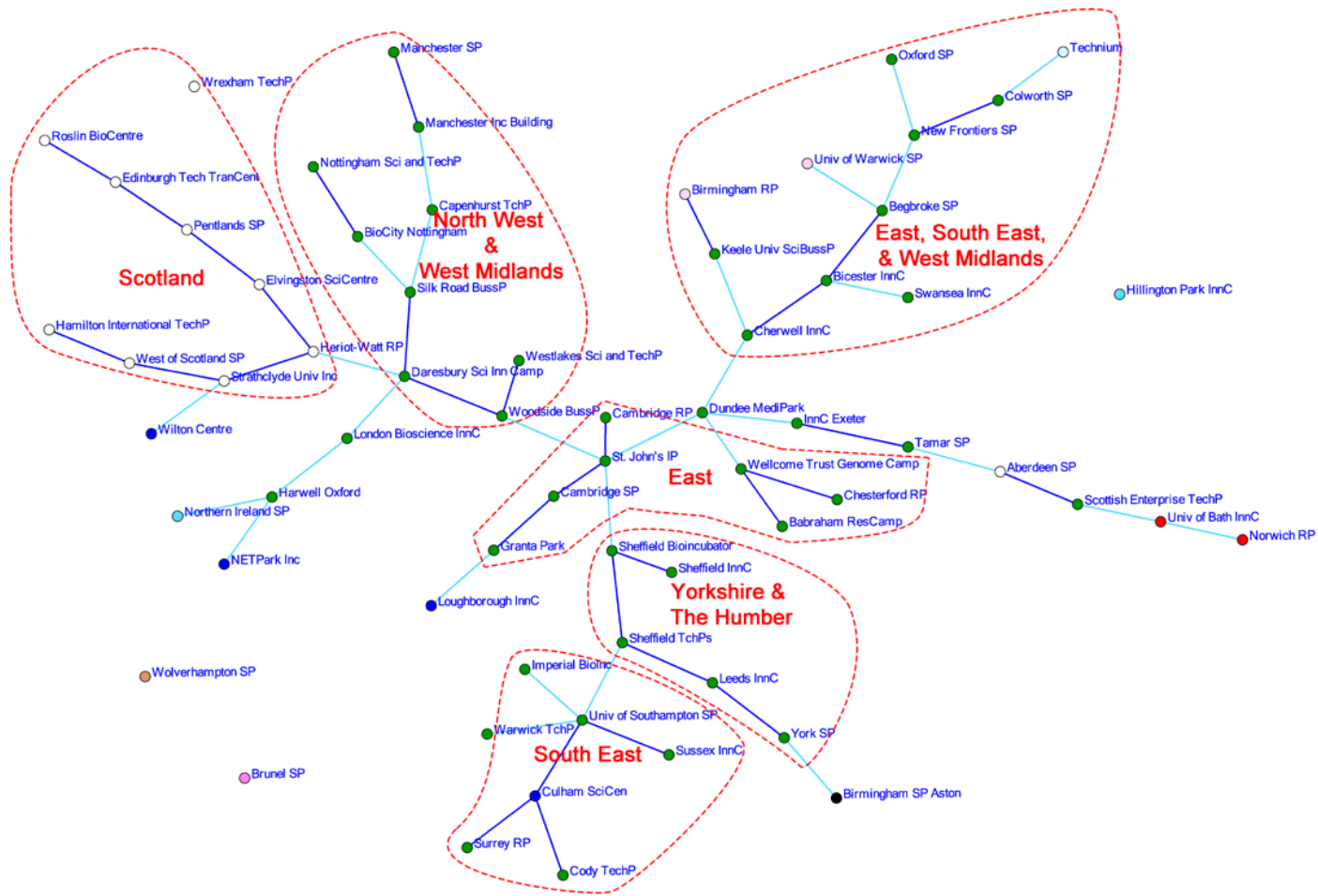

**Fig. 5.3** Uniparitite network of UK SPs defined by the similarity of equivalent off-park organisations that co-collaborate with each pair of SPs (normalised with Pearson correlations). 65 Nodes; 60 Edges. 2-mode modular structure: 12 Modules. 31 (52%) Intra-regional links (blue), 29 (48%) Inter-regional links (sky blue); 44 (73%) Intra-modular, 16 (27%) Inter-modular links. The node colours represent the different modules

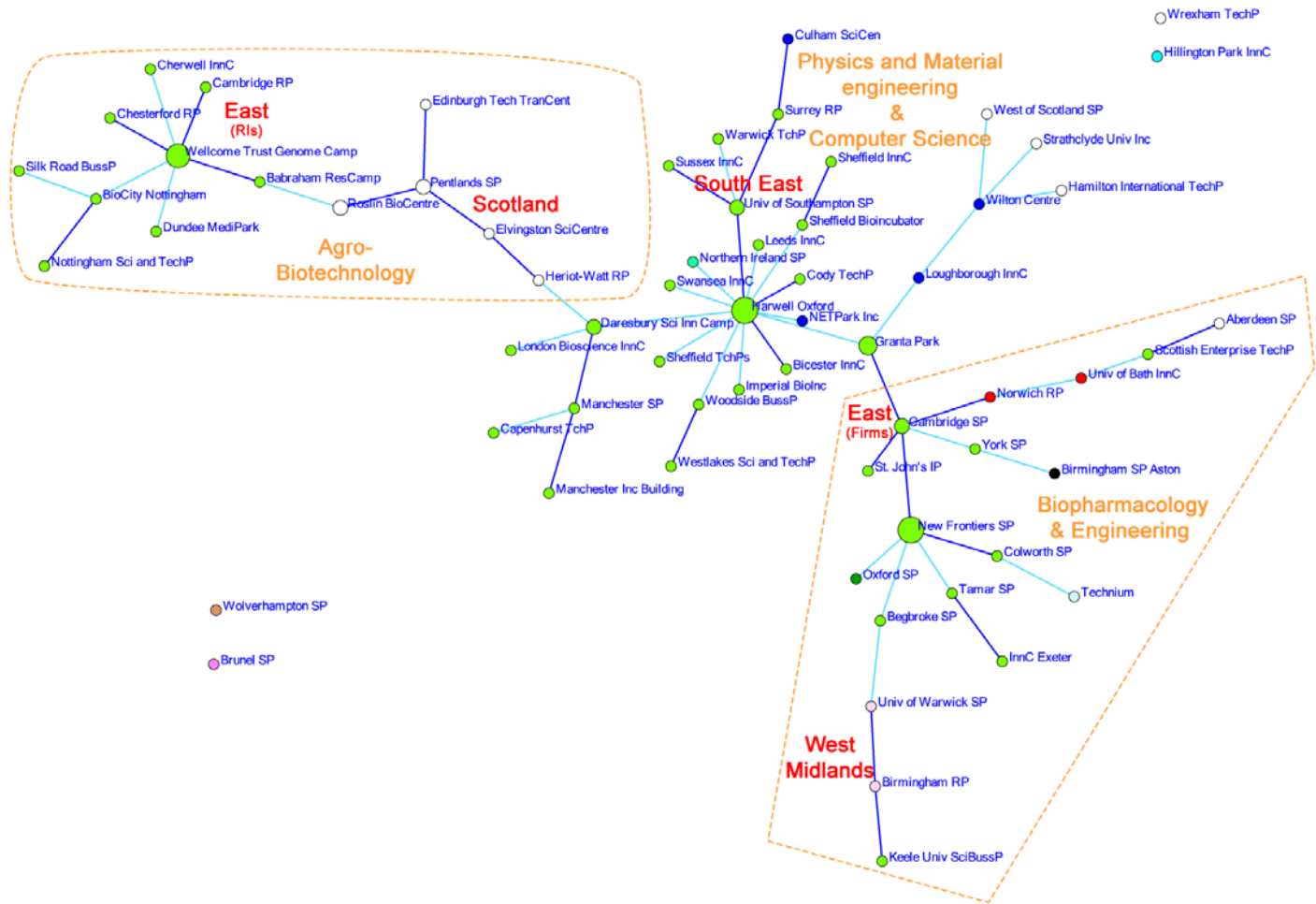

**Fig. 5.4** Unipartite network of UK SPs defined by the similarity of equivalent off-park organisations that co-collaborate with each pair of SPs (Covariance). 65 Nodes; 60 Edges. 2-mode modular structure: 12 Modules. 27 (45%) Intra-regional links (blue); 33 (55%) Inter-regional links (sky blue); 43 (72%) Intra-modular; 17 (28%) Inter-modular links. The node colours and size represent the different modules and roles, respectively.

### *a Top partnerships between HEIs and SPs*

Table 5.1 lists the SPs which are the first and second most common partners of HEIs, and the HEIs which are the most common partners of SPs and the HEIs with the highest number of collaborations with on-park organisations. The three top SPs are the main partners of 47% of the HEIs. This high concentration could be because these SPs host important RIs and companies. The number of main partners of HEIs from the same regions is much less (37%) than from outside the region (63%), however the local partnerships are more intense and generate more collaboration (74%). Similarly, in the case of the second most important partners, the collaborations are concentrated (50%) around four SPs, and the number of main partners in the same region is much less (36%) than outside (64%). For SPs, the top ten HEIs are the main partners of about 50% of the SPs and establish 50% of the U-I links. These HEIs are highly research-intensive and the top three are from the three main regional agglomerations of the SP movement (East, South East, and Scotland) (see Chapter 4), confirming the important role of HEIs in the development of research intensive industries associated with SPs. In contrast to HEIs, SPs' main academic partners are located in the same region (54%) and outside to a similar extent (46%), and at least 78% collaborate with one HEI from the same region.

**Table 5.1** Top collaborative partners of the collaboration between HEIs and SPs.

| HEIs' main SP partners | | | | | | SPs' main HEIs partners | | | | |
|---|---|---|---|---|---|---|---|---|---|---|
| First main partner | | (n=86) | Second main partner | | (n=82) | First main partner | | (n=141) | Partners collaborations | (n=5,090) |
| Harwell Oxford | 19 | 22% | Norwich ResPark | 16 | 20% | U. of Nottingham | 12 | 9% | U. of Cambridge | 449 | 9% |
| Norwich ResPark | 13 | 15% | Harwell Oxford | 9 | 11% | U. of Edinburgh | 10 | 7% | U. of Edinburgh | 309 | 6% |
| Wilton Centre | 9 | 10% | Wilton Centre | 8 | 10% | U. of Cambridge | 9 | 6% | U. of Oxford | 300 | 6% |
| Cambridge SciPark | 4 | 5% | Tamar SciPark | 7 | 9% | U. of Oxford | 8 | 6% | Imperial College London | 226 | 4% |
| Surrey ResPark | 4 | 5% | New Frontiers SciPark | 5 | 6% | U. College London | 7 | 5% | U. of Nottingham | 222 | 4% |
| Pentlands SciPark | 3 | 3% | Pentlands SciPark | 3 | 4% | U. of Manchester | 7 | 5% | U. of Manchester | 216 | 4% |
| Scottish Enter.TechPark | 3 | 3% | Begbroke SciPark | 2 | 2% | U. of Southampton | 6 | 4% | U. of Glasgow | 196 | 4% |
| Tamar SciPark | 3 | 3% | Heriot-Watt ResPark | 2 | 2% | U. of Sheffield | 5 | 4% | U. College London | 177 | 3% |
| Technium | 3 | 3% | Wrexham TechPark | 2 | 2% | U. of Strathclyde | 5 | 4% | U. of Birmingham | 173 | 3% |
| U. of Warwick SciPark | 3 | 3% | Aberdeen SciPark | 1 | 1% | U. of Warwick | 5 | 4% | U. of Sheffield | 161 | 3% |

The Kruskal-Wallis non-parametric ANOVA test was applied to formally test the relationship between HEIs' quality level, based on four quality groups (top, high, medium, and low) (HEFCE-OSI, 2009), and the number of collaborations with SPs. This shows that there is a significant difference between the four groups of HEIs ($H(3) = 38.877$, $p < .05$), and Jonckheere's test reveals a significant trend in the data, as the lower the HEI's quality level, the lower its median collaboration count ($J = 208.5$, $z = -6.8$, $r = -.81$). This finding indicates that there is strong relationship between the quality of HEIs and greater collaboration with SPs. To follow up this finding a Mann-Whitney test was used to compare the significance of the differences between the four quality groups, along with a Bonferroni correction that indicates that the critical level of significance is .0167, instead of .05. The test reveals that the level of collaboration is significantly higher for top HEIs in comparison with high quality HEIs ($U$=18.5, $z$ = -2.87 $p$ < .01, $r$= -.5, $\overline{x}$ $rank$=26.4 $vs.$ 14.2) and respectively between high versus medium quality HEIs ($U$=77, $z$ = -4.88 $p$ < .01, $r$= -.67, $\overline{x}$ $rank$=37.5 $vs.$ 16.9), although the collaboration level between medium and low quality HEIs does not differ significantly ($U$=78.5, $z$ = -2.262 $p$ > .01, $r$= -.37, $\overline{x}$ $rank$=22.09 $vs.$ 13.14).

### 5.3.3.    Firms-HEIs-RIs collaboration network

SPs, as U-I interfaces, are designed to be one of the main driving forces for the capitalisation of public research (Hauser, 2010), although there is no evidence about the structural role that HEIs play in the collaborative network associated with SPs or the properties of these research-intensive networks. For this reason, this section analyses the structure of collaborations between firms, HEIs, and RIs located either on or off the park sites in the UK. The modular structure and functional roles are identified with the help of a stochastic approach (Guimera & Amaral, 2005) based upon pairs of organisations co-authoring more than six publications. This threshold is to focus on the core of the network. The visual complexity of a diagram with 111 nodes connected by 300 bidirectional links representing 4,085 collaborations makes it necessary to apply multidimensional scaling for node positions and a shortest path spanning tree for connections between nodes. This only represents the most important links between each pair of actors. Thus, the groups and positions of the actors are identified based on all the links, while the representation and further analysis only considers the shortest path spanning tree.

Figure 5.5 represents the backbone of the collaborative structure associated with the UK SP movement. This diagram has more HEIs (47%) than firms (36%) and RIs (21%), resulting in similar off- (52%) and on-park (48%) representations. This network is formed by 6 groups and the modularity $M$ of the best partition is 0.386, while the average modularity $<M>$ of the randomizations of the network is 0.328 (S.D. 0.004), which is significantly modular. The greatest module has 33 (orange) nodes, primarily academic institutions (66%). It is characterised by research production in a wide range of areas. The top subject category is *condensed matter physics*; *biochemistry;* and *electronic, optical and magnetic materials*. The second one, with 26 (white) nodes, is dominated by south eastern organisations (50%) and by many firms (46%) researching *biochemistry*, *molecular biology*, and *genetics*. The third one (25 blue nodes) is characterised by organisations based in the East of England (32%) and having the highest research productivity. Its research interest is in *genetics*, *biochemistry*, and *food science*, having a similar research profile to the second module. The fourth module (17 yellow nodes) is dominated by organisations based in Scotland (59%) which investigate *veterinary medicine*, *immunology*, and *parasitology*.

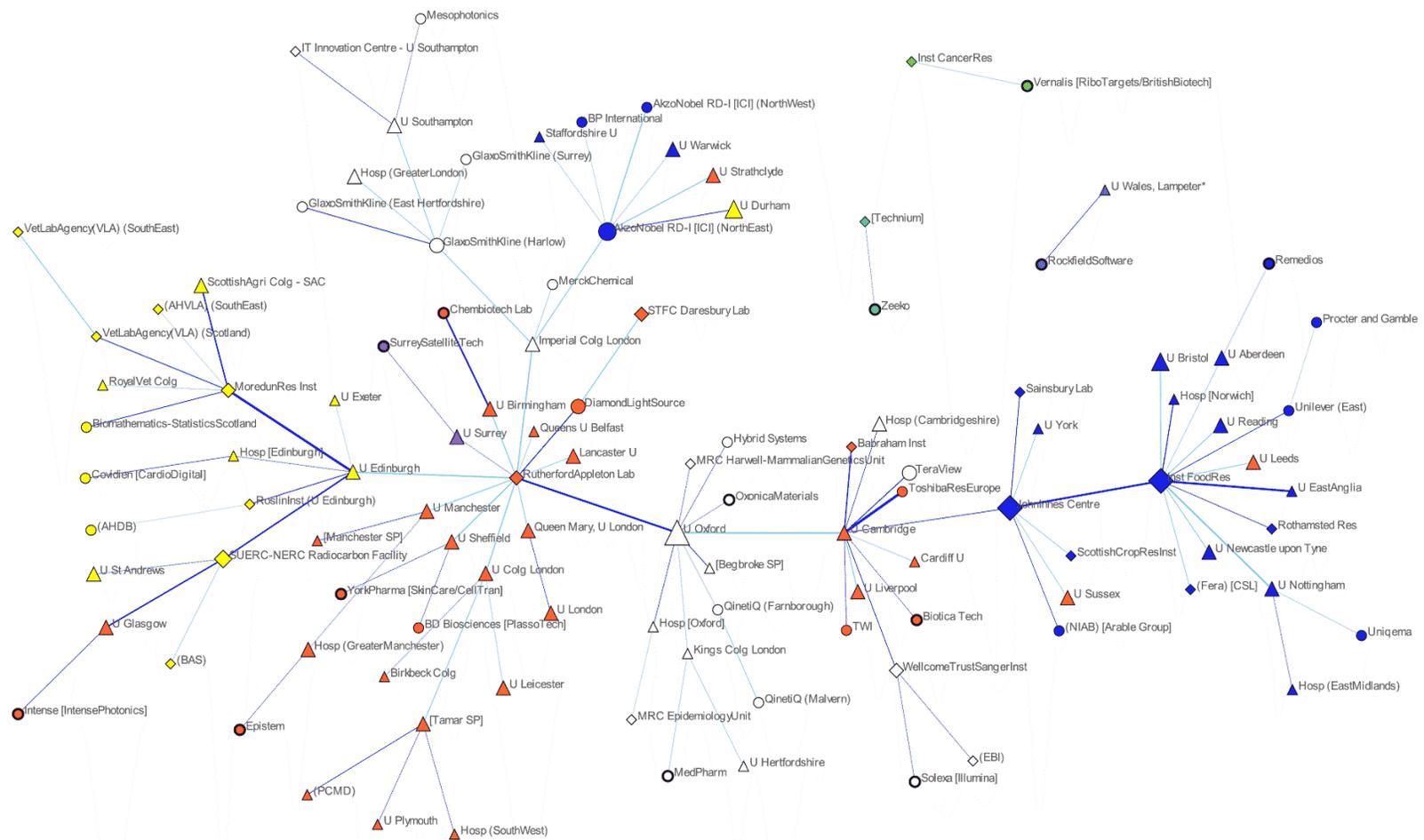

**Fig. 5.5** Shortest path spanning tree of the collaborations ≥ 7 among on-park or off-park HEIs (52 triangles), RIs (23 diamonds), Firms (23 circles), and Spin-offs (13 bold circles). 111 nodes; 107 Edges (1,913 collaborations represented). Modular structure: 7 Modules. 57 (53%) Intra-regional links; 50 (47%) Inter-regional links. The links opacity is proportional to the collaborations rate. The node colours and sizes represent the different modules and roles, respectively.

In terms of roles, 4 out of 7 potential roles were identified by the stochastic analysis. 61% are ultra-peripheral nodes (R1); 32% are peripheral nodes (R2); 0.4% are satellite connector nodes (R3); and the *Inst. of Food Research*, *John Innes Centre*, and the *Univ. of Oxford* are the three connector hubs (R6). The representation collocates two of the connector hubs, *Inst. of Food Research* and *John Innes Centre*, to the right side of the network. Both BBSRC institutes belong to the same module (blue) and specialise in food and health, and plant and microbial science, leading sub-networks of the closest partners. The sub-network led by the *Uni. of Cambridge* connects these two hubs with the third one, *Uni. of Oxford*, which in turn is linked to the rest of the network through the *Rutherford Appleton Laboratory*. This laboratory mainly supports research in materials and space science, building a multidisciplinary core with close ties to the academic community. Overall, the regions best represented are: East of England (22%), South East England (21%), and Scotland (14%). The number of intra-regional (53%) and inter-regional (47%) links is similar, but the number of collaborations for the intra-regional links (61%) is higher than the number of collaborations for the inter-regional links (39%). The pivotal role of HEIs is illustrated by their involvement in more than three quarters of the ties (*HEIs-RIs (32%), HEIs-firms (26%), and HEIs-HEIs (19%)*), and in 79% of all the collaborations (*HEIs-firms (42%), HEIs-RIs (23%), and HEIs-HEIs (14%)*), which makes them the main partner and the links between on- and off-park organisations the most common in terms of edges (70%) and frequency (72%).

To further illustrate the importance of the universities embedded in the R&D activities network of the SP movement, the main centrality measures (Wasserman & Faust, 1994: 169-90) were calculated on the original network. A central and intermediary role was quantified by having on average a much higher *degree centrality* and *betweenness centrality* (7.6; 130.08) than RIs (5.30; 122.7) and firms (2.3; 31.2). Similarly, the highest and lowest *eigenvector* and *closeness centrality* (HEIs .08; 1162.5; PROs .04; 1946.8; firms .02; 1934.4) shows that the universities are near to the others, being a central source of collaboration and knowledge exchange. Conversely, firms play a peripheral role in the collaboration network.

### 5.3.4.    HEIs ranking position vs. network centrality

The pivotal role of a small set of HEIs, as the main external source of research and interconnectivity of the SP movement suggests that there may be a significant association between the research excellence of research institutions and their position in the network as the result of private sector collaboration. This comparison can help to confirm whether HEI quality is also an important factor in the interaction with on-park organisations.

**Table 5.2** Statistical comparison of HEI quality level (top, high, medium, low) and functional position in the (≥2) HEI-Firm-RI collaboration network.

| | Kruskall-Wallis | | | Jonckheere | | |
|---|---|---|---|---|---|---|
| | $H$ | $df$ | $p$ | $n$ | $J$ | $z$ |
| Role | 26.9 | 3 | <.05 | 61 | 200.5 | -5.2 |
| Degree | 34.1 | 3 | <.05 | 61 | 123.5 | -6.2 |
| Betweenness | 30.8 | 3 | <.05 | 61 | 147 | -5.9 |
| Closeness | 33.9 | 3 | <.05 | 61 | 1023.5 | 6.2 |

Asymp. Sig. (2-tailed)

A Kruskal-Wallis test was used to examine the relationship between HEI quality level, based on four different groups (top, high, medium, low) calculated from research spending, academic research staff, 2001 RAE average quality score, and research intensity (HEFCE, 2009), with the three main centrality measures (*degree, betweenness, and closeness*) and the functional role obtained from the 61 HEIs embedded in the HEIs-Firms-RIs network with more than one collaboration (see Figure 5.6). The results show that there is a significant difference between the HEI four quality groups (see Table 5.2), and a Jonckheere's test reveals an overall significant trend in the data: as HEI quality level decreases the median collaboration count decreases. For *closeness centrality*, there is an opposite trend as a low value indicates closeness to the others actors in the network, giving a positive trend (see Table 5.2). This finding indicates that there is a strong association between the quality of HEIs and a more central and intermediary role in the collaboration network. To follow up this finding a Mann-Whitney test (see Table 5.3) was used to compare the differences between the four HEI quality groups, along with a Bonferroni correction that indicates that the critical level of significance is .0167, instead of .05. The test reveals that the centrality indexes of top and high quality HEIs are significantly different but not in terms of the functional roles. This shows that universities' level of collaboration with firms and RIs is significantly higher for top HEIs. The comparison of the high versus medium quality group differs significantly in terms of both centrality measures and functional roles, showing that high quality HEIs are more important partners. However, the third comparison, medium against low, does not show any statistical difference, finding that these two groups play a similar peripheral role. Similarly, the application of these tests to other centrality measures (*bonacich's*

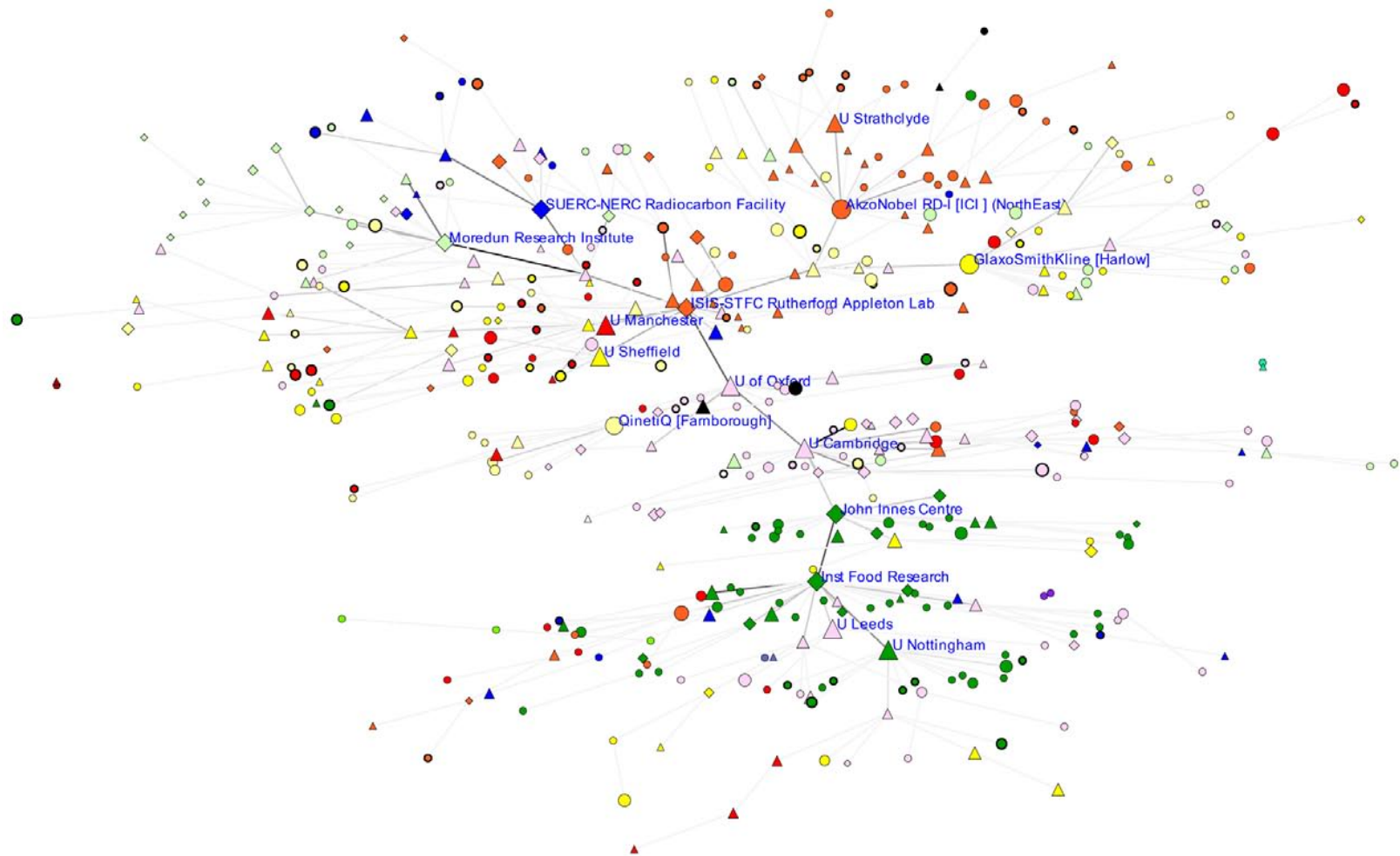

**Fig. 5.6** Shortest path spanning tree of the collaborations ≥ 2 among on-park or off-park HEIs (triangles), RIs (diamonds), Firms (circles), and Spin-offs (bold circles). 439 nodes; 433 Edges (2,867 collaborations represented). Modular structure: 15 Modules. Link opacity is proportional to the collaboration rate. Node colours and sizes represent the different modules and roles, respectively.

*centrality, authority, hub, harmonic closeness, reach, k-step reach centrality, eigenvector, local eigenvector*) gives the same results.

**Table 5.3** Pairwise comparison among the four quality groups and the functional position in the network.

| Group 1 vs. Group 2 | U | Z | p | r | Centrality index | Mean rank |
|---|---|---|---|---|---|---|
| Top vs. High | 9 | -3.439 | >.0167 | -0.57 | *Degree* | 32 vs. 15.8 |
| | 14 | -3.227 | >.0167 | -0.54 | *Betweenness* | 31.2 vs. 15.9 |
| | 13 | -3.268 | >.0167 | -0.54 | *Closeness* | 5.7 vs. 21.1 |
| | 47 | -1.897 | *n.s.* | -0.32 | *Role* | 25.7 vs.17.1 |
| High vs. Medium | 81.5 | -4.616 | >.0167 | -0.03 | *Degree* | 34.8 vs. 15.2 |
| | 107 | -4.15 | >.0167 | -0.58 | *Betweenness* | 33.9 vs. 16.4 |
| | 82.5 | -4.585 | >.0167 | -0.64 | *Closeness* | 18.3 vs. 37.6 |
| | 105 | -4.291 | >.0167 | -0.6 | *Role* | 34 vs. 16.3 |
| Medium vs. Low | 25 | -0.68 | *n.s.* | -0.14 | *Degree* | 294 vs. 31 |
| | 19 | -1.21 | *n.s.* | -0.24 | *Betweenness* | 300 vs. 25 |
| | 21.5 | -0.962 | *n.s.* | -0.19 | *Closeness* | 274.5 vs. 50.5 |
| | 33 | 0 | *n.s.* | 0 | *Role* | 286 vs. 39 |

Asymp. Sig. (2-tailed). Significant value of .0167

U=Mann-Whitney value; Z=standard deviation value; p=significant value; r=effect size value

Overall, this suggests that HEIs play a functional role for the SP movement as a source and bridge of new ideas and innovation. This role is determined by the HEI quality level that is in turn strongly associated with their central positions in terms of collaborations. Thus, HEIs' collaborative links with on-park organisations increase according to their quality level, particularly among top and high quality ones, although this central position decreases for medium and low quality HEIs, with the latter institutions seeming to play a peripheral role in research transference.

### 5.4 Discussion

The potential of research and technology to generate wealth and well-being has led to closer interactions between science and social, economic, and political domains, producing more contextualized and socially robust knowledge production (Nowotny et al., 2001). The pivotal role of academia in research-intensive industries (Cockburn & Henderson, 1998a) is closely related to companies' efforts to accumulate the necessary knowledge and skills to identify, assimilate and apply knowledge generated outside the organisation (Cohen & Levinthal, 1990). In this context, SPs are expected to provide the conditions to facilitate this knowledge transfer. However, there has been little evidence of R&D activities within on-park organisations, the entrepreneurial impact of SPs on the academia sphere, or the role of public science in the research production of on-park organisations. The analysis of research collaboration associated with the SP movement, therefore, provides a better understanding of the degree of U-I engagement to secure the competitiveness of knowledge-intensive industries.

### 5.4.1.    On-park collaboration network

The weak density network (see Figure 5.1) suggests that inter-firm links across SPs are not stronger than the collaboration found outside of previous studies (Miotti & Sachwald, 2003; Tijssen & Korevaar, 1997). This on-park collaboration network includes six main modules that represent a small fraction of the R&D activities of five different regions. This modular structure shows that the interactions are not necessarily established on a physical proximity basis. The inter-institutional partnerships at regional and national levels are established more often than within the same park despite tenants from the same SP cooperating up to three times more often than tenants in the same region or country, this close and productive cooperation reflects collaboration among RIs rather than cross-fertilisation between U-I or firms.

Firms occupy ultra-peripheral positions and are connected to the network by their collaboration with a few large, research-oriented nodes, such as RIs, that occupy a more central position. As expected, collaboration between on-park firms, mostly academic spin-offs and subsidiaries, is rare and relies on large pharmaceutical companies and RIs to form small collaboration communities. A similar pattern has also been pointed out by Soetano and Jack (2011). On the other hand, the group of central nodes has the function of supporting firms and links the network together. Here, the pivotal role of the big pharmaceutical companies could be explained in terms of their strong knowledge needs and economic resources, as most of the interactions between big pharmaceutical companies and other companies, based on licensing agreements, involve research and product development that gives them the first mover advantage for cutting edge technologies, and at the same time the possibility to attract new innovators and concentrate market share (Arora & Gambardella, 1994; Gay & Dousset, 2005: 1468). Similarly, the pivotal role of RIs could be explained by RIs' missions to support industry as well as the need for high-growth and knowledge-intensive industries on SPs to access external research and technology (Gay & Dousset, 2005: 1457). RIs are the major connectors within and across modules, and support research-intensive firms in knowledge creation despite RIs and industry collaboration being influenced by the market, and not generally leading to scientific production (Crow & Bozeman, 1998; Noyons, et al., 1999). Indeed, one of the reasons for the low network density could be the reliance on RIs to cover the knowledge needs of the tenants. This shows to what extent SPs, in some cases alongside RIs, are not able to foster significant on-park collaboration, with the university being the main external partner that brings together firms, RIs, and SPs across the country.

### 5.4.2.    SP and off-park collaboration network

As expected, the interactions between on-park and off-park organisation create a denser network (see Figures 5.2) in comparison to the ties between parks (see Figure 5.1). This density is the result of the inclusion of HEIs and consequently higher rates of research production. The collaboration between SPs and off-park organisations shows that HEIs are clearly the main external research partner for SP tenants. However, the HEI-SP collaboration only represents 14% of the total research production, which is similar to the low level of HEI-SP interaction found in qualitative studies previously (Felsenstein, 1994; Vedovello, 1997). The central role of academia in the collaboration network results in higher levels of centrality measures in comparison with other external partners. Interestingly, these external links are mainly concentrated around a small set of high-quality HEIs and are mainly established on an inter-regional basis. This suggests that even for on-park organisations, physical proximity is only a significant factor in conjunction with knowledge source research quality (Arundel & Geuna, 2004). However, these physical connectivity patterns should be considered as purely indicative because of the diverse sectors of activities of the on-park industry and the different types of physical U-I interaction involved (Abramovsky & Simpson, 2011).

The transformation of the SP and off-park collaboration network into a SP-SP network uncovers groups of SPs that are strongly interconnected by having the same off-park partners (see Figures 5.3 and 5.4). This finding confirms that research partnerships do not necessarily follow physical proximity (Laursen et al., 2011). Instead, the cognitive proximity tends to play a more important role in the interactive learning and knowledge creation implied in the co-authorship of research papers (Boschma, 2005). Similarly, the quality of the knowledge and expertise from the external partners are more important factors in the acquisition of external research and technology (Arundel & Geuna, 2004; Felsenstein, 1994). The SP-SP network shows two main trends based on a global or local method of analysis. First, the global approach shows that SPs based in the same regions are more likely to be closely interconnected as a result of sharing more similar groups of partners to some extent (see Figures 5.3). Second, the local approach shows that the degree of partner similarity between each pair of SPs follows a similar research subject interest (see Figures 5.4). This implies that physical proximity is less important than mutual research and technological interest and suggests that short physical distances are not necessary when interconnections are mainly based on cognitive proximity and the capacity of actors for learning and creating new knowledge. However, physical proximity at the same time stimulates social proximity, defined in terms of social interaction between agents at micro-level based on friendship, trust building and committed relationships (Boschma, 2005). Moreover, it also, favours local collaborations to some extent as shown by the higher degree of intra-regional connections, although they are weaker. The thematic network is formed by three different groups of parks that are led by three hubs that represent three different industrial sectors. These different groups of interconnected parks seem to be associated with the three clusters (East, South East, and Scotland) with the closest proximity to a science base  and where science and technology parks have the greatest impact in the country (Birch, 2009; Cooke, 2001b).

### 5.4.3.    Firms-HEIs-RIs collaboration network

The network established by the most collaborative organisations associated with the UK SP movement illustrates the centrality and important role of research producers in comparison with the peripheral positions and limited number of research-intensive firms. This structure (Figures 5.5) shows that a small number of highly innovative firms have built networks with other on-park firms or research producers to access intangible resources, such as knowledge and ideas (Bozeman, 2000; Rothaermel, et al., 2007; Soetanto & Jack, 2011). Only AkzoNobel and GlaxoSmithKline have access to a diverse portfolio of academic partners, confirming the closer links between large firms and other knowledge producers (HEFCE, 2010a; Vedovello, 1997).

On the other hand, the increasing dependence on public science as a result of intense competition in cutting-edge industries (Cohen et al., 2002; Godin & Gingras, 2000) is also shown in this study by  the central role of HEIs in the network. The central position of academic research suggests, based on other findings (Miotti & Sachwald, 2003), that this network is driven by technology and knowledge transfer, as well as the generation of new products and patents. East, South East, and Scotland are the best represented regions, showing the relationship between the SP movement and the life science and biotechnology industry (Sainsbury, 1999) and the development of research-intensive clusters (Cooke, 2001b). The innovation capacity of the organisations in East and South East (Kasabov & Delbridge, 2008) resulted in central positions and significant integration with other organisations in the collaboration network. This reflects the success of these regions in knowledge generation, technology transfer, and the exploitation of scientific research (Cooke, 2001b; Kasabov & Delbridge, 2008; Looy, Debackere, & Andries, 2003). The network also illustrates the peripheral role of the emerging biotechnology cluster in Scotland found by Leibovitz (2004). This peripheral position of Scottish organisations suggests a less central role in the global innovation system and, particularly, in the life sciences and biotechnology, in comparison with the impact of the East and South East regions in terms of outstanding research production and experience in technology transfer and commercialisation in research-intensive sectors (Birch, 2009; Cooke, 2001b; Kasabov & Delbridge, 2008).

### 5.4.4. HEIs' ranking position vs. network centrality

Higher quality HEIs fill a more central and important role in the collaboration network (see Table 5.2 and 5.3). Based on the centrality indexes calculated, HEI quality strongly associates with the degree of centrality, activity, power and influence, which are key sources of research collaboration and operational impact. The pairwise comparison of the quality groups shows that the quality level difference is significantly higher between the top and high, and between the high and medium quality levels. Quality is not a significant collaboration factor among medium and low quality HEIs, however, perhaps due to a similar peripheral positions and limited impact.

In terms of functional role, there is only a significant difference between the positions of high and medium quality HEIs. Presumably, most organisations associated with the SP movement rely on top and leading research institutions for the co-creation or acquisition of external knowledge. This supports the literature suggesting that the reputation, expertise, relevance and excellence of the research producer are important for establishing partnerships with  firms (Abramo et al., 2009; Beise & Stahl, 1999; Clarysse, Tartari, & Salter, 2011; Felsenstein, 1994; Fukugawa, 2006; Laursen et al., 2011; Vedovello, 1997). In addition, this is in line with Links and Scott's (2003: 1350) assertion that "university R&D activity is an instrument useful in predicting, in a benchmarking sense, what impact to expect from its science park involvement". Thus, the propensity for establishing U-I links is partly determined by the quality of the academic knowledge, which should therefore be one of the main factors to consider when establishing and assessing research-intensive support infrastructures. However, the relationship between research excellence and U-I links could be influenced by more investments from top universities in capitalizing their research (Geuna & Muscio, 2009).

Overall, these results should be interpreted with caution as this study has some important limitations. First, even though most U-I links tend to be informal (Bozeman, 2000; Soetanto & Jack, 2011) and the networking activities in this study are informal research links generated from intangible resources (e.g., knowledge and ideas) (Calero, et. al., 2007), inter-institutional collaboration only represents a quarter of the total on-park publication output, of which more than half is U-I. This indicates that even though HEIs may have an increasingly important role as a source of new ideas and innovation for firms, their cooperation with firms is still limited in comparison with the collaboration between firms with intermediaries or with other firms (Cosh & Hughes, 2009). Second, this study ignores the international research network of on-park organisations (Tijssen, 2009), despite this collaboration being more likely to boost innovation (Miotti & Sachwald, 2003). However, this study is limited to national networks because UK SPs aim to exploit the national knowledge base to support growth of local research-intensive industries. Third, U-SP interaction involves links with either industry or RIs, and cannot be considered as always being a U-I interaction. Fourth, the collaborations analysed here take place between co-authors at individual level and the aggregation of these interactions into inter-organisational does not necessarily represent any institutional engagement. Finally, joint U-I research as a proxy for knowledge transfer only sheds light on one dimension of a multi-dimensional interaction. Thus, the interpretations should be considered with caution due to the lack of data making it possible to determine the institutional setting and conceptualise the interactions that take place among these organisations (Bergenholtz & Waldstrøm, 2011: 543). For example, firms and low research quality universities may tend to translate a much lower proportion of their collaborations into formal academic publications.

### 5.5 Conclusions

The UK government has recently announced policies to develop further enterprise zones and support infrastructures near HEIs to promote partnerships with local enterprises and attract international inward investments with the goal to attain a global status in this regard (Wilson, 2012). This strategy assumes that the capacity of SPs to add economic value to academic research and technology turns the interaction between universities and SPs into a strategic partnership and driving force for socio-economic development. However, policy-makers' expectations contradict the rather uneven SP performance so far (Quintas et al., 1992; Donald Siegel, Westhead, et al., 2003; Westhead & Storey, 1995) and the insignificant impact of SPs on HEI technology transfer or research and technology output (see Chapter 4). There has been, however, a lack of evidence of the structure of research-based collaborations within SPs. This study, therefore, attempts to obtain a first understanding of the R&D collaborations generated at different levels across the UK SP movement. This approach is fundamental because even though U-I interaction is widely accepted as an indicator of the intermediary and operational success of SPs, the structural organisation of these ties has not been analysed.

In answer to the first research question, the weak modular structure of the on-park collaboration network provides the properties to understand the role of the different types of nodes in relation to the different regional communities where three main modular properties can be identified. First, there is a limited amount of interaction between on-park organisations, which are mainly firms and RIs that have ultra-peripheral (R1) and peripheral (R2) roles, respectively. Second, collaborations are mostly at regional and national levels, and this suggests that the SPs included in this study have not increased networking opportunities between their tenants. Third, the network relies on collaborations between anchor tenants, such as large pharmaceutical companies and RIs, as a result of the demand for external research and technology. However, the central role of RIs and firms could hinder the development of a bibliometric manifestation of this on-park collaboration network as the interaction among firms and between firms and RIs is less likely to result in papers in comparison with the interaction of these actors with HEIs.

In answer to the second research question, the analysis of SPs with external partners suggests that HEIs are the main sources of external knowledge and play a central role in the network. These interactions centre on a small number of prestigious and high quality HEIs, and follow an inter-regional and cognitive proximity basis. When the interconnections among SPs are analysed according to the degree of similarity of their different off-park partners, physically closer SPs cluster together. On the other hand, when the comparison is only based on the main partners of each pair of SPs, SPs with similar thematic interests cluster together. This suggests that the underlying thematic and cognitive network structure is built on stronger and more important ties. The tenants of the SPs mainly collaborate with external partners because of the quality and relevance in the knowledge and expertise they can offer, forming a network that is interconnected because of similar interests in the thematic profile and knowledge base of the external partners. These indirect interconnections among parks produce three sub-networks, which associate with three important research-intensive clusters in the UK.

In answer to the third research question, the collaboration network associated with the SP movement illustrates the central role of HEIs and RIs, which are the main sources of knowledge and competitive advantage for highly innovative firms. Conversely, firms only represent 36% of the actors and have a peripheral role in the collaboration network. The central and peripheral role played by academia and the private sector suggests a collaborative structure formed by knowledge and technology transfer. The network structure also helps to get a better understanding of the potential role of the support infrastructures in the growth and development of research-intensive clusters.

In answer to the last research question, there is a fairly strong relationship between HEI quality and importance in the collaboration network. However, there is no significant difference between medium and low quality HEIs. A comparison of the functional roles occupied by HEIs in the network reveals that only the group of high quality HEIs tend to have more important positions in comparison with medium quality HEIs. Since the quality and prestige of academic institutions seem to be important factors for closer U-I partnerships, public policies should consider the level of research excellence of HEIs associated with SPs in efforts to strengthen knowledge transfer between research producers and the private sector. In addition, the difficulties of SPs in fostering knowledge transfer in less dynamic environments suggests that policymakers' efforts should focus on strengthening and supporting cooperation between SPs and leading research-research intensive HEIs.

Overall, this study shows that the research collaboration embedded in intermediary infrastructures represents a very limited share of on-park research production. These scarce interactions have HEIs as the main partner, occupying a central position in the networks, which highlights the importance of academic institutions in the acquisition of external knowledge and competence of the on-park industries. HEIs' reputation and quality level is, however, a key factor in the U-SP collaboration and cross-fertilization as it significantly affects the degree of interaction with on-park firms. Thus, the higher the quality level of the research institutions, the more likely they are to establish stronger links with on-park organisations with a significant increase according to the quality level of HEIs considered as top- or high quality.

The increasing role of HEIs in providing a world-class research base in a multi-dimensional network of mutual dependence with businesses and government makes it necessary to understand each party's priorities and capabilities (Wilson, 2012). This study, therefore, provides a better understanding of the interaction between HEIs and on-park firms as a key factor to encourage a thriving entrepreneurial culture and competence. HEIs and parks can develop more realistic objectives based on the prestige and quality of the hosting HEIs as well as the dynamics of the area where the associated parks are embedded. Furthermore, this can also help both HEIs and firms to identify potential partners and gain access to new knowledge, expertise, and resources. However, it is still necessary to further investigate whether other proxies, such as patents or research contracts, provide a similar collaborative network structure between universities and SPs, and whether the collaborative ties of off-park firms also tend to be agglomerated with a small number of HEIs. This will further uncover the level of impact of SPs on facilitating the access to academic partners. In addition, the inter-regional and thematic connectivity patterns found in the interaction of support infrastructures with external partners makes it necessary to undertake analysis at regional level and more fine-grained studies in the future.

In summary, the results suggest that that inter-organisational collaboration represents a very limited share of on-park research production. These scarce interactions have HEIs as the main partners, occupying a central position in the networks. It highlights the importance of academic institutions in the acquisition of external knowledge and competence of the on-park industries. HEI research quality is a key factor for U-SP collaboration and cross-fertilization as it significantly affects the degree of interaction with on-park firms. Thus, the higher the quality of the research institutions, the more likely they are to establish stronger links with on-park organisations with a significant increase for HEIs considered to be top- or high quality.

The next chapter further studies the potential role of the University in the development of the SP movement and to what extent HEIs' close interactions with SPs leads to a more entrepreneurial academia.

# Chapter 6: Impact of science parks on HEIs' third mission activities

## 6.1 Introduction and goals

This chapter describes the final bibliometric study of the thesis. It explores whether SPs, or similar support infrastructures, are the right tools to help HEIs with R&D and knowledge transfer. It determines the time required by SPs to promote research activities and the factors that may influence this; and if a HEI's R&D performance is helped by formal relationships with SPs or other commercialisation activities. The analysis is based upon publications produced by on-park firms (Section 3.3.2), as well as patents and qualitative and quantitative data (Section 3.3.3) from national HEIs with collaborative ties with 92 SPs. Statistical analyses (Section 3.3.4) are used to assess specific related hypotheses. This study assesses some aspects of whether UK SPs can bridge between academia and on-park firms. In particular, the following research questions are addressed:

- How long after SPs are established do they start to promote research creation and cross-fertilization with academia?
- Which factors help R&D production and U-I collaboration in SPs?
- Do HEIs (and RIs) with a formal relationship –hosting, partnering, both or none - with SPs have a greater capacity to produce R&D and collaborate with the private sector or do other factors have a greater influence on U-I research interactions?

## 6.2 Results

### 6.2.1 Survival Analysis

Survival analysis is a statistical technique that helps to identify how long it takes for an event to happen and determines which causes help the event to happen. First, Kaplan-Meier survival analysis technique (Field, 2009; Kleinbaum & Klein, 2005) is applied to measure the median time taken for each SP to publish its first publication and its first U-I publication. Second, Cox regression (Field, 2009; Kleinbaum & Klein, 2005) is used to measure the extent to which the time to this research output and U-I collaboration within SPs is influenced by their age, type, establishment decade, UKSPA membership, and geographical region. The analysis deals with 162 SPs or SP—like research-oriented infrastructures identified as part of the address of tenants producing research papers, including UK Science Park Association (UKSPA) full members. However, as what is measured is the interval between the year of SP establishment and the year of its first publication, the lack of evidence for an establishment year resulted in the exclusion of 70 SPs from this analysis. The creation years of the SPs were obtained from the index designed by the UKSPA, World Alliance for Innovation (WAINNOVA), as well as the website and emails to the management teams of the infrastructures. Of the remaining 92 SPs, 26 had four or less publications by 2010, and are considered to be non-research active, and are censored from the analysis because they had not reached a minimum research activity.

### a. Kaplan-Meier analysis: time to first publication

The Kaplan-Meier analysis estimates that the median time from the inception to the first publication of the 92 analysed SPs is 3 years (95% CI 1.4 - 4.6 years). Since there are various types of tenants and collaborations within SPs, the median time was also estimated for a first publication specifically generated by (1) on-park firms, (2) RIs, (3) collaborations between on-park firms and on-park HEIs or RIs (i.e. *on-park U-I collaboration*), and (4) collaborations between on-park firms and off-park research organisations, or off-park firms with on-park research organisations (i.e. *on/off-park U-I collaboration)*. On average, the time to publish by on-park firms (4 years; 2.3 to 5.7 years) is the same as for RIs (4; 1.4 to 6.6), while *on-park U-I collaboration* (7; 4.6 to 9.3) takes longer to occur in comparison with *on/off-park I-U collaboration* (5; 3.3 to 6.7). The time difference between the four estimates is only 3 years and the number of events considered in the estimations is 65, 30, 13, and 66 respectively and the differences are statistically significant if the median of one group is not in the 95% confidence interval of another. The results suggest that firms are the major and earliest on-park research producer and the *on/off-park U-I collaboration* as the most likely and earliest type of cross-fertilisation between both sectors. Nevertheless on-park RIs only exist within some SPs and this can artificially lengthen the median because SPs with no on-park RI will never have an on-park U-I collaboration.

**Typology of Infrastructures.** The degree of involvement in R&D activities is significantly higher for research-oriented parks in comparison with commercially-oriented parks. However, there are diverse services, goals, and relationships with academia among the research-oriented parks, which suggests that there may be a difference in terms of the time to a first publication and collaboration. The six groups of infrastructures are: (1) Incubator[12], (2) Technology Park[13], (3)

---

[12] An *incubator* helps young and newly founded innovative firms to establish cooperative relationships with a broad range of economic actors and focuses on compensating for the resource deficit to ensure

Research Campus, (4) Research Park[14], (5) Science Park[15], (6) Science & Innovation Centre[16]. These have a relatively similar average time between the third (1 year; no 95% CI calculated) and first group (1; .0 − 2.5), and also between the fourth (2; .0 − 4.6), fifth (2; .53 − 3.5), and second group (3; .0 − 6.2), while the sixth group needs a considerably longer period of time (14; No 95% CI calculated). The log-rank test finds a statistically significant time difference (log-rank $p<.01$) across the groups with an overall median of 3 years, as the significant value is less than .05. In the case of *on/off-park I-U collaboration*, there is also a significant difference across the groups of parks ($p<.015$). There are only three groups with a shorter time than the overall median of 5 years, (3) Research Campus (1; No 95% CI calculated), (1) Incubator (1; .8 − 1.9) and (4) Research Park (1; .0 − 4.6), while the other groups need longer periods of time; fifth (5; 3.3 − 6.7), sixth (7; 1.8 − 12.1) and second (8; .53 − 15.5).

**Establishment of Infrastructures.** The increasing need for industry to develop in-house research capabilities and to engage in open innovation has influenced the public policies, characteristics and goals of new SPs over time in several waves (Bruneel, et al., 2012; Hansson, et al., 2005; Squicciarini, 2009). Whether the goals and evolving management strategies underpinning the creation of SPs influence further research intensification was therefore assessed. Comparing SPs created during the 70s & 80s, 90s, and 00s indicates a slight overall reduction in the median

---

entrepreneurial stability, sustainable economic growth and long-term business survival (M Schwartz & Hornych, 2008).

[13] A *technology park* is "a zone of economic activity composed of universities, research centres, industrial and tertiary units, which realise their activities based on research and technological development", and maintains strong links to large firms and the public research infrastructure at both national and international level (Saublens, 2007:56).

[14] *Research park or campus* are terms often used in the United States and are broadly defined as similar to science parks in the sense that both host a majority of tenants which heavily engage in basic and applied research and have formal associations with HEIs (A. Link & Link, 2003; A. N. Link & Scott, 2007).

[15] *Science parks* are research-based infrastructures with the following general characteristics: formal and operational linkages with HEIs or public research organisations (PROs); supporting the formation and growth of knowledge-intensive commercial businesses; active engagement in the transfer of science-based technologies and business skills (UKSPA, 2003).

[16] A *science & Innovation park or centre* is a development which it does not necessarily have operational links with a higher educational institution (Saublens, 2007).

time to their first publication (4 years; 95% CI 1.5 - 6.5 years), (5; 1.3 - 8.7), (2; 0.2 - 3.8), respectively, although this may be influenced by changing publishing practices and publication speeds. However, the differences across the three decade groups are not statistically significant (log-rank $p<.75$). A systematic reduction is observed in the average time when only the output of firms is considered (70s & 80s=7; 90s=5; 00s=2 years; $p <.042$), PROs (8; 2; 1; $p<.00$), *on-park U-I collaboration* (21; 7; 6; $p<.002$), and *on/off-park U-I collaboration* (9; 4; 2; $p<.009$). These significant values suggest that later SPs required less time to start producing research and collaborating.

**Spontaneous vs. policy-driven Infrastructures.** The early output of parks from the 70s & 80s compared to the outputs of parks from the 90s (4 vs. 5 years) may be due to the creation of the first parks in the UK to exploit the underlying dynamism of certain geographical areas (Castells & Hall, 1994). These are called *spontaneous clusters* by Chiesa and Chiaroni (2005:214-17). These exploit pre-existing concentrations of key conditions, such as a strong existing public scientific base, and an entrepreneurial culture among academics. In contrast, *policy-driven clusters* are triggered by the direct actions of policy-makers providing financial support and allocating research organisations to respond to an industrial crisis or to foster the development of a cutting-edge industry. The properties of these two types of clusters have also been applied to describe the characteristics of parks (Huang et al., 2012). Thus, when the infrastructures with research production before the year of their formal establishment are compared, these *spontaneous parks* on average publish during the first year (0; 0.0 to 0.0) versus 5 years (5; 2.5 to 7.5) ($p <.000$) for *policy-driven parks*. In terms of collaboration, the output as result of *on/off-park U-I collaboration* also shows a positive performance among *spontaneous* parks versus *policy driven* parks (1; .0 to 2.24 vs. 6; 3.87 to 8.13) ($p<.000$).

### b. *Cox regression: Predictors of time to publish and establish U-I collaborations*

Many factors can affect research production and U-I cross-fertilisation (Geuna & Muscio, 2009; Santoro & Chakrabarti, 2002). A Cox regression analysis was performed to determine if infrastructure typology, age, decade of establishment, UKSPA membership, and region could influence the time to start to produce research and transfer knowledge. In the case of research output, infrastructure type (p=.047) is the only significant predictor (Table 6.1). Among infrastructures, the categories Research Park & Campus (78%) and Science Park (75%) tend to publish first, using the Science & Innovation Centre as group reference, as illustrated by Figure 1. In the case of on/off-park U-I collaboration, infrastructure type (p .012) and age (p .013; 95% CI 0.81 − 0.98) are also predictive of the time from establishment of the infrastructure to U-I collaboration (Table 6.2). The types of infrastructures that tend to need less time to foster U-I cross-fertilisation are Research Parks & Campuses, and Science Parks with 92% and 67% chance to establish earliest U-I collaborations, respectively, using the Science & Innovation Centre as group reference (Figure 6.2). Interestingly, the statistical evidence for the predictor age shows that the new infrastructures increase the probability of establishing U-I collaborations by 47% with the younger infrastructures being more capable of promoting earlier knowledge transfer.

The significance of the infrastructure typology led to further investigations of interactions but without any further statistically significant results. This suggests that infrastructure type influences the time to publish and collaborate regardless of age, establishment decade, UKSPA membership or region. Even though information was collected regarding park sizes (number of tenants, incubated firms, and R&D technology institutions) and research properties (output, number of U-I collaborations, research-productive tenants, and type of tenants), and statistically significant influence was found for the number of incubated firms, U-I collaborations, and type of tenants, in shortening the time to publish or collaborate. These are excluded, however, because these cumulative variables are not likely to represent the circumstances of these infrastructures at the time they became research-active.

**Table 6.1** Cox regression for prediction of time to publication.

|  | Hazard Ratio (95% CI) | P |
|---|---|---|
| Region |  | 0.402 |
| East Midlands | 0.23 (0.04-1.34) |  |
| East of England | 0.98 (0.27-3.60) |  |
| London | 0.34 (0.07-1.63) |  |
| North East England | 0.36 (0.06-2.20) |  |
| North West England | 0.57 (0.15-2.16) |  |
| Northern Ireland | 0.27 (0.03-2.53) |  |
| Scotland | 0.74 (0.22-2.48) |  |
| South East England | 0.73 (0.23-2.35) |  |
| South West England | 1.91 (0.45-8.19) |  |
| Wales | 1.37 (0.31-6.08) |  |
| West Midlands | 0.95 (0.25-3.58) |  |
| Yorkshire and the Humber | 1.00 |  |
| Age | 0.96 (0.90-1.03) | 0.264 |
| Establishment decade |  | 0.845 |
| 70s & 80s | 1.23 (0.28-5.50) |  |
| 90s | 0.84 (0.92-0.39) |  |
| 00s | 1.00 |  |
| Typology |  | 0.047 |
| Incubator | 2.52 (0.95-6.71) |  |
| Technology Park | 1.52 (0.55-4.19) |  |
| Research Park & Campus | 3.78 (1.29-11.13) |  |
| Science Park | 3.09 (1.32-7.24) |  |
| Science & Innovation Centre | 1.00 | |
| UKSPA membership | | 0.578 |
| Full-member | 1.19 (0.65-2.18) | |
| Non-member | 1.00 | |

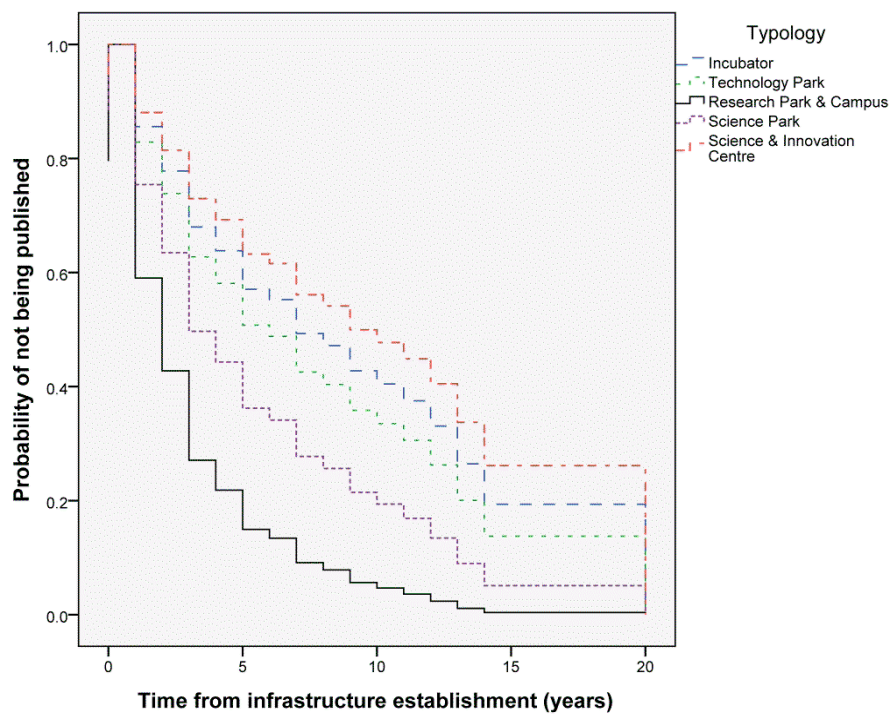

**Fig. 6.1** Proportion of publications not being published by infrastructure type

**Table 6.2** Cox regression for prediction of time to (on-park or off-park) U-I collaboration.

| | Hazard Ratio (95% CI) | P |
|---|---|---|
| Region | | 0.468 |
| East Midlands | 1.28 (0.21-7.94) | |
| East of England | 0.89 (0.17-3.06) | |
| London | 0.53 (0.09-3.06) | |
| North East England | 3.03 (0.59-15.72) | |
| North West England | 0.34 (0.04-3.25) | |
| Northern Ireland | 1.81 (0.511-5.41) | |
| Scotland | 0.92 (0.26-3.34) | |
| South East England | 2.32 (0.46-11.78) | |
| South West England | 3.89 (0.77-19.73) | |
| Wales | 0.98 (0.25-3.56) | |
| West Midlands | 1.47 (0.68-3.17) | |
| Yorkshire and the Humber | 1.00 | |
| Age | 0.89 (0.81-0.98) | 0.013 |
| Establishment decade | | 0.300 |
| 70s & 80s | 1.32 (0.24-7.31) | |
| 90s | 0.63 (0.23-1.76) | |
| 00s | 1.00 | |
| Typology | | 0.012 |
| Incubator | 1.70 (0.43-6.72) | |
| Technology Park | 1.17 (0.35-3.86) | |
| Research Park & Campus | 11.56 (2.66-50.35) | |
| Science Park | 2.39 (0.85-6.76) | |
| Science & Innovation Centre | 1.00 | |
| UKSPA membership | | 0.326 |
| Full-member | 1.47 (0.68-3.17) | |
| Non-member | 1.00 | |

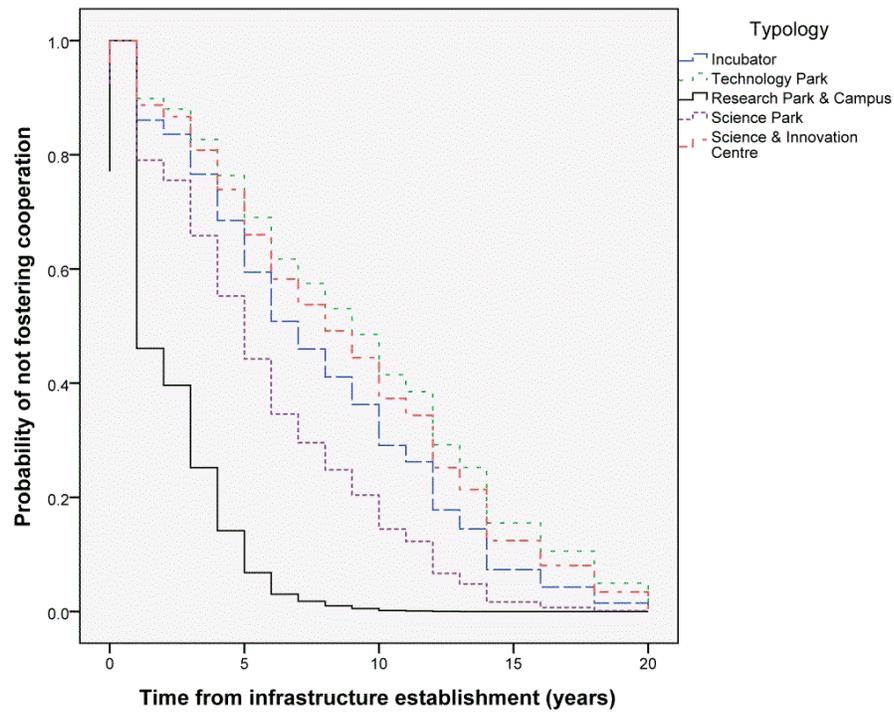

**Fig. 6.2** Proportion of on-park or off-park U-I collaboration not being published by infrastructure type.

### 6.2.2 SPs and commercial activities: impact on HEI R&D and U-I interactions

This analysis considers 78 HEIs and RIs to find to if a formal relationship with a SP positively associates with a greater entrepreneurial identity and research and technological capacity in general. This is measured by the publications co-authored with on- and off-park firms, and the overall publication and patent outputs.

**Table 6.3** Spearman correlations between different types of publication for HEIs and RIs (n=78).

| Dependent var. | 1 | 2 | 3 | 4 | 5 |
|---|---|---|---|---|---|
| 1. Patents | 1 | | | | |
| 2. Publications | .528** | 1 | | | |
| 3. On-park publications | .485** | .789** | 1 | | |
| 4. I-U on-park publications | .444** | .789** | .977** | 1 | |
| 5. I-U off-park publications | .577** | .839** | .747** | .715** | 1 |

** Correlation is significant at the 0.01 level (2-tailed).

Table 6.3 shows that there is a high correlation between four of the five variables examined, and there is a moderate correlation between patents and the other four variables. An exploratory factor analysis confirmed that variables are highly related with the exception of the patent data (see Table 6.4). The strong relationship between the different publication-based variables indicates that the collaboration with on- and off-park firms depends on the research activity, and probably the quality, of the HEIs. In contrast, patents, based on more applied research, might reflect a HEI's technological capacity more. The correlation between patents and publications is comparable to previous studies (Looy, et al., 2006), and is consistent with publications being preconditions for patenting despite reflecting two different dimensions: basic and applied research (Fabrizio & Di Minin, 2008; Shelton & Leydesdorff, 2012; Wong & Singh, 2009). Thus, there are only two important variables; patents and publications, representing the technology and overall research capability of HEIs. The selection of all publications rather than on-park or U-I on-park publications is because the number of publications is widely used in assessments and facilitates further benchmarking.

***Impact of HEI associations with SPs***. To find whether a formal relationship with a SP associates with any significant differences in U-I collaboration and R&D production in general, a Kruskal-Wallis test was used to assess whether patents and publications are related to the supportive mechanisms that HEIs provide (through the *HEI* itself, in collaboration with a *partner*, *both*, or *none*) to access to either a *Science park*, *On-campus incubator*, or *Incubator in the locality*. Table 5 reports that there are no significant differences.

**Table 6.4** Factor analysis of the five dependent (key) variables for 78 HEIs and RIs.

|  | Component | |
|---|---|---|
|  | 1 | 2 |
| Patents | 0.058 | 0.998 |
| Publications | 0.936 | 0.094 |
| On-park publication | 0.968 | 0.038 |
| I-U on-park publications | 0.969 | 0.052 |
| I-U off-park publications | 0.951 | 0.065 |

Extraction Method: Principal Component Analysis.
Rotation Method: Varimax with Kaiser Normalization.

**Table 6.5** Kruskal-Wallis test for the influence on patents and publications of aspects of the formal relationship between 78 HEIs and SPs.

| Grouping Variable* |  | Patents | Publications |
|---|---|---|---|
| Science park | $H(3)$ | 3.613 | 2.122 |
| accommodation | d.f. | 3 | 3 |
|  | Asymp. Sig. | 0.306 | 0.547 |
| On-campus | $H(3)$ | 6.554 | 7.509 |
| incubators | d.f. | 3 | 3 |
|  | Asymp. Sig. | 0.088 | 0.053 |
| Other incubators in | $H(3)$ | 0.460 | 0.673 |
| the locality | d.f. | 3 | 3 |
|  | Asymp. Sig. | 0.928 | 0.880 |

The significance level is .05.
*Groups: Hosting, Partnering, Both, and None.

***Impact of other factors.*** Technology transfer takes many forms, from informal links to formal research projects and research collaborations and may be fostered by a battery of policies to create right conditions to commercialise academic research. SPs are only one of the policies that can be implemented (Storey & Tether, 1998). Thus, the overall inability of the SP movement to nurture stronger U-I links here suggests that there may be other factors that might have a greater effect on promoting technology transfer and I-U collaboration. Recently, the *higher education – business and community interaction survey* (HEFCE, 2009) has systematically been gathering data regarding HEIs' third mission activities, covering most UK research institutions and facilitating further comparisons and analysis. This data source was used to obtain a set of eight qualitative and quantitative variables commonly used to study the intensity of U-I interactions. The eight qualitative variables are binary (Yes/No) and are divided in two groups: Strategy and Infrastructure. The first group focuses on the areas where HEIs are making the greatest contribution to economic development and the second group focuses on whether HEIs (on their own or in collaboration with an external partner) offer various facilities. The eight quantitative variables are in Table 6.7.

**Table 6.6** Relationships between publications and patents and third stream strategy and support (Mann-Witney tests) for 78 HEIs and RIs (U=Mann-Whitney value; Z=standard deviation value; p=significant value; r=effect size value).

| Strategy | Depen. variable | U | p= | z | r | n (no vs. yes) | Mean Rank |
|---|---|---|---|---|---|---|---|
| Research collaboration with industry | Patents | 365.5 | 0.00 | -3.568 | -0.40 | 78 (29 vs. 49) | 27.6 vs. 46.54 |
| | Publications | 375 | 0.001 | -3.469 | -0.39 | 78 (29 vs. 49) | 27.93 vs. 46.35 |
| Technology transfer | Patents | 586.5 | 0.10 | -1.628 | -0.18 | 78 (34 vs. 44) | 34.75 vs. 43.17 |
| | Publications | 531 | 0.03 | -2.187 | -0.25 | 78 (34 vs. 44) | 33.12 vs. 44.43 |
| Supporting SMEs | Patents | 428 | 0.27 | -1.096 | -0.12 | 78 (61 vs. 17) | 40.98 vs. 34.18 |
| | Publications | 390 | 0.12 | -1.555 | -0.18 | 78 (61 vs. 17) | 41.61 vs. 31.94 |
| Spin-off activity | Patents | 113 | 0.16 | -1.418 | -0.16 | 78 (73 vs. 5) | 38.55 vs. 53.4 |
| | Publications | 119 | 0.20 | -1.295 | -0.15 | 78 (73 vs. 5) | 38.63 vs. 52.2 |
| **Infrastructure/IP** | | | | | | | |
| Business advice | Patents | 114.5 | 0.17 | -1.388 | -0.16 | 78 (5 vs. 73) | 25.9 vs. 40.43 |
| | Publications | 92 | 0.07 | -1.846 | -0.21 | 78 (5 vs. 73) | 21.4 vs. 40.74 |
| Entrepreneurship training | Patents | 169 | 0.78 | -0.275 | -0.03 | 78 (5 vs. 75) | 42.2 vs. 39.32 |
| | Publications | 144 | 0.43 | -0.785 | -0.09 | 78 (5 vs. 75) | 31.8 vs. 40.03 |
| Seed corn investment | Patents | 501 | 0.08 | -1.73 | -0.20 | 78 (25 vs. 53) | 33.04 vs. 42.55 |
| | Publications | 578 | 0.37 | -0.905 | -0.10 | 78 (25 vs. 53) | 36.12 vs. 41.09 |
| Venture capital | Patents | 380 | 0.09 | -1.677 | -0.19 | 78 (17 vs. 61) | 37.23 vs. 47.65 |
| | Publications | 348 | 0.04 | -2.064 | -0.23 | 78 (17 vs. 61) | 36.7 vs. 49.53 |

Regarding the qualitative variables, Table 6.6 illustrates that only two variables related to HEI strategy have a significant relationship with R&D production. First, HEIs that consider research collaboration with industry as their main contribution to economic development have more patents ($U$=365.5, $p$=.000, $r$=-.40, $\overline{x}$ rank=27.6 $vs.$ 46.54) and publications ($U$=375, $p$=.001, $r$=-.39, $\overline{x}$ rank=27.93 $vs.$ 46.35). Second, HEIs which actively support technology transfer have significantly more publications ($U$=531, $p$=.029, $r$=-.25, $\overline{x}$ rank=33.12 $vs.$ 44.43). HEIs that claim to support SMEs extensively produce less R&D, while the few HEIs mainly involved in spin-off activities may have more R&D output although this difference is not statistically significant. On the other hand, the comparison of the four infrastructure variables reveals that HEIs providing these services have comparable R&D production with those not providing them. Only HEIs with a supply of venture capital are able to significantly raise the level of publications ($U$=348, $p$=.004, $r$=-0.23, $\overline{x}\ rank$=36.7 vs. 49.53) but not of patents.

**Table 6.7** Relationships between publications and patents and entrepreneurial activities for HEIs and RIs (N=78).

| Independent variable | Dependent var. | Corr. Coeff. | Sig. (2-tailed) |
|---|---|---|---|
| Collaborative research [a] | Patents | 0.337** | 0.003 |
| | Publications | 0.493** | 0.000 |
| Contract research [a] | Patents | 0.451** | 0.000 |
| | Publications | 0.709** | 0.000 |
| Consultancy [a] | Patents | 0.180 | 0.115 |
| | Publications | 0.157 | 0.169 |
| Facility and equipment service [a] | Patents | 0.207 | 0.069 |
| | Publications | 0.106 | 0.357 |
| Intellectual property (Patents, software and non-software licenses, spin-offs) [a] | Patents | 0.342** | 0.002 |
| | Publications | 0.283* | 0.012 |
| Spin-off activities with some HEI ownership | Patents | 0.515** | 0.000 |
| | Publications | 0.529** | 0.000 |
| Spin-off activities not HEI owned | Patents | 0.254* | 0.025 |
| | Publications | 0.286* | 0.011 |
| Active patents | Patents | 0.491** | 0.000 |
| | Publications | 0.508** | 0.000 |

(a) Income
*. Correlation is significant at the 0.05 level (2-tailed).
**. Correlation is significant at the 0.01 level (2-tailed).

Regarding the quantitative variables (see Table 6.7), there is a strong correlation between HEI R&D production and the income generated from contract research, the number of active spin-offs with some HEI ownership, and the number of active patents. Commercial activity based on collaborative research only statistically significantly associates with publications. There are no relationships between R&D production and the other HEI main income-generating activities, such as consultancy, facility and equipment service, IP, and the number of active spin-offs, which are not HEI owned. The overall results of the qualitative and quantitative variables suggest that the income from research contracts significantly relates to R&D capacity. Nevertheless, with the exception of the spin-off activity with some HEI ownership, which is highly dependent upon leading academic research (Wright, et al., 2008) and the supply of venture capital, the academic focus on entrepreneurial activities oriented to supporting firms is not directly linked to HEI on-park U-I collaboration and R&D production.

## 6.3 Discussion and Implications

Given the high expectations for a more active involvement of the private sector in research processes and knowledge exchange with academia, there is a need to uncover the role that different public policies, such as SPs, are playing in supporting industry to absorb and exploit the academic R&D (Storey & Tether, 1998), as well as in promoting entrepreneurship among Universities.

### 6.3.1. Time to first publication

Overall, the Kaplan Meier analysis shows that the main on-park research producers are the research-intensive firms, having an average time to publish and collaborate of three and five years, respectively. In terms of collaboration, the high rate of *on/off-park U-I collaboration* indicates that a process of collaboration between high-tech firms and universities, as the main off-park partners, is the most common U-I interaction. On the other hand, scarcity of on-park RIs limited the presence of on-park knowledge producers, leading to a late and limited output: scarce *on-park U-I collaboration*. This also reflects that RIs are industry-related and tend to focus on providing technical support and advice to local industry, leading to strong links with medium and low-level innovative firms (Soetanto & Jack, 2011). Nevertheless, the outcome of market- and product-oriented interactions is not necessarily codified into publications (Hayati & Ebrahimy, 2009; Noyons, et al., 1999).

### 6.3.2. Typology

The multivariate analysis confirms that the *Research Park & Campus* and *Science Park* groups have 78% and 75% increased likelihoods to publish compared to the slowest group, *Science & Innovation Centres*. Generally, both groups are expected to have strong ties with HEIs. However, they differ in the sense that research parks and campuses mostly host tenants that are heavily engaged in basic and applied research (A. N. Link & Scott, 2003; Saublens, 2007), and important RIs (Minguillo & Thelwall, 2013). The group with more probabilities to encourage prompt collaborations is again the *Research Park & Campus* group (92%), followed by the *Science Park* group (67%). This suggests that research parks and campuses may be the most adequate support infrastructures to develop cutting-edge industries, being most likely to provide the conditions to promptly support R&D activities and cross-fertilisation. This can be partially caused by the close links with research producers, as research parks and campuses concentrate the most public research, hosting the majority of RIs (Minguillo & Thelwall, 2013). Science parks are also successful in encouraging prompt U-I partnerships and two-way knowledge flows. In addition, if the probability to go from none patents or publications to one patent or publication is much lower than the probability to go from one to more (Squicciarini, 2009), there may be a relationship between prompt R&D activities and knowledge accumulation as both groups of parks are also the most research productive. It means that earlier R&D activities within SPs may make further publications more likely to occur as result of the knowledge accumulation of the residents and of the environment. In other words, prompt research production signals the research-orientation of the park (tenants) and it may be related to higher levels of research because the parks (tenants) are innovative and have an early start in the accumulation of knowledge.

Research campuses and incubators had a publication and collaboration starting time of only one year on average. The strong research bases within research campuses and the commitment to support academic spin-offs and new ventures by incubators (Barrow, 2001) could be the main reasons for this tendency.

In contrast, the weak research performance of the *Technology Park* group and the *Science & Innovation Centre* group is because the majority of tenants of the first group engage in applied research and development (A. N. Link & Scott, 2003) while the tenants of the second group do not tend to have operational links with HEIs (Saublens, 2007), being the least research-oriented group. Thus, an important aspect to consider here is how industrial involvement may cause detrimental effects on publication (Nelson, 2004). Yet, this result should be interpreted with caution as the differences between these intermediary infrastructures are not formally defined and are complex to identify due to the heterogeneous partners involved in the creation and operational procedures (A. N. Link & Scott, 2006, 2007; Löfsten & Lindelöf, 2002). They have different admission criteria and target markets, hosting a wide range of firms (Ferguson & Olofsson, 2004; Michael Schwartz, 2009).

### 6.3.3. Age

The systematic time reduction in outputs (from 7 to 2 years for firms and 8 to 1 for RIs) and in collaborations (from 9 to 2 years) between the first and last wave of parks is confirmed by the regression that identifies the park's age as a predictor. This indicates that the pioneer infrastructures are not doing as well as those established each year, because the newer parks have a 48% probability of building collaborations faster. Similarly, SP age has a negative effect on the performance of the tenants (Squicciarini, 2009), and tenants based on newer infrastructures, like incubators, make full use of the service portfolio (Bruneel et al., 2012). This suggests that the R&D activity is becoming more important for the SP movement and could depend on the increasing global competition for firms and the efforts of new innovation infrastructures to actively foster modern learning principles and open innovation processes among their tenants (Barrow, 2001; Hansson, 2007; Vedovello, 1997). Another important reason is the increasing interest among academics for establishing U-I links (PACEC, 2009).

### 6.3.4.    Spontaneous vs. policy-driven parks

There was a statistically significant difference between spontaneous and policy-driven parks in the time needed to become research active. The first are helped by the underlying dynamism in the local area, producing research from the first year, whereas the latter need on average 5 years to publish and 6 years to collaborate. Perhaps the networking effect of SPs is limited in less developed innovation systems (Bakouros et al., 2002; Vedovello, 1997), and the effort and time required by policy-driven parks to create adequate conditions only pays off in the long term, hence being an interesting solution for developing countries (Hung, 2012).

### 6.3.5.    Impact of HEI associations with SPs

The comparison of HEI access to SPs, on-campus incubators, or any incubator in the locality through their own *HEI*, in collaboration with a *partner*, a combination of *both*, or *none* showed that only HEIs hosting on-campus incubators have a significant association with publications. However, a follow-up pairwise comparison of the four access mechanisms (*HEI*, *partner*, *both*, *none*) revealed no significant differences between HEIs with formal relationships with on-campus incubators and those with none. Thus, HEI formal ties with different support infrastructures do not associate with higher levels of patents or publications. This finding contradicts the opinion of university administrators who consider that this relationship leads to slightly increased research outputs and little measurable impact on patents (A. N. Link & Scott, 2003). However, it is in line with studies that report that there is no significant difference between the various types of infrastructures in fostering U-I links (Fukugawa, 2006; Michael Schwartz & Hornych, 2010a). Other research has also found no strong evidence to support the idea that UK SPs foster stronger technology transfer links within or outside SPs, nor better performance for on-park firms (Quintas et al., 1992; Westhead & Storey, 1995; Westhead, 1997). The explanation for the very limited impact of SPs on U-I interactions could also be that tenants are attracted by the image and prestige of the site rather than pursuing stronger collaborative links with associated HEIs (Felsenstein, 1994; Vedovello, 1997). Despite the low rate of U-I interaction (HEFCE, 2010a), this national-level study is likely to overlook some successful examples of on-park U-I interaction, especially when these formal links are tightly related to the degree of development of the innovation systems where they operate (Bakouros et al., 2002), and there is uneven development of regional innovation systems across the country. Moreover, there are other important aspects, such as limited impact of the on-park research on the research community and low U-I collaboration rates that question the idea of SPs as the catalysts behind a knowledge-based development across regions (Minguillo & Thelwall, 2013).

### 6.3.6.    Other factors

SPs' failure to promote higher rates of collaboration between HEIs and on-park or off-park firms suggests that higher levels of technology transfer and U-I collaboration may be positively influenced by other strategic and infrastructure factors, or due to some of the main sources of academic revenues. This study shows that there are significant differences between research institutions that encourage research collaboration with industry, increasing the number of publications and patents, and encourage technology transfer, increasing the number of publications. HEIs that provide venture capital also tend to produce more publications. There is also a positive relationship between patents and publications and the income from contract research, the number of active patents, and the number of active spin-offs with some HEI ownership. Similarly, the income from collaborative research significantly correlates with publications, as pointed out by Gulbrandsen and Smeby (2005). Based on the framework used by Wright and his colleagues (2008), to examine the role of the U-I links involving R&D transfer, technology transfer in terms of joint R&D, contract research, academic spinning off and patents belong to the invention stage. These activities are frequently consequences of intense and long-term knowledge creation and exchange, and involve explicit and formal research transfer. This might partly explain the statistical association found between co-authored publications and levels of R&D outputs. The results of this study also suggest that the design of strategies and mechanisms to promote the growth and development of spin-offs and SMEs, as well as the capitalisation of research through consultancy, facility and equipment services, IP, and start-ups (spin-offs not directly related to the research community), are not significantly associated with R&D production. These activities are considered, by the same framework (Wright et al., 2008), to be not highly dependent on the accumulation of high quality knowledge and expertise, and are not expected to generate any formal R&D transfer activities due to the limited transfer of tacit knowledge involved.

There is a potential conflict between the level of income generated from joint research projects and research contracts with the academic research excellence of the university departments (D'Este, et al., 2013). These are two of the main activities where firms engage closely with HEIs (Soetanto & Jack, 2011). They are also the main public knowledge source (Cohen, et al., 2002) and private income source of R&D transfer activities, being typically undertaken for large businesses that can afford to access high quality research and relevant expertise (HEFCE, 2010a; Vedovello, 1997).

Technical facilities and laboratories, as one of the main reasons to locate on a SP, (Quintas et al., 1992; Soetanto & Jack, 2011; Vedovello, 1997) and licensing are less important and used among firms (Arvanitis, et al., 2008), and are therefore less unlikely to generate or be associated with a critical mass of research. Similarly, no correlation has been found between IP production and R&D expenditure for either on- or off-park firms (Westhead, 1997), suggesting that the revenues from IP activities are not directly related to research strength and are therefore not related with HEIs R&D output either. The embryonic stage of the overwhelming majority of protected inventions is the main reason for the low income and private interest because a very low proportion of technologies licensed are ready for practical or commercial use (Thursby, Jensen, & Thursby, 2001). This limitation is further illustrated by the lower returns in terms of consultancy, facilities and equipment, and IP, even though these interactions represent frequent modes of cooperation (Malo, 2009), in comparison with other research-based interactions which give high revenues for HEIs (HEFCE, 2010a).

Venture capital, despite being concentrated in a limited portfolio of ventures, is expected to give early financing for innovative ideas and high growth potential firms (Sunley, et. al., 2005; Wei & Wang, 2011), where knowledge is the main driving force in the formation process. Thus, in contrast to seed corn investment, which has a social perspective and is provided by public bodies to highly risky ventures in the early stages as a way to fill the gap left by private investments, venture capital is injected by private investors in growing firms with reduced risks and located in attractive sectors and at later stages.

The establishment of academic spin-off firms also have an impact on R&D production, in contrast to those which are not owned by HEIs, as the first group represents the process of taking research out of the laboratory and onto the SP, being based on the accumulation of technological knowledge (Clarysse, et al., 2011). This result is also supported by the fact that this activity is one of the two principal on-park U-I interactions found along with facilitating technology and research transfer (Lindelöf & Löfsten, 2004; Quintas et al., 1992). This academic firm activity also benefits patent production, impacting on research intensive industries, as the overall presence of U-I links respond positively to the propensity to patent (George et al., 2002; A. N. Link & Scott, 2003). The comparison between HEI R&D output and entrepreneurial activities uncovered a pattern that suggests that the greater tacit knowledge is required and the higher the level of innovation involved in the U-I interactions, the greater mass of R&D is likely to be disseminated and codified in the form of publications or patents.

This chapter has several limitations. It only focuses on one formal U-I interaction and ignores other factors, such as the provision of skilled graduates, the stimulation of networks, the formation of university spin-offs, and the development of new methodologies and instrumentation. Another important limitation is that the data used does not cover all the research generated within the SP movement because not all on-park organisations mention the name of the SPs where they are based as part of their affiliation address, and much research might not be published in Scopus journals. Company size is also an important factor in research production and collaboration capacity and needs to be taken into account in future studies. Nevertheless, although R&D production, as with any other indicator, is not able to provide a complete picture of the success of SPs and the U-I relationship, it can provide an insight to enhance the understanding of the relationships between HEIs and SPs. This benchmarking evidence can help HEIs to establish realistic targets according to the nature of their relationships with SPs, predict when cross-fertilisation could start to give results, and encourage third stream activities which benefit from active U-I interactions.

## 6.4 Conclusions

In response to the first research question, SPs on average promote tangible research outputs in the form of Scopus publications from about three years after their inception. The average time to create research for tenants is four years, while the fastest and most common type of cross-fertilisation is between on-park or off-park organisations with an average time of five years to start. Thus, the research carried out by firms and as a result of *on/off-park U-I collaboration* is the fastest and most common. Regarding the factors that benefit the R&D activities and U-I collaboration, the multivariate analysis finds that among five predictors, only the infrastructure type affects R&D activities, while the type and age of the infrastructures affect U-I collaboration. The most research active parks are research parks and campuses, followed by science parks. The first group of parks, along with incubators, are more likely to need less time to become research active and promote U-I cross-fertilisation processes in comparison with other types of intermediary infrastructures. The age of a park also significantly associates with the faster establishment of U-I partnerships, with the newcomers having a higher probability to promote a more effective processes of open innovation among their tenants.

In response to the second research question, there is no difference in R&D production or U-I collaboration between HEIs offering formal access to parks on their own, in partnership with an external organisation, in a combination of both mechanisms, or having no formal access. There is also no significant difference in R&D transfer among on-park firms compared with off-park firms (Quintas et al., 1992; Michael Schwartz & Hornych, 2010a), questioning the parks' positive impact on the academic and industrial sphere. In conclusion, there is no evidence that SPs are the right tools to strengthen U-I interactions and entrepreneurial activities for HEIs, or the performance of on-park firms.

The analysis of the relationships between R&D activities and the qualitative and quantitative variables that measure the degree of engagement in U-I collaboration and commercial activities revealed that there is a positive relationship between publications and patents with HEIs that actively support research collaboration and also between publications and institutional support for technology transfer and venture capital. Other determinants that also reflect activities depend on leading academic research, such as the income from U-I collaborative research, are also significantly associated with publication output. Similarly, variables that reflect the introduction of innovative goods and solutions into the market, such as the financial income generated from contract research and the number of active spin-offs with some HEI ownership and active patents, associate with publications and patents produced. Conversely, an academic focus on entrepreneurial activities oriented to support firms has no positive association with on-park U-I collaboration or R&D production, as illustrated, for example, by the supply of business advice, entrepreneurship training, seed corn investment, or institutional strategies to support SMEs and spin-off activities, as well as the revenues from consultancy, facility and equipment services, IP, and the number of active start-ups. Thus, activities that are more likely to be involved in the invention stage and be the result of close and long-term knowledge transfer significantly associate with HEI R&D capacity.

From a policy viewpoint, the main recommendation is that the different types of SP need to be re-assessed because establishing a support infrastructure in partnership with HEIs does not necessarily increase technology transfer. Policies should also encourage academics to participate in the commercial activities that are most associated with increased R&D production, as the main evaluation proxies for academics, because it could help to overcome the barriers that inhibit U-I interactions. The promotion of commercial activities with a high research basis is an interesting way to build platforms where both communities can actively engage and academics can increasingly gain a more entrepreneurial identity. Further research is certainly needed to investigate the structural organisation and role played by HEIs and RIs in the networks formed by collaborations with on-park firms. It is also important to determine if the research accumulation and prestige of research institutions affect collaborations with on-park firms and other firms. A similar approach applied on an international comparative basis could also provide evidence of the impact of SPs on HEIs among developing and developed countries. In addition, for a more accurate study of the U-SP relationship, SPs that have formal relationships with specific HEIs should be identified to determine the degree of entrepreneurial performance of their hosting or partnering academic institution (or even faculties or schools) after the establishment of the SPs. This could help to confirm if the impact of SPs on universities increases over time (A. N. Link & Scott, 2003).

In summary, this chapter reveals that research parks & campuses and SPs are the infrastructures that are most likely to promote prompt R&D activities and U-I collaboration for their residents and newer parks seem to be the most successful at encouraging U-I interactions. HEIs' formal associations with SPs have no significant impact on the volume of patents or academic publications produced, nor on knowledge transfer in comparison to HEIs without formal associations with SPs. Moreover, indicators of third stream activities resulting from a strong research base, such as income from U-I collaborations and contract research and the number of active spin-offs and patents, strongly associate with the academic production of publications and patents.

# Chapter 7: The intermediary role of SPs on the web: A pilot study

## 7.1 Introduction and goals

SPs primarily promote informal links among the actors embedded in regional innovation systems, and informal information exchange is the most common channel of interaction between academia and industry (Bakouros et al., 2002; Quintas et al., 1992; Michael Schwartz & Hornych, 2010b; Vedovello, 1997). However, the complexity of these informal linkages, which often do not leave a scholarly publication or other public trace that could be investigated, makes it difficult to identify and quantitatively assess this informal dimension (Section 2.7). In theory, however, a webometric approach might be able to help the study of such interactions. It is thus important to compare how a web-based study, which could track down any interactions that leave a trace on the web, could complement and extend the previous studies based on the S&T dimension, and academic publications in particular (Chapter 4-6). This chapter, therefore, introduces a method based on link analysis (Section 2.5) to investigate the relational structure of the R&D support infrastructure associated with science parks, in order to determine whether a webometric approach can give plausible results. Three science parks from Yorkshire and the Humber in the UK were analysed with webometric and social network analysis techniques (Section 3.3.6). Interlinking networks (Section 3.3.8) were generated through the combination of two different data sets (in- and outlinks) extracted from three sources (Yahoo!, Bing, SocSciBot) (Section 3.3.7). This is the first study that applies a web-based approach to investigate to what extent the science parks facilitate a closer interaction between the heterogeneous organisations that converge in R&D networks.

This exploratory study has two main objectives: (1) to develop a method for collecting and analysing web-based networks for SPs to determine whether hyperlinks can reflect relevant offline interactions; and (2), to determine the validity and viability of the webometric approach for this purpose. These objectives can be summarised by the following research questions:

- Can webometric methods identify interactions between institutional sectors as well as interactions between various types of organisations associated with SPs?
- Are the main features of SP interlinking networks in line with findings of official reports and surveys which are essential sources of information to evaluate the R&D infrastructure in the UK?

In addressing these questions, the viability of link analysis to identify SPs' intermediary role is investigated. As key organizational innovation structures, SPs are used to analyse part of the web-based organization of the R&D support infrastructures developed in a given region but this study does not investigate specific R&D aspects of the networks associated with SPs, such as knowledge transfer (Siegel, et al., 2003b; Westhead & Storey, 1995), firm performance and formation (Löfsten & Lindelöf, 2002; Quintas et al., 1992) or best management practices (Autio & Klofsten, 1998), which are normally studied through other proxies. This study also assesses whether the web-based and offline characteristics reflect similar patterns in order to investigate whether the general picture provided by the web may shed new light on the innovation networks centring on science parks.

### 7.2 Methods

To obtain data on the SP link networks of the three SPs from Yorkshire and the Humber, the websites of the Advanced Manufacturing Park (www.attheamp.com), Leeds Innovation Centre (www.leedsinnovationcentre.com), and York Science Park (www.yorksciencepark.co.uk) were crawled with SocSciBot. It collected 215 site outlinks that were manually checked to identify the type of relationship with the three SPs and to eliminate spam. Some of the irrelevant websites identified were: maps.google.co.uk; twitter.com; nationalexpress.com; adobe.com; youtube.com; news.bbc.co.uk. It reduced the number of relevant websites to 183. Then, it was used three different tools to collect the in-links (search engine Yahoo!) and out-links (web crawler SocSciBot and search engine Bing) of these websites. The focus on studying the inter-connections between these websites made it possible to combine the in-links and out-links collected to create a more reliable and inter-connected bi-directional network. Finally, these networks were dichotomised, represented and analysed with the help of Social Network Analysis (SNA) and visualisation techniques. Different cohesion and centrality measures were calculated using UCINET and represented with NetDraw (for a fully explanation of the methods, see sections 3.3.6 - 3.3.8).

### 7.3 Results:  Data analysis

#### 7.3.1.    Structural analysis: comparing the IN & OUT data sets

The variety in the data set and the novelty of the approach applied to analyse the interactions calls for a comparison between the characteristics of the networks formed by the in- and out-data sets to understand the qualities of each one. Hence, the networks were analysed with cohesion measures to estimate the degree of integration of the websites, the size of the network, the cohesion of the websites and the level of mutual interaction between them.

*Inclusiveness* describes the number of websites that are integrated into the network and is the total number of nodes minus the isolated nodes (Scott, 2000). *Connectivity gap* is the total number of in- or outlinks established in the in- and out-data sets minus the number of in- or outlinks in each data set and gives the proportion of links which the data set needs to reach the maximum number of links obtained from the data sets. *Density* refers to the proportion of all possible connections that are actually present, and *reciprocity* indicates the proportion of relations (links) that are reciprocal (S Wasserman & Faust, 1998).

**Table 7.1** Structural cohesion measures for the in- and out-data sets for three SPs.

| 1) Advanced Manufacturing Park (AMP) | | | | | | |
|---|---|---|---|---|---|---|
| | **Inclusiveness** | **(%)** | **Ties** | **Connectivity Gap** | **Density** | **Reciprocity** |
| **in** | 26 | (0.79) | 63 | 0.46 | 0.06 | 0.32 |
| **out** | 33 | (1) | 92 | 0.21 | 0.09 | 0.26 |
| **both** | 33 | (1) | 117 | | 0.11 | 0.36 |
| 2) Leeds Innovation Centre (LIC) | | | | | | |
| | **Inclusiveness** | **(%)** | **Ties** | **Connectivity Gap** | **Density** | **Reciprocity** |
| **in** | 45 | (0.92) | 97 | 0.21 | 0.04 | 0.29 |
| **out** | 49 | (1) | 106 | 0.14 | 0.05 | 0.28 |
| **both** | 49 | (1) | 123 | | 0.05 | 0.36 |
| 3) Yorkshire Science Park (YSP) | | | | | | |
| | **Inclusiveness** | **(%)** | **Ties** | **Connectivity Gap** | **Density** | **Reciprocity** |
| **in** | 103 | (0.99) | 307 | 0.19 | 0.03 | 0.33 |
| **out** | 104 | (1) | 312 | 0.17 | 0.03 | 0.37 |
| **both** | 104 | (1) | 378 | | 0.04 | 0.42 |

In the case of the *Advanced Manufacturing Park* (*AMP*), the comparison between the networks based on the in- and out-data sets shows that the outlink network connects all the websites through more links and a better distribution of them. The in-network only integrates 80% of the websites and establishes 54% of the potential ties, forming a small and sparse network (see Table 7.1). The cohesion measures based on the *Leeds Innovation Centre* (*LIC*) suggest that the in- and out-network are similar, but it seems that the inter-inlinks can only build a smaller and more centralised network as a result of linking slightly less websites with a lower number of ties and similar reciprocity. On the other hand, the outlinks integrate all the websites and form a slightly more dense network. In the case of *York Science Park* (*YSP*), again the differences between the in- and out-network are very small. The proportion of the connectivity gap, density and integration rate of both parks is almost the same, with the only difference being the higher degree of reciprocity and cohesion of the out data set.

From the initial overview of the three SPs, both data sets (inlinks and outlinks) provide similar structures and the patterns shown by the first SP differ from those of the second and third SP. This could be caused by the lower number of websites as well as the industrial profile of the AMP. Despite the in-data set being much larger than the out-data set, the latter might be the most effective because it seems to include more links among community members, hence forming the most complete community structure. It is also important to note that the inclusiveness of the out-data set has to be interpreted with caution because the websites of the sample were collected from the external outlinks of the SPs' websites, and thus all the websites should be linked. The high connectivity gap (0.23), the high rate of reciprocity when both data sets are combined, and the poor reliability of the webometric data suggest that the combination of both data sets could be necessary to ensure the most complete results.

Pearson correlations were calculated to measure the similarity between the in- and outlinks generated by the interlinking networks based on each data set (Table 7.2). The outlink counts obtained from the two data sets correlate significantly (Pearson's $r$= 0.84, 0.98, 0.87), showing similar trends among the websites independently of the data set. In contrast, the correlations for the inlinks are moderate or high but vary from SP to SP (Pearson's $r$= 0.58, 0.73, 0.94). The reason for the moderate and strong correlations in the in- and outlinks of the *AMP* is the structural differences between both data sets. The in-data set of this SP is characterised by the low density caused by the few linked websites. However, the structural differences do not affect the outlinks to the same extent when the behaviour of the linking websites are compared because the role of the connectors in this type of network is monopolised by certain types of websites, such as intermediaries, hybrid organisations and research institutions. Therefore, the high concentration of active websites facilitates the comparison between outlinks, and in contrast the wide dispersion of frequencies across the linked websites could make it difficult to find similarities when there are significant structural differences in both data sets.

**Table 7.2** Pearson correlations between the in- and outlinks of each interlinking SP network.

|  | Inlinks | Outlinks |
| --- | --- | --- |
| 1) Advanced Manufacturing Park (AMP) | 0.58 | 0.84 |
| 2) Leeds Innovation Centre (LIC) | 0.73 | 0.98 |
| 3) Yorkshire Science Park (YSP) | 0.94 | 0.87 |

The concentration of outlinks by a central group of websites that links a significant part of the network can be illustrated with the help of the Gini coefficient which measures the inequality of a distribution. It has previously been used to evaluate the web visibility of innovation systems (Katz & Cothey, 2006). A coefficient of 0 expresses total equality and a coefficient of 1 indicates maximal inequality. Hence, the inlinks should give a value closer to 0 whilst outlinks with a higher concentration should give a value closer to 1. Table 7.2 shows the inequality in the capacity to establish links in these networks and how the distribution followed by the in- and outlinks in both data sets is consistent. In the case of the *AMP*, the high coefficient of 0.74 for the inlinks in the in data set shows that only few websites are linked while the coefficient of 0.63 in the out-data set shows a slightly lower concentration in the distribution of the inlinks due to the higher number of interconnected websites. This helps to understand the structural differences of both data sets and the moderate correlation for the inlinks for this SP. However, it is still necessary to study the local properties of the interlinking networks formed by both data sets and study the multilateral linkages.

**Table 7.3** Gini coefficients of the in- and outlinks of each data set.

| | IN-data set | | OUT-data set | |
|---|---|---|---|---|
| | Inlinks | Outlinks | Inlinks | Outlinks |
| 1) Advanced Manufacturing Park (AMP) | 0.74 | 0.85 | 0.63 | 0.86 |
| 2) Leeds Innovation Centre (LIC) | 0.38 | 0.85 | 0.38 | 0.84 |
| 3) Yorkshire Science Park (YSP) | 0.48 | 0.86 | 0.48 | 0.85 |

After a detailed analysis of the interconnections among the organisations of both structures, similar patterns of interaction were found, suggesting that both web data sets provide consistent structures for all three SPs. A visual overview of the structures (based on both data sets) is in Figure 7.1. Despite the consistency of the results, there are some connectivity gaps in either one or the other data set caused by a low overlap, and which may lead to the underrepresentation of certain weak ties and influence the visibility of certain types of organisations. This may result in a loss of information and lead to misleading conclusions from a micro-level analysis. For example, the in-data set provides better interconnectivity with academia, and the out-data set to government, while the links of industry are almost equally represented in both data sets. Therefore, the reliability problems of hyperlinks as an indicator and the inherent shortcomings of the data collection tools make it difficult to rely on only one data set for a webometric analysis since it can lead to inaccurate results (Thelwall, 2008a), especially in the study of dynamic and complex R&D networks where heterogeneous organisations converge and interact (Tijssen, 1998). Consequently, the combination of both data sets, collected from different sources, is recommended to obtain more robust and reliable interlinking networks.

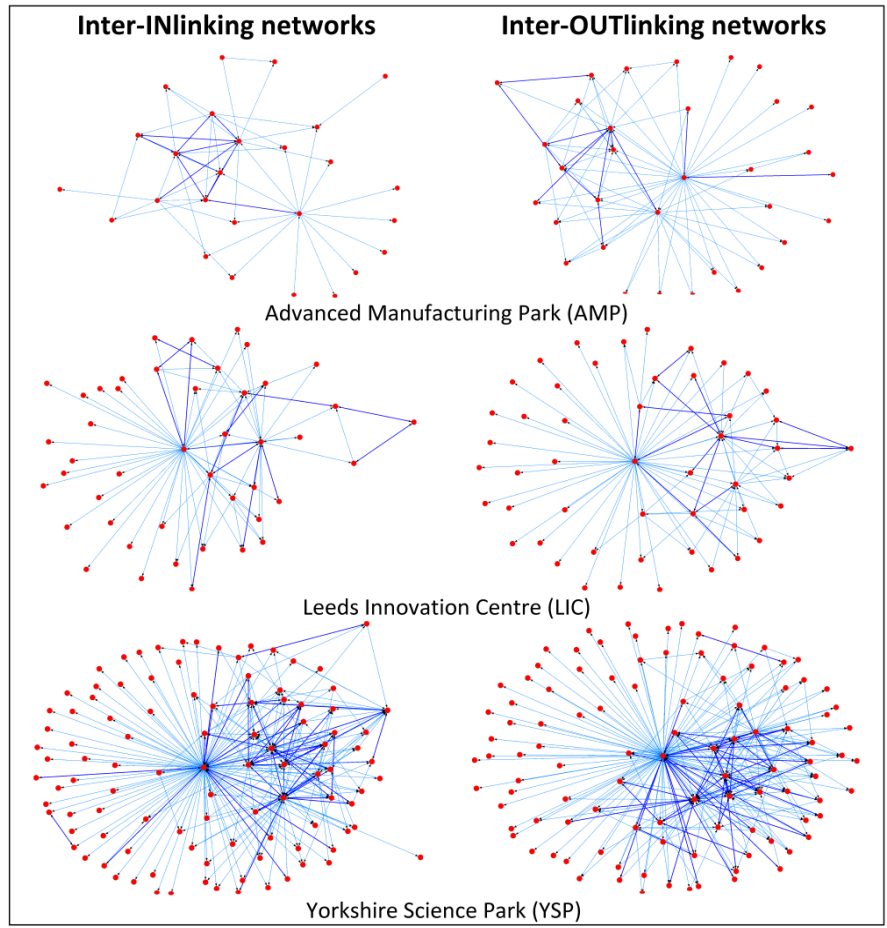

**Inter-INlinking networks**   **Inter-OUTlinking networks**

Advanced Manufacturing Park (AMP)

Leeds Innovation Centre (LIC)

Yorkshire Science Park (YSP)

**Fig. 7.1** Network diagrams of the in- and out- data sets for each SP analysed. The thicker lines indicate mutual linkages.

Since the evidence indicates that the joint use of both dimensions provides more robust and reliable structures, the next step is to achieve the two objectives of this exploratory study. The combined networks were therefore analysed in order to find out first, whether these web-based R&D infrastructures are able to reflect the linkages across the institutional sectors and the different types of organisations that interact within the SPs, and second, whether the web-based patterns coincide with those found by indicators that measure R&D activities.

### 7.3.2. Local analysis: combined interlinking networks

#### a. Interconnections between the institutional sectors and categories

The number of links between the sectors, and the average numbers of links associated with each organisation are reported in Table 7.4. Here the outlinks of the SP websites that were used as seed URLs are excluded to avoid redundant links. The academic sector, with the highest average out- and inlinks, is the most connected. The knowledge transfer function gives universities a central role, establishing links to public organisations that support the commercialization of academic knowledge and also to industry by means of firm-formation and consulting activities. The fact that academia links intensively but not in a reciprocal fashion is supported by previous findings (Garcia-Santiago & de Moya-Anegon, 2009). On the other hand, the high connectivity of government with itself and the low integration of industrial organisations also partly coincide and extend the findings of another webometric study (Stuart & Thelwall, 2006) where it was found that government is also well connected with itself, while only a little with universities and almost not connected with industry. Nevertheless, a classification according to three institutional sectors may be too broad because some central organisations are the result of partnerships between two or three sectors, being a combination of quasi-academic, -private or –public efforts.

**Table 7.4** Interconnections between the institutional sectors for all three SPs combined.

| Websites | Outlinks / Inlinks | Industry | Academia | Government | Total | Mean |
|----------|--------------------|----------|----------|-----------|-------|------|
| 122 | **Industry** | 26 | 13 | 43 | 82 | 0.67 |
| 12 | **Academia** | 26 | 20 | 45 | 91 | 7.58 |
| 52 | **Government** | 72 | 39 | 151 | 262 | 5.04 |
| 186 | Total | 124 | 72 | 239 | 435 | 2.34 |
| | Mean | 1.02 | 6.00 | 4.60 | 2.34 | |

In webometrics it is common that the number of links a website receives correlates with its number of webpages (Aguillo et al., 2006; Ortega & Aguillo, 2008). Therefore, larger websites should obtain more links in the network. To observe if website size has a direct impact on the interlinking networks, Spearman correlations were calculated for the in- and out-degrees and the number of webpages of the organisations. The values (in-degree/webpages=0.46 and out-degree/webpages=0.46) show that there is a significant moderate correlation and thus a tendency for larger web sites to be more central in the network. Therefore, the interconnectivity and position of the organisations within a R&D context may be partly influenced by their size on the web. This is a limitation of the webometric approach.

**Table 7.5** The five organisations with the highest centrality measures for the three SPs.

| 1) Advanced Manufacturing Park (AMP) | | | | | |
|---|---|---|---|---|---|
| **Organisations** | **InDeg** | **Organisations** | **OutDeg** | **Organisations** | **Betweenness** |
| yorkshire-forward.com | 10 | attheamp.com | 31 | attheamp.com | 164.5 |
| ec.europa.eu | 8 | amptechnologycentre.co.uk | 20 | yorkshire-forward.com | 147.6 |
| twi.co.uk | 8 | yorkshire-forward.com | 14 | amrc.co.uk | 124.1 |
| amrc.co.uk | 7 | amrc.co.uk | 9 | amptechnologycentre.co.uk | 59.7 |
| materialise.com | 7 | ec.europa.eu | 7 | ec.europa.eu | 37.6 |
| **2) Leeds Innovation Centre** | | | | | |
| leeds.ac.uk | 13 | leedsinnovationcentre.com | 47 | leeds.ac.uk | 660.5 |
| yorkshire-forward.com | 6 | leeds.ac.uk | 18 | leedsinnovationcentre.com | 629.4 |
| hm-treasury.gov.uk | 6 | connectyorkshire.org | 9 | connectyorkshire.org | 189.4 |
| connectyorkshire.org | 6 | yorkshire-forward.com | 8 | yorkshire-forward.com | 121.3 |
| europa.eu | 5 | europa.eu | 4 | ukcrn.org.uk | 48.2 |
| **3) Yorkshire Science Park** | | | | | |
| york.ac.uk | 24 | york.ac.uk | 37 | york.ac.uk | 1467.7 |
| businesslink.gov.uk | 20 | sciencecityyork.org.uk | 31 | businesslink.gov.uk | 897.0 |
| yorksciencepark.co.uk | 18 | businesslink.gov.uk | 19 | sciencecityyork.org.uk | 835.3 |
| york.gov.uk | 13 | york-england.com | 19 | yorkshire-forward.com | 326.8 |
| sciencecityyork.org.uk | 12 | yorkshire-forward.com | 18 | york-england.com | 249.4 |

### *Advanced Manufacturing Park (AMP):*

Located in Rotherham, the AMP is a joint venture between Yorkshire Forward and UK Coal. The park is designed to host manufacturing companies which specialise in precision manufacturing and advanced material technology processes (Advanced Manufacturing Park, 2010; Pullin, 2006). The AMP forms a network of 33 websites connected through 115 links, and is the smallest of the three SPs (see Figure 7.2). The in- and the out-data sets provide 59 and 94 links respectively, and a combination of both gives an increase of 56 (93%) links for the in-data set and 21 (22%) for the out-data set. In the core of the network there are eight central organisations, including the regional development agency (RDA) Yorkshire Forward, with the highest in-degree and the third highest out-degree (see Table 7.5). Other important organisations are the SP (atteamp.com) and the AMP Technology Centre, which fills the role of an incubator. These intermediaries have the highest out-degrees, building a bridge between businesses and the core of the network. The main organisation responsible for the creation and diffusion of knowledge is the Advanced Manufacturing Research Centre (AMRC), a research centre established in partnership between the University of Sheffield, who claim to be world-leaders in aerospace supply chains, and government offices, and the TWI Technology Centre (twi.co.uk). The AMRC and TWI are two leading research organisations and attract funding from contracts with public and private sector, and EU programmes (Hauser, 2010). The other three central support structure organisations are: Business Link, which delivers publicly funded business support products and services designed to help new businesses; the EU-Regional Policy (ec.europa); and the Manufacturing Advisory Service (mas-yh.co.uk), which delivers free and grant-funded advice as well as practical assistance to assist manufacturing businesses. These core organisations link important international giants and intensive R&D consultants. The central role of private and public sector funding in the development of the new ventures is illustrated by the collaboration between the business developer LIFE-IC, the RDA, and the incubator of the SP, with a spin-off from the University of Leeds, Inertius.

**Fig. 7.2** Inter-linking network of the Advanced Manufacturing Park (AMP). (*A colour version is available at: http://home.wlv.ac.uk/~in1493/11/fig-3.jpg*)

There are three features that should be noticed; first, the central position of the research centre AMRC is caused by its strong formal ties with the consultants TWI and Fripp Design, and the business supporters MAS and Business Link. This sub-network is a collaboration that brings research and technology to improve the competence of local industry. Moreover, the AMRC is also supported by the RDA and the EU Regional Development Fund (ERDF). Second, only the three consultants with most intensive R&D activities benefit from research contracts and alliances which bring together the knowledge, resources, expertise, and needs from three sectors. In contrast, the other consultants may be slightly isolated because they usually work in short term projects and commercialize a particular service, and thus do not require the same level of collaboration and resources (Bessant & Rush, 1995). Third, even though most of the businesses commercialise innovative products only eight firms are linked with government agencies, and four have direct links to technology producers and consultants.

**Leeds Innovation Centre (LIC):**

The LIC was established by the University of Leeds in 2000, and hosts a variety of companies, public organisations and university spin-offs and creates and attracts new healthcare and bioscience companies through its research, facilities and investments (Leeds Innovation Centre, 2010). LIC forms a network of 49 nodes connected through 123 links, and the initial 90 links provided by the in-data set are increased in 33 (37%) while the 106 links of the out-data set constitute an increase of 17(16%) (Figure 7.3). In the interlinking network three different subgroups can be identified: the peripheral subgroup on the left is basically formed by service-based firms and consultants, and by public organisations on the right, whilst the central subgroup is tied together by academia, a transfer office, business developers and technology-based firms and university spin-offs. Hence, the interplay between the sectors occurs in the middle of the network and has the University of Leeds as the key connector. The other seven organisations in the core are: the RDA, three government institutions, the SP, and the business supporter Investors in People. Cross-fertilisation and knowledge transfer could be represented in the dynamic sub-network formed by the public seed capital facilitator, Connect Yorkshire, with the university which ties the intellectual property and trademark company, Techtran (techtrangroup.com), as well as bio companies such as Photopharmica, and the university spin-offs, Instrumentel, LUTO Research, Chamelic and Tracsis. Finally, the two nodes formed by the Business Centre Association (bca.uk.com) and the University of Leeds Careers Centre (careerweb.leeds.ac.uk) are less integrated than expected, since they both offer a range of services intended to help young entrepreneurs.

**Fig. 7.3** Inter-linking network of the Leeds Innovation Centre (LIC). (A colour version is available at: http://home.wlv.ac.uk/~in1493/11/fig-4.jpg)

The R&D network is characterised by the leading position of the University and its ability to bridge the gap between the institutional spheres, connecting five spin-offs (instrumentel.com; tracsis.com; luto.co.uk; photopharmica.com; evidence.co.uk), business developers, support and public organisations, and health organisations. Being the first UK university to set up a dedicated technology transfer function (Lambert, 2003), the University's entrepreneurial identity is also materialized through strong and formal ties with two consultants. In 2002 it became the first university in the UK to outsource its technology commercialisation activities to Techtran (IP Group, 2011). This partnership helps the university to identify IP with high commercial potential and offers seed capital and strategic support services for new ventures. Consequently six of the nine knowledge-based firms identified in the park have spin-off from the University of Leeds.

### York Science Park (YSP):

YSP is a joint venture between the University of York and private investors and is situated on the campus of the University to stimulate technology transfer with the knowledge-based enterprises located in the SP (UKSPA, 2012b). Its tenants are related to bio- and health science and IT, having a similar profile to the LIC SP. YSP has the most heterogeneous and largest network with 104 websites and 378 links. The in- and the out-data sets provide 283 and 312 links respectively, and the combination of both produces an increase of 95 (34%) links for the in- and 66 (21%) for the out-data set. Thus, in order to facilitate the representation and analysis of this comprehensive network, the implicit external links of the SP to all the websites were eliminated and only the incoming links of the SP were taken into account. This reduces the number of websites to 75 (72%) and the number of links to 275 (73%). In the new network, three brokers link different sub-networks together (see Figure 7.4). The most important broker is the University of York that intermediates between a group of knowledge-based firms and spin-offs, and various consultancy offices, public and non-government organisations with the central support structure organisations and business developers. The second broker is the business developer Science City York (SCY) that builds a bridge between the firms with the University and the central R&D structure. The third broker is the SP which is connected by firms, and organisations related to the UK SP movement and support organisations. The number of links received by the SP is also significant because this differs from the patterns shown by previous SPs. The other eight organisations in the core of the network are: support structure organisations and local authorities (Business Link, the RDA, York-England and City of York Council) that attract new investments and provide economical resources, specialised advice and support for the network. These supporters allow the establishment of partnerships and the delivery of funds oriented to other central intermediaries and networking organisations such as Higher York (a partnership between local Higher Education Institutions to offer professional training), the local Chamber of Commerce, Business Link Yorkshire, and York Professionals. The interactions between these three sectors act as drivers for the whole network and are in turn linked to close partners. The central position of the intermediaries and support organisations reflects the importance of these hybrid nodes in the development of the whole network and its fundamental role in innovative contexts.

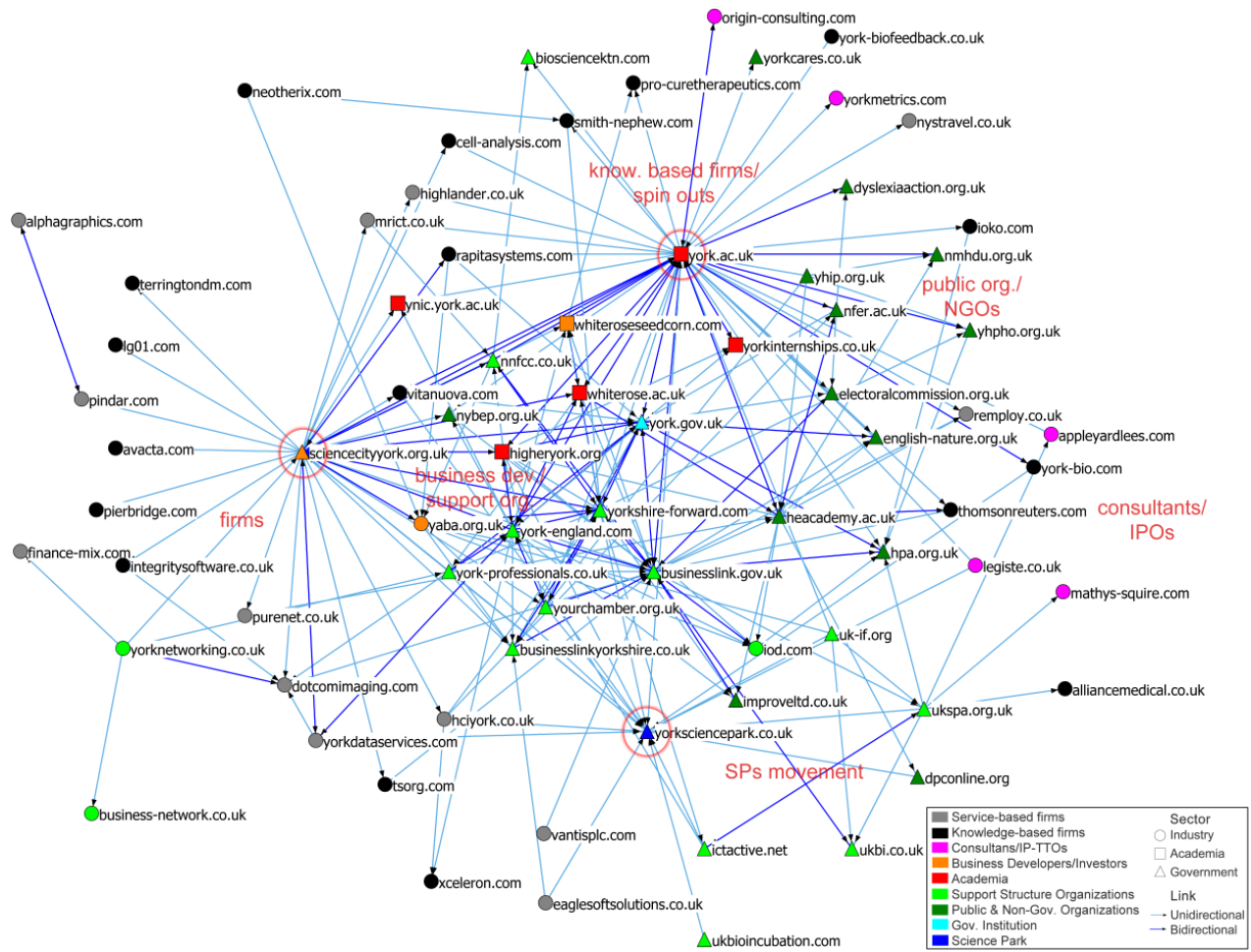

**Fig. 7.4** Inter-linking network of the York Science Park (YSP). (A colour version is available at: http://home.wlv.ac.uk/~in1493/11/fig-5.jpg)

The supportive infrastructure is again tied together by the active role of the University in partnership with the other institutional sectors. This collaboration facilitates the commercialisation of academic knowledge and technology through consultancy and firm-formation activities, which are in line with the University's emerging social and entrepreneurial mission. The collaboration between the University and support structure organisations, especially the RDA, has set up the seed capital investor White Rose Technology Seedcorn Fund (whiteroseseedcorn.com), and business developer SCY, and has also supported the investor Yorkshire Association of Business Angels (yaba.org.uk), which together promotes the creation and growth of business through public supported mentoring and facilitating investment funds. This has a significant impact on the local fast growing knowledge-based industry (Lambert, 2003). This R&D structure may be the main reason that nine university spin-offs are based on the SP. There are seven from the University of York (yorkmetrics.com; cell-analysis.com; origin-consulting.com; rapitasystems.com; xceleron.co.uk; pro-curetherapeutics.com; cybula.com) and two from Leeds (avacta.com; tissueregenix.com). The similar number of service-oriented and knowledge-based firms in the SP could be the result of a wide range of firms that are supported by the SCY, which supports both types of firms. Finally, the consultancy services offered by the University could be dived into internal activities, through the research contracts with independent bodies and NGOs, and external activities, through quasi-academic businesses.

**7.4 Discussion**

The representation of the links established between the various types of organisations provides a fast and broad overview of the R&D networks, identifying the interactions and key organisations which could be expected to be associated with SPs. This helps to answer the first research question regarding the capacity of webometric methods to uncover interactions between institutional sectors and between various types of organisations associated with SPs. The analysis of the local properties of the interlinking networks shows that the SPs tend to be organised according to the academic, industrial and governmental sectors. The degree of interaction between the groups reveals that some of the firms and some governmental organisations tend to occupy peripheral positions. These two groups are tied together by the third group which occupies the centre of the network. The central sub-network is led by academia and support structure organisations which support the rest of the network. In this overlapping area the groups of support, knowledge generation and knowledge exploitation interact to foster an innovative environment. The close interactions among research organisations, which form the knowledge generation subsystem, and how this group is linked by the companies, which constitute the knowledge exploitation subsystem, reflects the importance of the interdependence and interactive learning between the two subsystems in innovative contexts (Coenen, et al., 2004). The cross-sectoral interactions are also facilitated by hybrid organisations, such as business developers, which act as the main interface with the private sector and promote the exploitation of academic knowledge through the creation and support of knowledge-based industry. The central role of the universities and research centres may reflect the degree of engagement of academia in these networks associated to support infrastructures, and where SPs are expected to be used as platforms to foster collaboration and cross-fertilisation (Etzkowitz, 2008). On the other hand, the SPs also play an important role in the network, connecting businesses with the other central organisations. Thus, the analysis of the three SPs uncovered interesting patterns which can give insights into how the organisations in these innovative social circles engage on the web, and how these links may be proofs of offline relationships.

In order to answer the second goal of this study, it is necessary to find out whether the key features observed in the three web-based R&D networks correspond to the actual features of R&D infrastructure developed in the region of Yorkshire and the Humber. For this purpose, reports and surveys commissioned and produced by the Higher Education Funding Council for England (HEFCE) and the Department for Business, Innovation and Skills (BIS) were consulted and used for the comparative analysis. Despite the similar structures formed by *LIC* and *YSP* differing from the small network formed by the industrial *AMP*, there are some interesting shared properties, such as the prominent role of the research producers, the public support and commitment, the knowledge transfer fostered by the networking capabilities of business developers, and knowledge exploitation through contract research, consultancy services and creation of new ventures. In the *AMP* the research centre *AMRC* is the anchor tenant and forms the core of the network, showing how this flexible and integrated research facility works in collaboration with government, academia and industry to provide the network with the advanced technology to compete at an international level (Technology Strategy Board, 2010). In *LIC* the *University of Leeds* is the key connector and a local organizer, using the SP as the ideal quasi-academic platform to commercialize its academic research and technology through knowledge transference, firm formation and consultation activities (Lambert, 2003). Unlike the *LIC*, the *University of York* not only uses the SP to interact more easily with industry and government and nurture a dynamic environment. *YSP* is also used as a tool to embrace both an active academic entrepreneurial activity to increase the competitive advantage of the local industry as well as a wider social mission designed to collaborate and support the civil society, as indicated by the number of public organisations and NGOs based on the park. As Etzkowitz (2006) argues "[York] University has moved into a more central role both in the York region and in the larger society as a generator of new business firms... These developments have led to a new format of academic and business organisation, bringing together elements of each at a common site. ... It has successfully replaced declining industries and is now positioning itself to become the core of the city-region's economy".

Second, the efforts of the local and regional authorities to restructure the traditional industry into a knowledge-based one could be reflected by the central position of support structure organisations such as RDA, *Business Link*, ERDF, whose aim is to support the infrastructure as strategic drivers of regional development (Hauser, 2010). This hands-on approach supposes the delivery of the highest public investments in the UK to develop a R&D infrastructure (BIS, 2009), and goes hand in hand with one of the highest levels of academic spending in R&D (Lambert, 2003) with collaboration between government and academia as the main driving force behind the networks. This collaboration has produced networks which are able to provide the capital, advice, and resources through early-stage investment facilitators and business developers, fostering a growing knowledge-based industry. Consequently the networking capabilities and hybrid roots of business developers stimulate higher cohesion between the institutional spheres. In the case of *LIC* and *YSP*, both have different knowledge transfer models. *LIC* takes advantage of the University's strong links with a private IP office and focuses on licensing activities and supporting the formation of promising spin-offs, while *YSP* has a public-university oriented approach that focuses on setting up a wide range of businesses (Lambert, 2003). This difference in both approaches can be identified by the selected number of knowledge-intensive spin-offs that have flourished despite the limited presence of business supporters in *LIC*, while *YSP* promotes the development of a wider range of knowledge-based and services-oriented firms which are supported by a larger public infrastructure. Hence, there is a system that only invests in deals which suppose low-risk and high returns, and another that invests to obtain immediate results in the form of firm and job creation, independently of the quality and sustainability of the ventures in the long term (Nottingham BioCity, 2010). Apart from this, the role of publicly backed investments in the establishment of business developers and in the provision of redirected funds observed in the web-based networks may match the increasingly representation of the public sector in the capital market, which participated in 42% of all venture capital deals in 2009 and 68% of all early-stage deals in 2008 in the UK (Pierrakis, 2010).

Fourth, it seems that only most research intensive consultants are able to link research producers and other organisations. In the case of the *AMP*, the high number of R&D consultants and the pivotal role of the *AMRC* in the diffusion and application of new advanced technology show that the SP functions as a seedbed for the industry in joining and engineering development (House of Commons, 2011; Technology Strategy Board, 2010). However, only most R&D intensive consultants and research consortiums are able to liaise with the institutional spheres and develop technology transfer programmes (Bessant & Rush, 1995). This fact is observed in the *AMP* and *YSP* where the consultants with a research basis and academic roots are involved in a more complex and institutionalized processes, integrating more. Nevertheless, the weak connectivity of the consultants may depend on the reticence to make commercial relationships public or on establishing links with clients outside the SP, which makes it necessary to study the external visibility of these organisations to determine the value of their services in the national and international arenas. Here it is worth noting that the incomes of the academic sector, regarding business and community interactions, are mainly obtained through non-commercial research contracts whereas consultancy, IP, and the use of facilities and equipment provide marginal incomes, and the public and third sector organisations are the major clients and provide the highest incomes in the region (HEFCE, 2010b). Consequently, the interactions identified in the form of research contracts between the *University of York* with public and third sector organisation based on *YSP* proved to be representative.

Finally, university spin-offs are the other identified forms of knowledge exploitation. As expected, in these structures the considerable number of spin-offs suggests a strong research basis and the conditions achieved by the R&D infrastructures. However, although the collaborative process to establish and develop knowledge-based firms leads to better integration than for service-oriented firms, the private sector still occupies a peripheral position. This may be caused by the lack of private investments in the region (Nottingham BioCity, 2010) and the lack of linkages among firms located in SPs (Quintas et al., 1992; Suvinen et al., 2010), as well as the inadequacy of direct hyperlinks to detect commercial ties (Stuart & Thelwall, 2006; Vaughan & You, 2006). Nevertheless, an important fact is that the proportion of collaboration among firms engaging in innovation activities is only 23%, and the three less frequent partners as well as sources of innovation information are furthermore consultants and private R&D institutes, Universities, and Government or public research institutes, respectively (BIS, 2009). Therefore, regardless of the application of direct or indirect links it could still be difficult to find signs of cooperation, including among innovating enterprises.

Overall, although a web link study can help to track how the institutional sectors and organisations associated with the SP movement interact on the web, and these web-based patterns may be similar to off-line ties it is important to highlight that webometric evidence cannot be used for the assessment of SPs. The main limitations of hyperlinks as indicators of offline phenomena are the early-stage of development of webometric tools and techniques applied beyond academic links (Thelwall, 2010b), search engines bias and barriers to collecting data (Thelwall et al., 2012; Vaughan & Yang, 2012) and complex interpretation (De Maeyer, 2012) because of both incoming and outgoing hyperlinks are potentially social ties with meaning that can vary in relation to the contexts in which a set of off-line actors interact on the web.

## 7.5 Conclusions

This exploratory study introduced a web-based approach based on interlinking networks within a SP-based community. Given that the interconnections within a group of websites should derive the same structure regardless of the use of in-links or out-links, both dimensions were collected from three different sources and were used jointly in an attempt to obtain a more robust and reliable structure. A structural comparison confirmed that both dimensions provide similar structures, because there was a significant correlation between the links generated by both data sets. It was also found that the out-dimension, collected from Bing and SocSciBot, with half of the total links of the in-dimension, collected from Yahoo!, provided more cohesive structures. In addition, the overlap between Bing and the crawler SocSciBot was only 4%.

- The first research question asked: *Can webometrics methods identify interactions between institutional sectors as well as interactions between various types of organisations associated with SPs*? The combined interlinking networks show that the SPs tend to be organised according to the academic, industrial and governmental sectors, and that they are primarily interconnected by a central sub-network which is led by academia and support structure organisations that support the rest of the network. These web-based networks probably reflect collaboration between academia and government to generate an infrastructure that hosts heterogeneous organisations which are dedicated to exploiting the local research basis in order to promote knowledge-intensive industry in the region.

- The second research question asked: *Are the main features of SP interlinking networks in line with findings of official reports and surveys which are essential sources of information to evaluate the R&D infrastructure in the UK*? The analysis of the main features shared by the web-based R&D networks seems to reflect the potential strengths and weaknesses of the policy measures and conditions described by different surveys and reports on the R&D infrastructure developed in the region. According to the evidence based on traditional indicators the web-based network may reflect the prominent role of the research producers and the commitment of the public support through various support structure organizations. Both sectors make significant economic efforts and closely interact to foster networks that are able to provide capital, advice, and resources. The difference in the private and university-public knowledge transfer models followed in the SPs may also be illustrated by the type of firms that are established and by the size of the support infrastructures developed. In addition, the significant presence of the public sector identified also matches its representation in the capital market. The knowledge exploitation mechanisms observed are contract research, consultancy services and creation of new ventures. The evidence found seems to confirm that only the most research intensive consultants are able to interact with research producers and other organisations, and that the universities' research contracts are primarily obtained from public and third sector organizations. The considerable number of spin-offs indicates a strong research basis and the conditions achieved by the R&D infrastructures, while its low degree of linkage among firms located in SPs corroborates previous studies.

The novel method introduced is able to extend previous webometric attempts that study R&D networks and provide a broad overview of a complex and highly institutionalised innovation infrastructures developed in the studied region and that could be further analysed by other approaches. This web-based approach may facilitate the study of a complex underlying structure that needs to be assessed by studies that focus on the different aspects and particular actors that are embedded in these technology and innovation centres. The first findings suggest that it may be useful to investigate the social and entrepreneurial activities of the university, large technology and innovation centres, clusters, regional innovation strategies, and dynamic systems with a high institutional heterogeneity. However, due to the exploratory nature of this chapter as an early type of webometric study, these findings are merely indicative and additional research is required. A clear limitation of this approach is only taking into account the hyperlinks between the websites of the sample and not using information about specific inter-organization connections in the SPs studied. This means that the bias introduced by the use of external connections through spam and irrelevant connections is ignored, but at the same time these internal links may not reflect all the potential relations established so the results should be taken as illustrative rather than exhaustive. An example of this limitation is the *AMP*, which represents a successful model that has been exported to 9 other countries (House of Commons, 2011; Technology Strategy Board, 2010) but the low number of links with industry found may suggest that the analysis is only able to uncover a small fraction of the technology transfer and innovation processes which take place in Sheffield. Consequently, questions about the role of the *University of Sheffield* and about the impact of the research centre and consortiums outside the SP remain unanswered. The low visibility of organisations that establish commercial ties, such as firms and consultants, may be also affected by the use of direct links, suggesting that the use of indirect links could be also necessary and undermining the validity of the web approach.

Finally, the analysis of various types of organisations may lead to a bias towards the public and academic organisations which tend to have large websites and then be more visible. Moreover, the traditional classification (industry-university-government) is not always appropriate in R&D networks, because it is too broad to show the hybridization process of the institutional sectors to establish some important central organisations. Future studies should therefore evaluate the R&D networks constructed based on direct and indirect links to determine the potential similarities and differences of both approaches, design a framework to identify and evaluate the behaviour of the institutional agents on the web, and design a method to identify interactions that could be relevant in the external environment of the SPs.

# Chapter 8: The science park actor framework

## 8.1 Introduction and goals

This final empirical chapter introduces a structured analysis of science parks as arenas designed to stimulate institutional collaboration and the commercialization of academic knowledge and technology, and the promotion of social welfare (Section 1.8). A framework for the key actors and their potential behaviour in this context is introduced based on the Triple Helix (TH) model and related literature. A web link analysis was conducted to build an inter-linking network (Section 3.3.7; Section 3.3.8) to map the infrastructure support network through the web interactions of the organisations involved in York Science Park (Section 3.3.6). A comparison between the framework and the diagram shows that the framework can be used to identify most of the actors and assess their interconnections. The web patterns found correspond to previous evaluations based on traditional indicators and suggest that the network, which is developed to foster and support innovation, arises from the functional cooperation between the University of York and regional authorities, which serve as the major driving forces in the trilateral linkages and the development of an innovation infrastructure.

The previous chapter introduces a web-based attempt to investigate the infrastructure of SPs using link analysis that shows strong TH cooperation and identifies universities as central in the networks (Section 2.2.1.). In order to gain deeper insights into the configuration of SPs it is necessary to carry out a systematic and structured analysis. Therefore, the purpose of this chapter is twofold: (1) to design a new framework, the SP actor framework, based on the TH model that lists the key actors that should be involved in the SPs and identifies their missions, functions, and potential interactions; and (2) to compare this framework with the network created by the hyperlinks among the actors within the SP to determine if the links reveal potential offline behaviours and whether the patterns identified are in line with the results obtained with traditional indicators. The latter makes sense because hyperlinks can reflect social interactions and represent underlying social structures (Reid, 2003). These two objectives are addressed by the following research questions:

- Can the organisations in the SP actor framework be identified in a hyperlink network generated by the web sites of organisations associated with York Science Park?

- Do the links between organisations in hyperlink network reflect the potential behaviour of the different types of actors described in the SP actor framework?

## 8.2 The SP Actor Framework

A number of different types of actors, listed below, are often found in organizational innovative environments (Etzkowitz, 2008) although this list does not include the full range of actors that could be present (Howells, 2006). The presence or absence of particular actors in a SP depends on its economic development strategy, which is determined by the local and regional conditions and objectives, as well as the degree of cooperation between the institutional spheres (University-Government-Industry). Thus, the effectiveness of a SP with regards to the commercialization of research could not only be related to the establishment of certain intermediary organizations but also to the functionality and productivity of the interconnections in the network, as highlighted by Suvinen, Konttinen and Nieminen (2010). Collaboration within the network may facilitate the hybridisation needed to overcome the absence of certain types of actors, so the remaining actors, especially intermediaries, would take on the missing roles in order to fill the gaps in the innovation infrastructure (Etzkowitz & Leydesdorff, 2000). The intention with this framework is therefore not only to identify the key social actors and their behaviours but also to learn which functions and roles are important to create an innovative environment and to examine which actors could fill these roles.

### 8.2.1.    University

As an institution, the University interacts with industry and government to embrace a new social mission that expands its academic and scientific missions. It attempts to commercialize its research by connecting to industry and thereby attracts external funding, promotes employment and fosters regional strategic development. This entrepreneurial function requires the development of technology transfer capabilities in the form of consulting, patenting and licensing, and firm-formation activities that can lead to independent entities emerging from the university or at least having strong academic links. These firms are often located in dynamic and quasi-academic spaces like SPs.

**Role in network:** The University is the driving force in a SP and is expected to occupy a central position in an innovative support structure. The entrepreneurial university keeps strong ties with hybrid actors rooted in academia that produce, capitalize, and disseminate knowledge, such as technology transfer offices, research centres, consulting organizations, and incubators. It is also expected to have direct connections with government and industry actors, such as regional development agencies (RDA), R&D units of large firms, start-ups, spin-offs, and other actors that support the development of the network and need research and advanced technology (Etzkowitz, 2008; Etzkowitz, et al., 2000).

### 8.2.2. Research centres

Research centres have different organisational characteristics but tend to depend on industry and/or government sponsorship to foster technology transfer between universities and firms. They are designed to span university-industry boundary and partly fulfil the research missions of universities and typically involve strategic alliances to achieve long-term goals. They may bring various intellectual, physical and organizational resources together within a single university or span several universities and non-academic institutions, such as government research institutions and firm laboratories, to engage in a more intense and interdisciplinary collaboration. A centre may host several research groups around a theme and for several purposes: (1) to attract more funding, (2) to access to better facilities and instruments, and (3) to undertake large-scale projects. Some of their main advantages are: (1) external use of academic laboratories, (2) increase of faculty consultants, (3) research collaboration, and (4) job creation for students.

**Role in network:** Research centres liaise with academia, industry, government, and the wider public to provide a space for collaboration between actors with diverse perspectives who are interested in pursuing applied knowledge. This concentration of resources and interests makes research centres important social actors that may serve a heterogeneous network of organisations that need highly specialized scientific knowledge and technology to resolve particular problems and increase their competitive advantage. Therefore, research centres are expected to occupy a central position in the network and to be linked with knowledge producers and consumers (Adams & Chiang, 2001; Boardman, 2009; Etzkowitz, 2008; W. Larsen & Fernandez, 2005; Lee & Win, 2004).

### 8.2.3. Consulting organizations

These are liaison offices that convert informal individual consulting into a professional and organized group activity. They identify technological needs in industry or government and put together the necessary resources to provide new customized solutions. Their strong university links suggest that this function could be filled by the university or a quasi-academic actor that connects the university with a group of consumers related to the private and public sector. Such consulting organizations can be either an economic arm of the university that brings external funding to the academic world, i.e. an internalised consulting model, or an independent business, hiring individual scholars for specific projects, i.e. an externalised consulting model.

**Role in network:** These actors organize and strengthen the interactions between researchers or research units, regional organizations and companies through consultation and research contracts. They represent the first steps to capitalize knowledge, and would therefore be expected to have a close interaction with knowledge consumers, while their connections to the university would depend on the degree of independency of the consultant. (Etzkowitz, 2008:95).

### 8.2.4.  Technology transfer offices

These have a dual search mechanism; an internal mechanism to identify and commercialize relevant research and technology produced by the university, and an external mechanism to identify potential markets for it and bring potential customers to the university. These offices are responsible for identifying, patenting, marketing, and licensing intellectual property in order to attract additional research funding via royalties, licensing fees, and research contracts.

**Role in network:** Technology transfer offices are important brokers of the innovation system which value, protect and sell university inventions. This commercial function could be realized by the university itself or an independent actor which may have links with intellectual property lawyers. These actors would usually have strong academic connections and are one of the main intermediaries between the university and a group of technology-based firms (Etzkowitz, 2008:37, 89; Siegel, Waldman, & Link, 2003).

### 8.2.5.  Incubators

University-related incubators are sponsored by the university, usually in partnership with other interested players. They are independent units intended to commercialize research, technology and intellectual property produced by the university in the form of new firms. This organizational support structure provides the space and added value to assist spin-offs and start-ups with consultation, business services, straightforward funding, subsidized space, access to university facilities, expertise and assistance of researchers and students, networking opportunities, and other services to encourage entrepreneurship and promote academic-industry collaboration. This improved environment should increase on-site firms' chances to survive and grow.

**Role in network:** The assisting function of incubators in the growth of university spin-offs makes it possible to identify an entrepreneurial sub-network intended to promote new ventures. The role of this actor primarily consists of establishing relationships between the university and entrepreneurs and to create ties to investors as well as public and private hybrid actors in order to support early-stage firms (Barrow, 2001; Etzkowitz, 2002).

### 8.2.6. Investors

Venture capital (VC) is at the heart of the firm-formation process and SPs with an incubation facility should create an ideal organized environment for investors who can take advantage of innovative ideas and high growth potential ventures in attractive sectors to invest with reduced risks. The injection of funds and resources by investors plays an important role at all stages of the firm-formation process. Early-stage investment provided by business angels and seed capital investors (see below) are the most common form of investment while their involvement in the network is basically related to the particular number of firms in their portfolios.

**Role in network:** Early stage investments are typically provided by public and private actors that are connected to new and high growth potential companies. Despite the importance of these actors in the network, they may have few hyperlinks because interactions tend to be intense but informal between investors and companies in the SPs. The intensity of the hyperlinks may exhibit two patterns according to the degree of collaboration between both parties in relation to the stage of development of the firms: (1) strong and informal ties in the case of business angels and early stage ventures which are necessary to get the business off the ground; and (2) weak and formal ties between VCs and growing firms which are necessary to accelerate the growth of businesses (Barrow, 2001:110).

A **business angel** is defined by the European Business Angels Network (2010) as "*a private individual who invests part of his/her personal assets in a start-up and also shares his/her personal business management experience with the entrepreneur*". These are informal suppliers of high risk capital which adopt functions similar to business incubators, spending time in mentoring early-stage and growing business. The rise of business angel networks (BANs), in which groups of angels cluster together to pool investment and expertise, increases their investment capacity and confidence through a larger collaboration network (Barrow, 2001). On the other hand, **seed capital** is usually offered by the government, universities, and corporations in which the technology of the new firm has been produced. Government and university VC focuses on a social perspective and supports both long-term and research-based projects, or less-favoured fields and less venture capital-intensive regions through funds, research grants, subsidies or indirect loans to promote economic growth. Early-stage funding is intended to fill the gap left by private investments, which are oriented to later stages, reduced risk and short-term financial returns. A combination of private-public sponsorship may also operate at this stage (Etzkowitz, 2006; 2008:122-36).

### 8.2.7. Government agencies:

These can directly or indirectly encourage organized TH collaborations and sometimes operate through quasi-government agencies. They provide a regulatory environment and also act as a public venture capitalist to increase innovation and regional competitiveness (Etzkowitz, 2008). An innovation policy should have the capacity and autonomy to be driven by institutions at local, regional, central, and supranational levels, to design adequate infrastructures, examples of government agencies operating in the UK include the *European Regional Development Fund* (ERDF), the *Regional Development Agencies* (RDAs), and the *Business Link*.

**Role in network:** At a practical level the region is the key space for economic governance and RDAs are the major mechanism to facilitate central policies. They provide financial support to the innovation infrastructures through the establishment of intermediaries and partnerships with the university and key actors to exploit the accumulated research and to support the growth of knowledge-based businesses as significant sources of employment growth and sustainable development (Webb & Collis, 2000).

### 8.2.8. Knowledge-based firms:

These are firms, such as university spin-offs, spin-ins, start-ups, and R&D units, which have emerged from, or at least are closely associated with, a university or another knowledge-producing institution. The formation process of high-tech and knowledge-based firms requires multiple resources and support, and the cross-fertilization between technical and business skills, which is embedded in collaborative relationships that may include other firms in strategic alliances and actors from the university and government. These efforts of businesses to overcome their limited in-house knowledge capacity to become more innovative through the process of harnessing external knowledge and skills in conjunction with their internal R&D to commercialise new technological advances is considered as an open innovation process (Chesbrough, 2003), and requires an intensive cooperation that is supported and mediated by the innovation network infrastructure.

**Role in network:** The commercialization of research embodied in firms is the engine of innovation strategies. Knowledge-based firms are expected to be the central economic actors in the interactions occurring through networks across institutional spheres. They may be linked with the university, governmental and private investors, research groups, incubators, and other universities' economic arms (Etzkowitz, 2006, 2008).

### 8.2.9.  Service-based firms:

These are firms which tend to operate with an incremental perspective toward product development, utilizing new combinations of existing technologies to solve a problem or provide a service (Etzkowitz, 2008:54). Start-ups and small- and medium-sized firms (SMEs) can be attracted to SPs for the competitive advantage of a prestigious location and a network of potential customers to market their services to. In this group there are also recruiting, accounting and marketing firms, lawyers, and web-designers.

**Role in network:** Being market-oriented, these firms are likely to establish competitive relationships and neither engage in advanced research nor develop collaborative relationships with the members of the network. This lower degree of reliance on the innovation infrastructure to perform their activities suggests that they could be more isolated than knowledge-based firms.

## 8.3 York Science Park

The research questions were addressed with a case study of York Science Park. This SP was chosen due to its location in a region with significant R&D investments undertaken by the higher education sector, especially by the universities of Leeds, Sheffield, and York as well as the RDA Yorkshire Forward and the ERDF. These efforts are oriented towards the development of a regional innovation system to support the firm formation and university-industry links as a means to reverse the subsequent decline of traditional industries and bring about economic dynamism in the region (Dabinett & Gore, 2001; Huggins & Johnston, 2009:234-6). Furthermore, York SP provides the largest web-based network among the SPs in Yorkshire and the Humber region (see Chapter 7), which makes it suitable for a deeper structural analysis.

In order to generate the hyperlink network for York Science Park, this study draws on the same data which was collected in May 2010 in relation to the previous chapter (Section 3.3.6) and the following process was used (Section 3.3.7 and Section 3.3.8). The interlinking network of York Science Park has 104 actors (including the York SP) that were connected by 378 links, but to facilitate the representation and analysis of this comprehensive network the implicit external links of the SP to the 103 actors were eliminated and only the incoming links to the SP were considered. This eliminated the actors that were only linked by the SP, reducing the number of actors to 75 (72%) and the number of links to 275 (73%). The number of actors from each category and the interconnections between the categories are in Table 8.1.

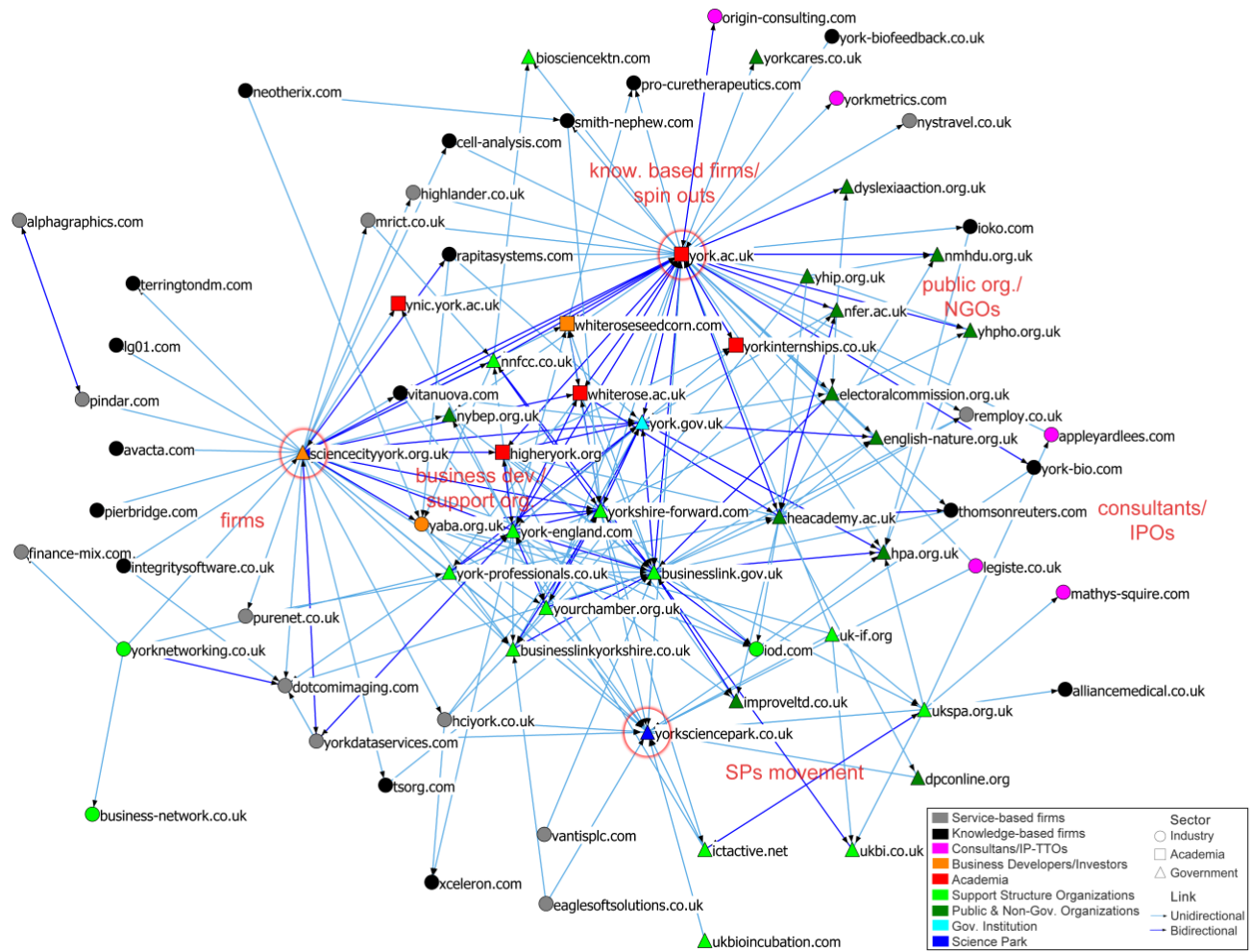

**Fig. 8.1** Inter-linking network of York Science Park (YSP). (A colour version is available at: http://home.wlv.ac.uk/~in1493/issi-11/fig-1.jpg)

**Table 8.1** Links between the categories found in York Science Park.

| Actors | Categories | Service-based firm | Know.-based firm | Consultants/IP-TTOs | Business Dev./Invest. | Academia | Support Struc. Org. | Public & Non-Gov. Org. | Government | Science Park | Total - outlinks | Mean |
|---|---|---|---|---|---|---|---|---|---|---|---|---|
| 24 | Service-based firm | 3 | 1 | | 1 | | 7 | 2 | | 3 | 17 | 0.7 |
| 30 | Knowledge-based firm | 1 | 1 | | 3 | 3 | 3 | 2 | 2 | 2 | 17 | 0.6 |
| 9 | Consultants/IP-TTOs | | | | | 3 | | | | 1 | 4 | 0.4 |
| 3 | Business Developers/Investors | 7 | 10 | | 3 | 6 | 9 | 1 | 1 | 2 | 39 | 13.0 |
| 5 | Academia | 4 | 8 | 2 | 5 | 8 | 11 | 11 | 2 | 2 | 53 | 10.6 |
| 17 | Support Structure Organization | 5 | 2 | 2 | 12 | 12 | 45 | 12 | 5 | 7 | 102 | 6.0 |
| 14 | Public & Non-Gov. Organizations | 1 | 1 | | | 8 | 5 | 12 | 3 | 1 | 31 | 2.2 |
| 1 | Government | | | | 1 | 2 | 6 | 3 | | | 12 | 12.0 |
| 1 | Science Park | | | | | | | | | | 0 | 0.0 |
| | **Total - inlinks** | 21 | 23 | 4 | 25 | 42 | 86 | 43 | 13 | 18 | 275 | |
| | **Mean** | 0.9 | 0.8 | 0.4 | 8.3 | 8.4 | 5.1 | 3.1 | 13.0 | 18.0 | | |

IN

OUT

## 8.4 Results and Discussion

The infrastructure around the SP builds a network with three main brokers forming a triangle that connects together the business service and support structure organizations situated in the middle with different sub-networks (see Figure 8.1). With the highest in-degree (24) and out-degree (37) the *University of York* is the most important actor and connects the central actors with spin-offs, knowledge-based companies, consulting offices, and third sector organizations. The second most important broker is the business developer *Science City York* (SCY), which acts as an intermediary between the industry and the central business services and support structure organizations. The third broker is the SP, which connects the heart of the network with some firms and actors related to the development of the SPs in the UK.

Analyses of the presence, or the lack of presence, of the actors listed in the framework as well as the impact or gap of their ego-networks are described below.

### 8.4.1. University:

As the most central actor, the University of York is directly connected to 44 (59%) of the interconnected actors and is therefore at the core of the network. Knowledge exchange with the private sector is manifested through connections with nine knowledge intensive companies from the Bio, Health, and IT industry and links with four University spin-offs (yorkmetrics.com; cell-analysis.com; origin-consulting.com; rapitasystems.com). The extensive interactions with the public sector are divided in two groups; one central group with nine intermediaries and support structure organizations, and another group with eleven nodes formed by independent bodies and organizations within the third sector. The enterprise initiatives of the University are also reflected through three consulting offices, two of which are spin-offs and the last is an IP office (appleyardlees.com). The presence of the University in the SP is extended through 5 other actors: the White Rose university consortium (whiterose.ac.uk) that promotes collaboration between the universities of Sheffield, Leeds, and York, its seed capital investor White Rose Technology Seedcorn Fund (whiteroseseedcorn.com), York Neuroimaging Centre (ynic.york.ac.uk), the Student Internship Bureau (yorkinterships.co.uk), and Higher York (higheryork.org), which is a partnership between higher education institutions to offer training and consultancy for local businesses. In addition, the central position of the three academic actors that aim to exploit academic knowledge and technology shows how they collaborate with support structure organizations to attract public funding and create the adequate conditions for new businesses to flourish. The prominent role of the university as the main knowledge producer in this knowledge-based network confirms a two-way flow of influence between the university and the other actors, becoming the liaison between the institutional spheres and intermediaries. The SP is used as the arena to carry out outreach activities that promote an entrepreneurial involvement and identity which are critical for a sustainable economic and social development in the region (Etzkowitz, et al., 2000; Godin & Gingras, 2000).

### 8.4.2. Research centres:

*York Neuroimaging Centr*e (ynic.york.ac.uk) is a research facility established by the University of York to produce multidisciplinary research and serve the demands of the University, the National Health Services (NHS) and industry. However, its level of integration within the SP is very low.

### 8.4.3. Consulting organizations:

Six actors can be identified whose main activity is consultancy but only three are connected to the network. The increased demand by industry and government for technology and customized solutions has led to the establishment of two spin-off companies rooted in the Department of Computer Science, namely *YorkMetrics* (yorkmetrics.com) and *Origin Consulting* (origin-consulting.com), and a third consulting organization called *Legiste* (legiste.co.uk). In addition, the links between the University departments and independent bodies and NGOs are also established by consultation and research contracts. These quasi-academic actors only link with the University, which could be because they are likely to have their customers outside the park, and these customers would not be identified by the link analysis method used here. Another reason could be that their relationships with customers are not strong enough to be advertised through hyperlinks – or are kept secret for commercial confidentiality reasons. On the other hand, the three consultants that do not engage in R&D activities are isolated: the electrical, franchise and educational consultancies. This shows that the University actively offers consultancy services and has formalized the commercialization of academic knowledge through spin-off companies, although the links of the consultants with potential customers are not evident.

### 8.4.4.  Technology transfer offices:

The network includes three IP & Trademark offices: Appleyard Lees (appleyardlees.com), Mathys and Squire (mathys-squire.com), and the isolated Murgitroyd & Company. Although the first office works with the University, they have a peripheral role in the network without being connected with a spin-off or enterprise. The low connectivity of this group might call into question the need for studying the external relationships of the actors to determine whether certain actors could be investigated with the help of hyperlinks. Nevertheless, the reason behind their low activity might be the low proportion of enterprises that apply for a patent (2.5%) or register a trademark (5.2%) in the region (BIS, 2009), and the negative effect on the licensing activities as a result of the overemphasis on the generating spin-offs, regardless of their quality, driven by the availability of public funds that see the new firms as a source of employment (Lambert, 2003:58-62).

### 8.4.5.  Incubators:

Despite the lack of an incubator, the network contains many new ventures. This role is filled by the University in an active collaboration with its partners, including York SP, the business developer SCY, and the White Rose consortium. This reflects the importance of the structural collaboration between university and government to establish intermediary organizations that are capable of encouraging technology transfer and firm-formation.

### 8.4.6.  Investors:

The Yorkshire Association of Business Angels (yaba.org.uk) and White Rose Technology Seedcorn Fund (whiteroseseedcorn.com) are responsible for injecting capital and commercial expertise into new businesses. The first of these organisations connects business angels with entrepreneurs looking for finance and mentoring. The second organisation invests in early stage commercial opportunities based on new technology emerging from the universities of York, Leeds and Sheffield until they are ready for later stage investors. As risk capital providers, they receive funding from the RDA and academia and are closely interconnected with the University and various business supporters, while their direct links with businesses are limited to only three spin-off companies. These few ties could be the result of the often informal relationships between early-stage investors and new firms. Another important actor is the business developer Science City York (SCY) (sciencecityork.org.uk), which was established by York City Council and the University of York to ensure York's economic regeneration and build a reputation as a centre of scientific and technological excellence. It works in partnership with the University and support structure organizations to stimulate the creation and growth of business through public supported mentoring, facilitating investment funds within bioscience, IT, and the creative industries, and contributing to a growing local knowledge-based industry (Lambert, 2003:72). As might be expected, SCY fosters firm-formation activities around the University and builds a bridge between companies and support structure organizations at the heart of the network. This group of actors is interconnected with key actors and shows the importance of seed-stage investments in this context as well as the difficulties to identify the ties at this stage of development. Furthermore, it illustrates the efforts of the public sector to encourage the collaboration required for an open innovation process.

### 8.4.7. Government agencies:

There are two groups of organizations with public roots; one formed by business support organizations which occupy a central position, and another formed by third sector organizations and independent bodies which are more peripheral, on the left side of Figure 8.1. The few actors that are involved in the development of SPs are situated towards the bottom, as are the networking organizations. The largest group provides economic resources, specialised advice and support for the network. It is led by the support organizations Business Link (businesslink.gov.uk), which delivers publicly funded business support products and services designed to help new businesses, the regional development agency (RDA) Yorkshire forward (yorkshire-forward.com), York England (york-england.com), which supports regional businesses attracting new investment for the region, and City of York Council (york.gov.uk). As might be expected, these public actors at the heart of the network encourage co-operation among the nodes through partnerships and funds to key brokers, and are surrounded by the York SP, business service support organizations such as Your Chamber, Higher York, Institute of Directors (iod.com) and networking organizations (york-professionals.co.uk; yorknetworking.co.uk; business-network.co.uk).

The second group exhibits intensive cross-fertilization between the University and independent bodies and NGOs. This relationship is established through partnerships and research contracts which bring problems and needs into the University and at the same time allow the organizations to take advantage of the accumulated knowledge to provide efficient advice and services to policy-makers and the wider community. These actors have local offices or are based in the park and are linked to university departments that are judged as world-leading and top-ranked by the last Research Assessment Exercise (RAE, 2010), such as the Biology department, which is linked to *Natural England* (english-nature.org.uk); Health Economics and Science, which is linked to Yorkshire and Humber Public Health Observatory, Yorkshire & Humber Improvement Partnership, Health Protection Agency and National Mental Health Development Unit (yhip.org.uk; yhpho.org.uk; hpa.org.uk; nmhdu.org.uk); Social Policy and Social Work, which is linked to the National foundation for research (nfer.ac.uk); and the department of Psychology, which is linked to the Higher Education Academy and Dyslexia Action (heacademy.ac.uk; dyslexiaacition.org.uk). Finally, there are also a few actors that work to promote the development of SPs and business incubators (ukspa.org.uk; ukbi.co.uk; uk-if.org; iactive.net). The range of outreach activities shows that the third mission of the University of York is not limited to an entrepreneurial approach, but also embraces a wide social function assisting policy-makers and civil society. The balance between profitable and social activities in the network could give the York SP a science shop's identity, it  means places that offer citizens, non-governmental organisations (NGOs), municipalities, and sometimes small and medium enterprises free or very low-cost access to scientific and technological knowledge in a wide range of issues related to environment, health, education, labour, law and housing (Fischer, et al., 2004). However, there are key aspects such as the secure access to public and infrastructural funding and the level of institutionalisation of the SP that distinguishes it from traditional science shops.

### 8.4.8.  Firms:

There are 30 actors which can be identified as knowledge-based firms and 24 service-based firms in the SP. The first group is more connected, having 18 (60%) firms linked to the network, and has the most interconnectivity with the business developer SCY and the University. On the other hand, the second group has 13 (54%) firms linked to the network. Due to the significant impact of SCY and the University among both groups of firms, with 24 (77%) of the firms linked to one or both, these are the main brokers between the  public and industrial sector. Furthermore, most of the businesses are within the three SCY clusters of interest (bioscience, IT, creative industries), while nine knowledge-based companies have spun off from the university, seven from the University of York and two from Leeds. However, despite the higher level of interlinking of the knowledge-based firms, the presence of the service-based firms is also significant. This could be a result of the range of firms that are eligible for support from SCY (including bio-related, graphic and web design companies) and might be driven by the efforts of public organizations to increase the number of firms and jobs created instead of their quality, as pointed out in the UK Life Science Start-up report (2010). In addition, the Yorkshire region, with the highest public expenditure in the UK (Robson & Kenchatt, 2010), distributes small amounts of money to many firms rather than funding just firms with high growth potential, hence taking a lower risk or not being able to identify promising firms. However, this financial support to a high number of service-based start-ups is unsustainable because the region was not able to attract private investments nor to form a science cluster (Nottingham BioCity, 2010).

The analysis of the network indicates that York SP provides a network characterized by the collaborative efforts between the university and local government to create conditions to allow knowledge-intensive businesses to flourish. It is interesting to observe how this is in line with the policy agenda and characteristics observed in York by the Lambert review: "The university's science park provides incubator space for new start-ups, while the council helps to link businesses with legal, financial and marketing professionals." (2003:72). As a producer of research, the university is the main source of firm-formation and knowledge-based services in the network and attracts both private and public actors as consumers. The functional university-government collaboration is supported by the important investments made by the RDA that fills its expected task, as the major stakeholder in university outreach activities, especially in relation to supporting SMEs and local communities (Woollard, Zhang, & Oswald, 2007:390). Interestingly, these efforts coincide with the fact that the hands-on approach of the regional government involves the highest public investment in the UK and the Universities in the region have some of the highest budgets in the UK, but at the same time the region also attracts one of the lowest levels of private investment (Lambert, 2003:65; Nottingham BioCity, 2010), as shown by the lack of involvement of the private sector in the interlinking network. This also suggests that the network might struggle to survive without public funding.

The university links the private, public and third sectors together and works in partnership with the RDA to set up important actors, namely the business developer SCY and the York SP, to obtain external funding and services to support new businesses and thus complement its own seed capital unit. The links also suggest that the university uses the SP to capitalize its research via consultancy and research contracts, and the establishment of knowledge-based spin-offs. On the other hand, other actors, such as the research centre and the consulting and IP offices, exhibit a low degree of activity within the SP. This could be the result of the inability of the hyperlinks to reflect their relationships accurately or the low engagement with the members of the network. In the previous chapter study of an industrial SP in the same region also showed that research centre and consultants have a low level of interconnectivity. Due to the low connectivity of these three types of actors, it could be interesting to analyze their links outside the SP to find out if their activity is focused on external relationships.

Even though the injection of private investments in the infrastructure of the network is limited, businesses take advantage of privileged links to the University and its partners as well as of a sub-network of actors dedicated to provide specialised business and financial support services. This promotes the growth of a high number of businesses in the park despite the lack of an incubator that supports this complex process, and thus highlights the key role of the hybridisation process among the actors to form a dynamic and flexible support infrastructure. This fundamental role is taken by the business developer SCY that, together with two investors, is one of the main intermediaries for public funding and services which allows spin-offs and SMEs to flourish. The impact of SCY leads to a close collaboration between the knowledge-based companies and the University, and the significant integration of service-based firms driven in part by aligning the aims of York SP and SCY to host and support companies with a wide range of profiles. Other actors taking advantage of the research and social commitment of the university are the independent bodies and third sector organizations which engage in partnerships that lead to knowledge exchange and application in the wider community, reinforcing the social function of the University of York.

### 8.5 Conclusions

- In answer to the first research question, the study shows that the *SP actor framework* can be used to identify eight of the nine types of actors represented in the York SP network. The lack of an incubator is interesting because the incubation process seems to be essential for the formation and development of new businesses and receives considerable attention in the literature, as one of the main actors in the innovative infrastructures. Nevertheless, the dynamic infrastructure established in York SP is still able to attract and generate new businesses and university spin-offs.
- In answer to the second research question, the links between the organisations coincide with the potential behaviour of only three of the actors described in the *SP actor framework*. The actors that best follow the expected patterns are the university that links all the different types of actors, the government agencies that provide the financial resources to promote university-industry collaboration, and the investors that collaborate with the university and public business supporters to support a couple of companies with a high growth potential. On the other hand, there are actors with a low connectivity to the network, such as the consulting organizations that are only connected with the university, although the consultancy services offered by the university are well represented. The knowledge-based and serviced-based firms occupy a peripheral position and are mainly connected to either the university or the business developer, respectively. Finally, the links of the research centre and the technology transfer offices do not follow the expected patterns.

Nevertheless, the connections are not exhaustive since there will be formal and informal connections between the actors which are not represented by hyperlinks. Moreover, the strong participation of the public actors might be biased due to the need to increase government transparency and provide electronic access to government information (Jaeger & Bertot, 2010). On the other hand, the few links among the companies might be attributed to the low rate of hyperlinks established by the private sector (Stuart & Thelwall, 2006; Suvinen et al., 2010). Likewise, other actors such as the consulting organizations and IP offices might tend to establish few internal links, suggesting that the external environment of the SPs should also be investigated. The major limitation of the method used in this study is that it only focuses on the intra-networking dynamics of the SP while it is also important to study how they influence the region and whether they are able to extend beyond the region. Therefore, future research should focus on studying the inter-networking dynamics of the SPs, operating through internal networks among SPs and actors from different SPs in the region, as well as the external-networking dynamics, operating through external networks with the actors within the SPs.

In spite of the inherent limitations of a webometric approach, the framework makes it possible to carry out a structured analysis of the actors involved in the SPs through the identification of key actors and their expected behaviour in terms of the robustness and dynamism embedded in the SP. However, the framework still needs to be applied to other networks to determine its capacity to understand and assess these dynamic structures. Moreover, SPs are often the result of trilateral partnerships that integrate heterogeneous actors embedded in particular socio-economic conditions which influence the mission and operational procedures that make each SP a unique social environment. This makes it difficult to draw general conclusions, and more studies are needed to realize how their R&D infrastructure is configured and observe if they reflect off-line characteristics in order to determine if the interlinking networks in this context may be used as weak benchmarking indicators (Thelwall, 2004b). The need for broadening the R&D indicators, the cost of data collection related to these studies (PREST/CRIC, 2006) and the broad platforms of interaction and communication in the network society (Castells, 2009) turn link analysis into an experimental indicator that can complement other proxies to assess the S&T capacity in a region through the mapping of the articulation of the knowledge infrastructure, the entrepreneurial role of the university and the degree of impact of the regional innovation strategy.

# Chapter 9: Conclusions

This doctoral thesis originated from the need to capture more socio-economic benefits from public science, and its main focus is to introduce a novel scientometric and webometric approach to investigate science parks (SP) as one of the most important mechanisms to enable research commercialisation and innovation. This approach examines two key aspects related to SPs' main mission: their capacity to foster greater research collaboration and interaction between academia and industry to exploit the underlying research base, and their networking ability to bring together the diverse organisations that operate in innovation systems and hence to facilitate open innovation.

This project seeks to extend the application of webometric and bibliometric indicators beyond traditional library and information science research areas and introduce new indicators to uncover new dimensions of SPs. This approach led to the employment of bibliometric indicators to investigate research within SPs. It also led to the development of a new method to detect and analyse interconnections within a R&D community on the web and to the design of a framework for analysing the web-based structure of R&D networks. In addition, it provides some empirical evidence about the development and main characteristics of the research structure of the UK SP movement. It also provides evidence to justify the use of web links to design weak benchmarking indicators, and thus demonstrates that webometric and bibliometric approaches can jointly be applied to obtain a deeper understanding of the R&D structures associated with science parks. This project can, therefore, contribute to the development of both knowledge domains, and is guided by the following key research questions:

(1) What insights can scientometric methods give into the role of public science and HEIs in the structure of R&D networks associated with SPs?
(2) Can web-based patterns reflect the configuration of R&D support infrastructures associated with SPs?

## 9.1 Answers to the research questions

In response to the first research question, the first three empirical chapters of this thesis show that bibliographic data can provide interesting information regarding the degree of research activities associated with the UK SP movement, the role of HEIs in the production of these research activities, and the extent to which the UK SP movement can benefit from HEI performance in terms of R&T production and third stream activities.

*The map of the R&D activities of the UK SP movement* (Chapter 4) *shows that:*

- R&D production is highly concentrated in terms of infrastructures and regions, and on-park firms strongly collaborate with academic partners beyond their local region. This is evidence against the success of two of the main goals of SPs: the regeneration of socioeconomically depressed areas, and the physical proximity between HEIs and industry.
- Three major agglomerations (East of England, South East and Scotland) concentrate the highest proportion of infrastructures and R&D production, and have distinct areas of research: (1) a public science-based specialism in food-biotechnology and bio-pharmacology, (2) a private science-based specialism in physics and material engineering, (3) a public science-based specialism in the agro-biotech sector.
- Science parks and research parks are more successful infrastructures in fostering cooperation and research production than science & innovation centres, business parks, technology parks, research campuses, incubators, innovation parks, and industrial parks.
- HEIs are the main external source of knowledge and collaboration for on-park firms.
- The research quality of publications authored on SPs is significantly higher than the average for the research areas, although its impact is not significantly higher than the national average.

*The structure of the R&D networks* (Chapter 5) *shows that:*

- HEI quality levels significantly associate with their degree of interaction with on-park firms: higher quality HEIs interact more with on-park firms.
- HEI central positions in the collaboration networks suggest that HEIs are the main source of external knowledge and competence for on-park firms.
- Inter-organisational collaboration produces a very limited share of on-park research production.

*The impact of SPs on R&T production and third stream activities* (Chapter 6) *shows that:*

- SPs' limited impact on the academic sphere indicates that SPs might not be the right policy tools to create U-I synergy across the country.
- HEIs' formal associations with parks do not strengthen their R&T capacity or U-I synergy.
- Research parks, research campuses and science parks are more likely to foster prompt publications and collaborations in comparison with infrastructures, such as, incubators, technology parks, science & innovation centres. Overall, newer parks are more likely to generate synergy quickly.

- R&T capacity is associated with factors involving U-I research collaboration, contracts, and spin-offs.
- HEIs' third stream activities resulting from a strong research base, such as income from U-I collaborations and contract research and the number of active spin-offs and patents, strongly associate with the academic production of publications and patents.

Overall, bibliographic-based patterns provide evidence of R&T activities that are widely accepted for the evaluation of research intensive organisations and give insights into R&D output, the diversity of the infrastructures which form the UK SP movement, and the inter-institutional collaboration that may take place on SPs. However the scarce R&D activities across the movement, excluding three main regional agglomerations characterised by driven-research industries, suggest that a systematic scientometric assessment could only be relevant for assessing a small part of the UK SP movement. This makes it necessary to conduct further studies on the research production of SPs across Europe and other developed countries. Only more studies at different scales can provide a fuller understanding of the potential application of this quantitative approach, and thereby assess whether it can be added to the battery of socioeconomic indicators used to evaluate SPs.

SPs are policy instruments oriented to reduce barriers to U-I collaboration and to create favourable conditions to boost technological innovation, and ultimately to promote socio-economic growth and sometimes also to regenerate economically disadvantaged areas. However, this thesis has shown that in regions with a limited mass of R&T the establishment of SPs should be re-thought since they are only able to promote U-I linkages in dynamic areas with an underlying accumulation of knowledge and expertise. Similarly, SPs' functions as mechanisms to bridge the gap between academia and industry to create favourable conditions and to strengthen information flows, mutual exchanges of ideas, innovation, and research commercialization are also called into question. This is because only a quarter of the total on-park publication output associated with the movement is the result of inter-institutional collaboration, and more than half of this is university and industry collaboration. Policies oriented to promote and support the development of SPs across the country should concentrate on establishing operational partnerships with top research institutions which have the knowledge and expertise to attract and collaborate with businesses, and support the development of research intensive industries. SPs may be the right instruments to capitalize the academic R&T and facilitate the U-I cooperation but only in research-intensive regions with already substantial signs of dynamism, as the quality of the regional environment seems to significantly affect the commercialisation of academic research (Casper, 2013).

On the other hand, the role of HEIs, especially top institutions, in the development of the UK SP movement is central as these knowledge producers are the main off-park partners and form the core of the U-I collaboration network. U-I collaboration also has significant associations between research excellence and stronger interactions with industry. Thus, the regions with higher rates of R&T activities associated with SPs also stand out because of hosting world-class institutions and comprehensive support structures that are likely to foster competitiveness and dynamism. This suggests that efforts to promote U-I linkages should take into account the quality of academic research, experience in transferring academic knowledge to industry and external factors in the localities and regions. Another interesting aspect that should be considered is how the gap concerning I-U links could be reduced. In this regard, the partnerships and third stream activities that are more closely related to research production should be strongly promoted to overcome the barriers and problems with U-I links. In the network society, HEIs are expected to support localities to tackle the difficult task of acting locally but engaging nationally and internationally (Witty, 2013: 6). However, only a more realistic U-I strategy, which also takes into account the nature of the academic research community, may help HEIs to have a more visible impact on local research-intensive industries and produce more balanced and robust economic development.

In response to the second research question, the main findings in the last two empirical chapters of this thesis suggest the extent to which web-based patterns might reflect the configuration of R&D support infrastructures associated with SPs.

*The structure of the web-based networks* (Chapter 7) *shows that:*

- Link analysis is not a reliable method for assessment exercises despite the new method introduced reducing the biases inherent in webometric studies.
- Interlinking-networks reflect to some extent the offline strengths and weaknesses of the policy measures and conditions described by surveys and reports.
- Industry-university-government collaboration has the University as its main interface.
- The core of the network is formed by research institutions, support structure organizations, and business developers. This central infrastructure is supported by the strategic partnership between higher education and public institutions, which are the main driving forces.
- Universities use SPs as a quasi-academic platform to commercialize their research and technology.
- The periphery of the network is formed by firms and governmental organizations.

*The SP actor framework applied to a web-based network* (Chapter 8) *shows that:*

- It is possible to identify the presence and potential behaviour of the institutional agents involved in R&D networks.
- Actors with the expected types of links are:
  (1) Universities that link all the different types of actors.
  (2) Government agencies that provide the financial resources to promote the university-industry collaboration.
  (3) Investors that collaborate with the university, public business supporters, and companies.
- Actors with fewer than the expected types of links are:
  (1) Consulting organizations that are only connected with universities.
  (2) Knowledge-based and serviced-based firms that occupy a peripheral position and are mainly connected to either universities or the business developers, respectively.

Despite the important limitations of this approach, it represents an interesting way to obtain a first and broad overview of the indirect impact of the intermediary organisations involved in SPs, as shown by previous studies (Howells, 2006; Phan et al., 2005). Therefore, this approach helps to some extent to tackle the problem of collecting quantitative data to analyse innovations networks. The increasing importance of the internet makes this method a promising approach for future studies of SPs. As Castells (2009) states, cyberspace allows the analysis of different institutional networks that coexist in the digital dimension and are interconnected through hyperlinks, building global digital networks of interactions, which transcend territorial and institutional boundaries. Thus, web-based patterns might provide a broad overview of the relational structure generated by groups of organisations that intend to reduce the gaps between institutional sectors. However, webometric methods and evidence still have to be considered as exploratory and cannot be employed for assessments and decision making processes.

Overall, the findings of the scientometric and webometric approach in this thesis are characterised by the narrow and formal evidence of the former approach, and the very broad and unreliable evidence of the latter approach. This difference, however, makes it possible to gain a better understanding of the multi-level and complex interactions embedded in innovation systems. Thus, an informetric approach represents an appealing analytical tool to be added to the interdisciplinary toolbox that is applied to understand innovation networks (Ozman, 2009), especially when quantitative evidence is necessary to back up diverse and rich qualitative evidence regarding innovation support infrastructures. Both approaches help to investigate two of the most important aspects in relation to SP missions: bridge-building between academia and industry, and bringing together the diverse organisations that operate in innovation systems to foster a dynamic and collaborative network. The main advantage of applying these two approaches is the ability to identify different types of actors involved and to uncover the network established by the set of broad interactions between them. The analysis of the effect of this social capital is important to better understand the role of underlying infrastructures to create value in innovative systems.

## 9.2 Limitations

As with most studies that primarily rely on only one type of indicator, the results obtained from an informetric approach are also likely to be limited and therefore, the results of this contribution must be interpreted with caution. The structured data used here to analyse R&D production is not able to provide a complete picture of the success of SPs and the U-I relationship, but rather a first insight to understand the R&D activities within SPs and the relationships between HEIs and SPs. Another important limitation concerning the method applied to collect the bibliographic data is that the data does not cover all the research generated within the SP movement because not all on-park organisations mention the name of the SPs where they are based as part of their affiliation address, and some research might not be published in Scopus journals, or may not be published at all. Similarly, the inherent limitations of webometric approaches make it difficult to consider the results as conclusive. A clear limitation of the webometric method applied is that it only takes into account the hyperlinks between the websites of the organisations linked by the SPs' websites. However, the inter-connections between these organisations probably do not reflect all the potential relationships established between them, nor all the relations fostered by SPs, so the results should be taken as illustrative rather than exhaustive.

## 9.3 Publications and Hyperlinks as measurement tools

Scientific publications represent the production of the public good or knowledge, and as Paula Stephan (2012:25p.) states, "*Scientists are motivated to do research by a desire to establish priority of discovery. But the only way that this can be done—that a scientist can establish ownership of an idea—is by given the idea away. ... The interest in priority motivates scientists to produce and share knowledge in a timely fashion.*" This priority system encourages scientists to make the research findings their own and at the same time creates a reward system that encourages the production and sharing of knowledge that leads to professional reputation and financial rewards (Whitley, 2000). The central role of knowledge sharing  in scientific communication as meant to reach greater diffusion and social impact is argued to be the main barrier for private researchers—who operate within market transactions, values and norms—to actively contribute to the development of public science (Bruneel et al., 2010; Godin, 1996).

Although the growing contextualisation of scientific research has led to a growing interaction with industry in certain fields, and the Mertonian view of science may underestimate the diversity of organisations and institutional orders within international higher education and public R&D sectors (Perkmann et al., 2013), publications are considered as an unreliable proxy to describe the research activity of private firms that need to be complemented with quantitative and qualitative evidence (Rafols et al., 2012). This limitation is related to the difference between industrial sectors and the strategic reasons for industry to openly publish, such as, (1) to send signals to stakeholders and attract investors, (2) to signal one's competencies and capabilities, (3) to defend against others' attempts to control particular areas of technology, (4) to attract customers and gain reputation (Cockburn & Henderson, 1998b; Hicks, 1995; Nelson, 1990; Rosenberg, 1990), (5) to market products in regulatory arenas and among practitioners (Polidoro & Theeke, 2012; Sismondo, 2009; Smith, 2005). On the other hand, industry also contributes to public science as a strategy to (1) effectively plug into the scientific community to strengthen their actively learning, absorptive capacity and (2) gain a first-mover advantage to access to new knowledge, human resources, and informal advice (Cockburn & Henderson, 1998b; Hicks, 1995). Other reasons may also be: (1) unintentional publishing, and (2) basic research as long-term investment (Rosenberg, 1990). However, the constant increase of publications produced by private researchers and the exponential growth of U-I collaboration in the last decade, as found in this study, makes it necessary to undertake further studies to gain a better grasp of the new reasons triggering this increase of private contribution to public science and U-I collaboration (Calero et al., 2007; Godin & Gingras, 2000; Godin, 1996; Perianes-Rodríguez, et al., 2011; Rafols, et al., 2012; R. J. W. Tijssen, 2009; R. Tijssen, 2012), and the influence of public research on industry (Arundel & Geuna, 2004; Cohen et al., 2002; Toole, 2012). This is important as the contextualisation and socialisation of sciences is stimulated  on one hand by the need of academia to actively contribute to a socio-economic development and generate external incomes to fulfil its third mission (Etzkowitz, et al., 2000), and on the other hand by the need of industry to access to an external knowledge base to strengthen the learning capability and the potential to innovate and remain competitive (Chesbrough, 2003).

In this study, the comparison of publications with other knowledge transfer and commercialisation activities confirms that publications, within research and technology intensive environments, can be considered as a research engagement activity along with joint research projects, research contracts and consultancies. These activities are aligned with knowledge-based collaborations and traditional research activities that pursue external resources to support academics' research agenda, which is primarily individually driven and differs from the academic commercialisation activities (Perkmann et al., 2013). However, the fact that U-I co-authorship is likely to be related to academic engagement rather than academic entrepreneurship does not reveal to what extent firms agree publishing to allow scientists to gain reputation and academic benefits. Therefore, scientific publications are considered as only the tip of the iceberg of U-I interaction, and it may represent one of the tools to track down the new development taking place. To better understand the rationale behind U-I collaboration and private research dissemination, it is still necessary to understand the determinants in this research process and the tangible and intangible outputs that leads to successful knowledge transfer, as well as the reasons to partially disseminate it in form of scientific publications. This makes it necessary to investigate what publications are representative in terms of knowledge transfer and research commercialisation for the private and academic sector.

Hyperlinks, on the other hand, are reflections of on- and off-line social interactions and their meaning varies according to the social context and relationship of the actors studied. The motivations and meanings behind hyperlinks found in other contexts (Bar-Ilan, 2005b; Harries, et al., 2004; Stuart, Thelwall, & Harries, 2007; Thelwall, 2003b; Vaughan, Gao, & Kipp, 2006; Wilkinson, et al., 2003) differ from the wide range of organisations and potential interactions established between them in a research and technology intensive environment. Chapter 7 and 8 shows that hyperlinks associated with support innovation infrastructures mostly reflect formal relationship between the different organisations involved, for example: the location of a firm in a support infrastructure, partnership between faculties with NGOs, partnerships between universities with intermediary organisations, the academic roots of spin-offs, or the support of intermediaries to on-park firms. The main goals and functions of the organisations interconnected can help to shed light of the motivations behind hyperlinks, as shown by the framework introduced in Chapter 8 and other studies (Kenekayoro et al., 2013), while a manual check of the web-pages interconnected can help to confirm the relationship. Nevertheless, this approach does not allow estimating the degree of importance and connectivity between the actors, leading to evidence that is difficult to quantify, and thus should be only used as illustrative.

### 9.4 Application for evaluation purposes

Despite the fact that the webometric approach is far from being considered indisputable and verifiable empirical information, and that further research is needed to understand the process involved around U-I co-authorships, the quantitative approach introduced here to study R&D activities and networking structures associated with support infrastructures can still provide efficient solutions for evaluation purposes. The two main advantages are the large scale and limited need for resources to be implemented, in terms of time, funding and human capital. Nowadays, the concept of economic efficiency is becoming more and more important, and the allocation of resources is crucial. In this way, the group of performance indicators, as measurements that simplify a phenomenon, should be oriented to provide the most comprehensive assessment of particular activities.

Therefore, this informetric approach represents an interesting methodology to gain a first insight into the research and networking performance of support infrastructures at local, regional, national and international level, as well as systematic monitoring and cross-institutional benchmarking. It may be applied as a first step in the evaluation process to identify and understand the population to be studied, and uncover interesting patterns that should be taken into account in the design and application of further quantitative and qualitative datasets and methods such as surveys or interviews. The application of qualitative methods is recommended to collect in-depth information that helps to validate and interpret the results. Only the application of this approach in combination with qualitative information and other quantitative indicators that measure the performance of the wide range of activities related to research commercialisation will make it possible to support the decision making process and in-depth monitoring.

### 9.5 Further research

The comprehensive study of the UK SP movement in this thesis shows that an informetric approach may uncover dimensions in the assessment of support infrastructures that are still hidden. However, further investigations into the webometric approach are still necessary because of the pilot phases of the webometric studies conducted here and the poor reliability of this type of unstructured data. Going forward, the most interesting goals that could be explored by means of web-based methods in relation to the networking nature of SPs are:

- To uncover the inter-networking dynamics between SPs, operating through internal networks among SPs and actors from different SPs in the region. This might provide an overview of the role of SPs in the promotion of regional systems of innovation.
- To uncover the external-networking dynamics operating through external networks (off-park organisations) with the on-park actors of the SPs. This might make it possible to represent the degree of integration of the SPs in the local and international arena. In addition, existing methods could be adapted to reduce the multidimensionality of large web-based networks and detect external community members.

- To further develop and refine the methodology, and to expand the analysis of SPs based in other vibrant regions in the UK, like East and South East England, as well as other SP movements across Europe and other developed countries. This would provide benchmarking standards for other SPs, guiding structural changes for more effective SP management and support of start-ups.
- To change the level of analysis from SPs to clusters to gain a deeper understanding of regional innovation systems.
- To explore the use of Twitter among supporting organisations and firms, and the degree of impact of this social media communication channel between both sets of organisations.

Overall, despite the fact that webometric techniques are generally speculative and exploratory, a web-based approach could potentially be used to systematically monitor and further investigate to what extent: (1) national and regional policies that support the growth of research-intensive industries lead to major structural changes over time; and (2) which inter-connections and structural positions of university spin-offs and start-ups associate with better performance and a higher chance to survive. Similarly, the structural position occupied by on-park firms in research collaboration networks can be used to investigate whether central positions, roles or strong links with knowledge producers might be associated with better performance and survival rates than similar off-park counterparts.

In terms of the scientometric approach, there are also interesting lines of research beyond the scope of this thesis that are worth pursuing.
- *Identifying research-intensive clusters*: the mapping of the SP movement at the national level only uncovers a small part of the exploitation of public science and how research-intensive industries might be fostered and emerge. It is, therefore, still necessary to undertake comparative and retrospective studies at regional levels that allow determining to what extent different localised research- and technology-agglomerations associated with SP movements might be able to underpin the development of science-based clusters. This will provide empirical evidence about how academic research and its ties with private research might be able to underpin the development of science-based clusters and then, systematically monitor their emergence and development. Here public science is considered as a fundamental socio-economic asset.
- *Industrial and technological classification scheme*: the increasing demand for effective knowledge transfer and the growth of R&T activities in certain industrial sectors suggests the use of scientometric indicators to the study of not only scholarly communities and public science. Consequently, it is necessary with a systematic effort to gain a better understanding of the private research community and the design of new scientometric indicators and tools to understand and asses it. A classification scheme is therefore needed that organises and delineates research production based on industrial and technological sectors rather than on scientific fields (Witty, 2013). The capacity to map the research production based on the characteristics of the public and private research communities and monitor the research base that enables technologies could generate a better understanding of the socialisation of science (Technology Strategy Board, 2012), the potential role of scientific research in underpinning innovation across different economic sectors and markets.

- *Re-thinking the scientometric assessment of knowledge transfer to industry:* the still limited evidence from scientometric indicators on the impact of academic research on innovation and socio-economic benefit may be due to the narrow and inadequate perspective taken to study this phenomenon (Caraça et al., 2009; Cooke, 2005b). The expectations about the central role of knowledge producers in the production and innovation system, based on a linear model of innovation like Mode 2 and the Triple Helix model (Etzkowitz, 2006, 2008; Nowotny et al., 2001), should be extended by the study of the learning ability and competence building process (Enkel, Gassmann, & Chesbrough, 2009; Lundvall, et al., 2002; Nonaka, 1997, 2006; Wenger, 2000). Therefore, to better understand the role of the University in boosting innovation and supporting the development of dynamic and competent environments, it is also necessary to undertake studies into how academic research, as extramural knowledge for firms, is absorbed, transformed, and learned within productive and innovative spaces. Only a deeper understanding of the impact of public science on the institutional learning process in a particular location and how it stimulates regional knowledge capabilities will provide a fuller understanding of public science as source of innovation and the development of the knowledge production system.

These topics are in line with the growing importance for further developing methods to investigate the central role of the research base, knowledge transfer to industry, and networking capabilities to promote sustainable socio-economic development in a globalised economy. It can help, for example, to monitor the British government's long-term strategy to stimulate innovation and growth through the rapid development of a national network of new elite technology and innovation centres, named "Catapults", which are designed to focus on a specific area of technology and expertise, and that bring together the business and research communities to rapidly transform great research into commercial successes (Technology Strategy Board, 2013).

**Appendix 1 - Illustration of potential collaborations between on-park and off-park organisations**

| Types of collaboration based on the physical location of on- and off-park organisations | | |
|---|---|---|
| **Organisational co-authorship** | **Organisational collaborations** | **Total physical collaborations** |
| **Paper A:** Firm a; Firm b<br>*Address:* New Frontier Science Park;<br>Granta Park | Firm a & Firm b = On-park / On-park<br>Firm b & Firm a = On-park / On-park | 2 = On-park / On-park |
| **Paper B:** Firm a; Firm c; HEI a<br>*Address:* New Frontier Science Park;<br>off-park; off-park | Firm a & Firm c = On-park / off-park<br>Firm a & HEI a = On-park / off-park<br>Firm c & Firm a = Off-park / on-park<br>Firm c & HEI a = Off-park / off-park<br>HEI a & Firm a = Off-park / on-park<br>HEI a & Firm c = Off-park / off-park | 2 = On-park / Off-park<br>2 = Off-park / On-park<br>2 = Off-park / Off-park |
| **Paper C:** RI a; HEI a; HEI b; Firm c<br>*Address:* Norwich Research Park;<br>off-park; off-park; off-park | RI a & HEI a = On-park / off-park<br>RI a & HEI b = On-park / off-park<br>RI a & Firm c = On-park / off-park<br>HEI a & RI a = Off-park / on-park<br>HEI a & HEI b = Off-park / off-park<br>HEI a & Firm c = Off-park / off-park<br>HEI b & RI a = Off-park / on-park<br>HEI b & HEI a = off-park / off-park<br>HEI b & Firm c = off-park / off-park<br>Firm c & RI a = off-park / on-park<br>Firm c & HEI a = off-park / off-park<br>Firm c & HEI b = off-park / off-park | 3 = On-park / Off-park<br>3 = Off-park / On-park<br>6 = Off-park / Off-park |